%% file: Crawford14.tex
\documentclass{emulateapj}
\usepackage{amsmath,natbib,graphicx}
\usepackage{epsf}
\input{epsf}
\usepackage{epstopdf}
\usepackage{multirow}
\usepackage{subfigure}
\usepackage{epsfig}
\usepackage{color}
\usepackage{lscape}
\usepackage{latexsym}
\usepackage{appendix}
\usepackage{threeparttable}
\usepackage{ifthen}
\usepackage{sistyle}
\SIthousandsep{,}

\DeclareGraphicsExtensions{.jpg,.pdf,.png,.eps,.ps}
\graphicspath{{FIGURES/}}

\newcommand{\degs}{deg$^2$}
\newcommand{\beq}{\begin{equation}}
\newcommand{\eeq}{\end{equation}}
\newcommand{\simleq}{{\raise.0ex\hbox{$\mathchar"013C$}\mkern-14mu \lower1.2ex\hbox{$\mathchar"0218$}}}
\newcommand{\simgeq}{{\raise.0ex\hbox{$\mathchar"013E$}\mkern-14mu \lower1.2ex\hbox{$\mathchar"0218$}}}
\newcommand{\tcmb}{\ensuremath{T_\mathrm{CMB}}}
\newcommand{\muk}{\ensuremath{\mu \mathrm{K}}}
\newcommand{\dommap}{\ensuremath{\Delta \Omega_\mathrm{map}}}
\newcommand{\dom}{\ensuremath{\Delta \Omega}}

\newcommand{\bl}{\ensuremath{\mathbf{l}}}
\newcommand{\bispv}{\ensuremath{B(\bl_1,\bl_2,\bl_3)}}
\newcommand{\bisp}{\ensuremath{B(l_1,l_2,l_3)}}
\newcommand{\wisp}{\ensuremath{W(l_1,l_2,l_3)}}
\newcommand{\bispe}{\ensuremath{\hat{B}(l_1,l_2,l_3)}}
\newcommand{\bispenu}{\ensuremath{\hat{B}(l_1,l_2,l_3,\nu)}}
\newcommand{\nl}{\ensuremath{N_{l_1,l_2,l_3;\Delta l}}}
\newcommand{\nlinv}{\ensuremath{\frac{1}{\nl}}}
\newcommand{\nlinvsh}{\ensuremath{\frac{1}{N_l}}}
\newcommand{\sumsh}{\ensuremath{\sum_l}}
\newcommand{\lrad}{\ensuremath{l_\mathrm{rad}}}
\newcommand{\blrad}{\ensuremath{\hat{B}(l_\mathrm{rad})}}

\newcommand{\bx}{\ensuremath{\mathbf{x}}}

\newcommand{\fnl}{\ensuremath{f_\mathrm{NL}}}
\newcommand{\wmap}{\textit{WMAP}} 
\newcommand{\btsz}{\ensuremath{B_\mathrm{\tiny{tSZ}}}}
\newcommand{\atsz}{\ensuremath{A_\mathrm{\tiny{tSZ}}}}
\newcommand{\aksz}{\ensuremath{A_\mathrm{\tiny{kSZ}}}}
\newcommand{\ttsz}{\ensuremath{T_\mathrm{\tiny{tSZ}}}}

\newcommand{\tpeak}{\ensuremath{T_\mathrm{peak}}}

\newcommand{\bclust}{\ensuremath{B_\mathrm{\tiny{clust}}}}
\newcommand{\acib}{\ensuremath{\alpha_\mathrm{\tiny{c}}}}
\newcommand{\mtwoh}{\ensuremath{M_{200}(\rho_\mathrm{crit})}}
\newcommand{\bigmass}{\ensuremath{8 \times 10^{14} M_\odot / h}}
\newcommand{\threshmass}{\ensuremath{3 \times 10^{14} M_\odot / h}}
\newcommand{\chisq}{\ensuremath{\chi^2}}
\newcommand{\dchisq}{\ensuremath{\Delta \chi^2}}
\newcommand{\chired}{\ensuremath{\chi^2_\mathrm{red}}}

\newcommand{\sigeight}{\ensuremath{\sigma_8}}

\def\sigeightmeas{0.787}
\def\sigeightunc{0.031}

\def\tsztemplnomask{-9.8}
\def\tsztemplhugemask{-7.5}
\def\tsztempldetmask{-2.4}
\def\tszmeasnomask{-5.2}
\def\tszmeashugemask{-4.4}
\def\tszmeasdetmask{-1.7}
\def\tszlradtemplnomask{-11.1}
\def\tszlradtemplhugemask{-8.3}
\def\tszlradtempldetmask{-2.6}
\def\tszlradmeasnomask{-5.9}
\def\tszlradmeashugemask{-4.9}
\def\tszlradmeasdetmask{-1.8}
\def\tszskewtemplspt{-53.3}
\def\tszskewtemplact{-64.6}
\def\tszskewmeasact{-34.4}
\def\dtszskewmeasact{3.3}
\def\dtszskewmeasactsv{7.7}
\def\cibmeasnomask{3.21}
\def\dcibmeasnomask{0.68}
\def\cibmeashugemask{3.51}
\def\dcibmeashugemask{0.70}

\def\atszbtsz{2.96}
\def\datszbtsznom{0.64}
\def\datszbtszbig{0.77}
\def\atszbtszrt{3.08}
\def\datszbtszrtnom{0.56}
\def\datszbtszrtbig{0.63}
\def\akszbtszrt{2.9}
\def\akszbtszrtnp{2.6}
\def\dakszbtszrtnom{1.6}

\def\dakszbtszrtnp{1.8}
\def\akszninefivebtszrtnom{5.6}

\def\datszrt{1.05}

\def\akszrtnp{0.3}
\def\dakszrtnp{3.3}
\def\akszninefive{6.7}
\def\datszbtszprojnom{0.4}
\def\dakszbtszprojnom{1.0}

\def\KICPChicago{1}
\def\AAUChicago{2}
\def\EFIChicago{3}
\def\ArtInstChicago{4}
\def\ArgonneHEP{5}
\def\UChicago{6}
\def\PhysicsUChicago{7}
\def\Argonne{8}
\def\NIST{9}
\def\McGill{10}
\def\Berkeley{11}
\def\Colorado{12}
\def\Davis{13}
\def\LBNL{14}
\def\Caltech{15}
\def\Michigan{16}
\def\Munich{17}
\def\ExcellenceCluster{18}
\def\MPE{19}
\def\CaseWestern{20}
\def\Minnesota{21}
\def\CfA{22}
\def\BCCP{23}

\begin{document}

\title{A measurement of the secondary-CMB and millimeter-wave-foreground bispectrum using 800 square degrees of South Pole Telescope data}

\shorttitle{SPT bispectrum measurements}
\shortauthors{Crawford et al.}

\author{
  T.~M.~Crawford\altaffilmark{\KICPChicago,\AAUChicago},
  K.~K.~Schaffer\altaffilmark{\KICPChicago,\EFIChicago,\ArtInstChicago},
  S.~Bhattacharya\altaffilmark{\KICPChicago,\ArgonneHEP},
  K.~A.~Aird\altaffilmark{\UChicago},
  B.~A.~Benson\altaffilmark{\KICPChicago,\EFIChicago},
  L.~E.~Bleem\altaffilmark{\KICPChicago,\PhysicsUChicago},
  J.~E.~Carlstrom\altaffilmark{\KICPChicago,\AAUChicago,\EFIChicago,\PhysicsUChicago,\Argonne},
  C.~L.~Chang\altaffilmark{\KICPChicago,\EFIChicago,\Argonne},
  H-M.~Cho\altaffilmark{\NIST},
  A.~T.~Crites\altaffilmark{\KICPChicago,\AAUChicago},
  T.~de~Haan\altaffilmark{\McGill},
  M.~A.~Dobbs\altaffilmark{\McGill},
  J.~Dudley\altaffilmark{\McGill},
  E.~M.~George\altaffilmark{\Berkeley},
  N.~W.~Halverson\altaffilmark{\Colorado},
  G.~P.~Holder\altaffilmark{\McGill},
  W.~L.~Holzapfel\altaffilmark{\Berkeley},
  S.~Hoover\altaffilmark{\KICPChicago,\PhysicsUChicago},
  Z.~Hou\altaffilmark{\Davis},
  J.~D.~Hrubes\altaffilmark{\UChicago},
  R.~Keisler\altaffilmark{\KICPChicago,\PhysicsUChicago},
  L.~Knox\altaffilmark{\Davis},
  A.~T.~Lee\altaffilmark{\Berkeley,\LBNL},
  E.~M.~Leitch\altaffilmark{\KICPChicago,\AAUChicago},
  M.~Lueker\altaffilmark{\Caltech},
  D.~Luong-Van\altaffilmark{\UChicago},
  J.~J.~McMahon\altaffilmark{\Michigan},
  J.~Mehl\altaffilmark{\KICPChicago,\Argonne},
  S.~S.~Meyer\altaffilmark{\KICPChicago,\AAUChicago,\EFIChicago,\PhysicsUChicago},
  M.~Millea\altaffilmark{\Davis},
  L.~M.~Mocanu\altaffilmark{\KICPChicago,\AAUChicago},
  J.~J.~Mohr\altaffilmark{\Munich,\ExcellenceCluster,\MPE},
  T.~E.~Montroy\altaffilmark{\CaseWestern},
  S.~Padin\altaffilmark{\KICPChicago,\AAUChicago,\Caltech},
  T.~Plagge\altaffilmark{\KICPChicago,\AAUChicago},
  C.~Pryke\altaffilmark{\Minnesota},
  C.~L.~Reichardt\altaffilmark{\Berkeley},
  J.~E.~Ruhl\altaffilmark{\CaseWestern},
  J.~T.~Sayre\altaffilmark{\CaseWestern},
  L.~Shaw\altaffilmark{\McGill},
  E.~Shirokoff\altaffilmark{\Berkeley}, 
  H.~G.~Spieler\altaffilmark{\LBNL},
  Z.~Staniszewski\altaffilmark{\CaseWestern},
  A.~A.~Stark\altaffilmark{\CfA},
  K.~T.~Story\altaffilmark{\KICPChicago,\PhysicsUChicago},
  A.~van~Engelen\altaffilmark{\McGill},
  K.~Vanderlinde\altaffilmark{\McGill},
  J.~D.~Vieira\altaffilmark{\Caltech},
  R.~Williamson\altaffilmark{\KICPChicago,\AAUChicago}, and
  O.~Zahn\altaffilmark{\BCCP}
}

\altaffiltext{\KICPChicago}{Kavli Institute for Cosmological Physics, University of Chicago, Chicago, IL, USA 60637}
\altaffiltext{\AAUChicago}{Department of Astronomy and Astrophysics, University of Chicago, Chicago, IL, USA 60637}
\altaffiltext{\EFIChicago}{Enrico Fermi Institute, University of Chicago, Chicago, IL, USA 60637}
\altaffiltext{\ArtInstChicago}{Liberal Arts Department, School of the Art Institute of Chicago, Chicago, IL, USA 60603}
\altaffiltext{\ArgonneHEP}{High Energy Physics Division, Argonne National Laboratory, Argonne, IL, USA 60439}
\altaffiltext{\UChicago}{University of Chicago, Chicago, IL, USA 60637}
\altaffiltext{\PhysicsUChicago}{Department of Physics, University of Chicago, Chicago, IL, USA 60637}
\altaffiltext{\Argonne}{Argonne National Laboratory, Argonne, IL, USA 60439}
\altaffiltext{\NIST}{NIST Quantum Devices Group, Boulder, CO, USA 80305}
\altaffiltext{\McGill}{Department of Physics, McGill University, Montreal, Quebec H3A 2T8, Canada}
\altaffiltext{\Berkeley}{Department of Physics, University of California, Berkeley, CA, USA 94720}
\altaffiltext{\Colorado}{Department of Astrophysical and Planetary Sciences and Department of Physics, University of Colorado, Boulder, CO, USA 80309}
\altaffiltext{\Davis}{Department of Physics, University of California, Davis, CA, USA 95616}
\altaffiltext{\LBNL}{Physics Division, Lawrence Berkeley National Laboratory, Berkeley, CA, USA 94720}
\altaffiltext{\Caltech}{California Institute of Technology, Pasadena, CA, USA 91125}
\altaffiltext{\Michigan}{Department of Physics, University of Michigan, Ann  Arbor, MI, USA 48109}
\altaffiltext{\Munich}{Department of Physics, Ludwig-Maximilians-Universit\"{a}t, 81679 M\"{u}nchen, Germany}
\altaffiltext{\ExcellenceCluster}{Excellence Cluster Universe, 85748 Garching, Germany}
\altaffiltext{\MPE}{Max-Planck-Institut f\"{u}r extraterrestrische Physik, 85748 Garching, Germany}
\altaffiltext{\CaseWestern}{Physics Department, Center for Education and Research in Cosmology and Astrophysics, Case Western Reserve University,Cleveland, OH, USA 44106}
\altaffiltext{\Minnesota}{Department of Physics, University of Minnesota, Minneapolis, MN, USA 55455}
\altaffiltext{\CfA}{Harvard-Smithsonian Center for Astrophysics, Cambridge, MA, USA 02138}
\altaffiltext{\BCCP}{Berkeley Center for Cosmological Physics, Department of Physics, University of California, and Lawrence Berkeley National Laboratory, Berkeley, CA, USA 94720}

\email{tcrawfor@kicp.uchicago.edu}

\slugcomment{Published in The Astrophysical Journal: 784 (2014) 143}

\begin{abstract}

We present a measurement of the angular bispectrum of the millimeter-wave sky in observing
bands centered at roughly 95, 150, and 220~GHz, 
on angular scales of $1^\prime \lesssim \theta \lesssim 10^\prime$ (multipole number 
$1000 \lesssim l \lesssim \num{10000}$). At these frequencies and angular scales, the main contributions 
to the bispectrum are expected to be the thermal Sunyaev-Zel'dovich (tSZ) 
effect and emission from extragalactic sources, predominantly dusty, star-forming galaxies
(DSFGs) and active galactic nuclei.
We measure the bispectrum in 800~\degs\ of three-band South Pole Telescope data, and 
we use a multi-frequency fitting procedure to separate the bispectrum of 
the tSZ effect from the 
extragalactic source contribution. We simultaneously 
detect the bispectrum of the tSZ effect at $>$10$\sigma$, 
the unclustered component of the extragalactic source bispectrum 
at $>$5$\sigma$
in each frequency band, and the bispectrum due to the 
clustering of DSFGs---i.e., the clustered
cosmic infrared background (CIB) bispectrum---at $>$5$\sigma$.
This is the first reported detection of the clustered CIB bispectrum.
We use the measured tSZ bispectrum amplitude, compared to model 
predictions, to constrain the normalization of the matter power spectrum 
to be $\sigeight = \sigeightmeas \pm \sigeightunc$ and to predict the amplitude of the 
tSZ power spectrum at $l = 3000$. This prediction improves our ability to separate
the thermal and kinematic contributions to the total SZ power spectrum. 
The addition of bispectrum data improves our constraint on the tSZ power spectrum 
amplitude by a factor of two compared to power spectrum measurements alone
and demonstrates a preference for a nonzero kinematic SZ (kSZ) power spectrum,
with a derived constraint on the kSZ amplitude at $l=3000$ of 
$\aksz = \akszbtszrt \pm \dakszbtszrtnom \ \muk^2$, 
or $\aksz = \akszbtszrtnp \pm \dakszbtszrtnp \ \muk^2$ if the default $\aksz > 0$ prior is removed.

\end{abstract}


\section{Introduction}
\label{sec:intro}

Measurements of the millimeter-wave sky are a rich source of cosmological information. 
Studies of the intensity of the cosmic microwave background (CMB) have provided much 
of the evidence for our current cosmological model \citep[e.g.,][]{komatsu11,hinshaw12}, and 
ever more sensitive 
and wide-ranging experiments (in terms of both sky coverage and range of angular scales 
probed) continue to improve our constraints on cosmological parameters 
\citep{reichardt09a,das11,keisler11,story13,hou14,sievers13}.  
Beyond revealing the properties of the primary CMB fluctuations (those generated at the
surface of last scattering), 
the high-resolution millimeter-wave sky maps generated by current experiments also enable the study of 
secondary anisotropies in the CMB---those due to interactions of the CMB photons with 
matter along the line of sight---as well as emissive foreground sources.  These secondary
signals also carry interesting cosmological information, probing different epochs of cosmic 
history \citep[e.g.,][]{lueker10,das11,reichardt12b}.

The primary CMB signal arising from fluctuations at the surface of last scattering is expected
to be very close to a Gaussian random field. The simplest models of inflation predict 
very small levels of non-Gaussianity \citep[e.g.,][]{acquaviva03}, and experimental results 
up to this point have been essentially consistent with these predictions \citep{komatsu09,komatsu11}.
Under the assumption of Gaussianity, all of the information about CMB intensity fluctuations 
is contained in the second moment of the distribution.  Hence, the power spectrum has historically
been the simplest and most useful statistic for characterizing CMB fluctuations and constraining 
cosmological models.  

Searches for non-Gaussianity in the primary CMB, as pointed out in, e.g., 
\citet{coles87}, have the potential to inform inflationary models.
These measurements involve constructing statistics that test the skewness of the 
maps, or, more generally, that depend on the three-point angular correlation function 
or its harmonic-space equivalent, the bispectrum. Such statistics can be tailored
for sensitivity to specific models for inflationary non-Gaussianity, which are then
parameterized with a single amplitude \fnl.  Several analyses of \wmap\ data have
produced conflicting constraints on the amplitude of a particular model---the so-called
``local model"---ranging from tentative
detections to limits consistent with $f_\mathrm{NL,local}=0$, but always with error
bars of $\delta \fnl \sim 20-30$ \citep[e.g.,][]{yadav08,komatsu11}. 
Analyses of higher-resolution CMB data over small patches of the sky have 
resulted in upper limits of order $f_\mathrm{NL,local} < 1000$ \citep[e.g.,][]{santos02,smith04}.
More precise constraints or measurements 
of \fnl\ will require a combination of large sky area and high 
sensitivity; limits from {\it Planck} are expected to be at the level of 
$\delta \fnl \sim 5$ \citep[e.g.,][]{yadav07}.  Progress will also depend on 
understanding the non-Gaussian behavior of secondary CMB anisotropies and foregrounds.

Secondary CMB anisotropies include gravitational lensing of the 
CMB \citep[e.g.,][]{hu00} and the thermal Sunyaev-Zel'dovich (SZ) effect \citep{sunyaev70}.
These signals are non-Gaussian in general, although they are often 
studied through their effect on the power spectrum. For example, detections of 
the effect of gravitational lensing in CMB data have often relied on the effect 
of acoustic peak smearing in the two-point function or power spectrum 
\citep[e.g.,][]{reichardt09a,das11,keisler11,story13,das13}, but more information can be extracted 
from CMB lensing when the four-point function is considered 
\citep[e.g.,][]{das11b,vanengelen12}.

The thermal SZ (tSZ) effect arises from the spectral distortion of the CMB through 
interactions with hot gas in galaxy clusters and provides an efficient means of
finding distant, massive clusters 
\citep[e.g.,][]{williamson11,planck11-5.1a,reichardt13,hasselfield13}. 
The tSZ power spectrum measures the mean squared signal from all clusters and
is a sensitive probe of the normalization of the 
matter power spectrum $\sigeight$,  with the tSZ power spectrum amplitude 
predicted to scale as $\sigeight^\gamma$, with $\gamma \sim 7$-$8$ 
\citep{komatsu02,reichardt12b}.
The tSZ power spectrum amplitude at few-arcminute scales ($l \sim 3000$)
has been constrained through recent measurements of the  
small-angular-scale power spectrum of the mm-wave sky
\citep[e.g.,][]{lueker10,das11,reichardt12b}. 
However, the tSZ power 
spectrum at these scales is dominated by high-redshift, low-mass 
clusters that are not well studied at other wavelengths \citep[e.g.,][]{holder02b}.
Significant modeling uncertainty for this
population of clusters complicates the translation of the measured tSZ power spectrum amplitude 
into a constraint on \sigeight\ \citep[e.g.,][]{lueker10}.  

An alternative approach to constraining \sigeight\ and
other parameters that affect the growth of  structure is to study just those galaxy 
clusters that can be individually detected in millimeter-wave maps through their tSZ signature. 
When redshifts are obtained for every cluster, this approach can constrain both
\sigeight\ and the equation of state of dark energy \citep[e.g.,][]{wang98,haiman01}.
The scaling of the observable (the number of clusters detected) with 
\sigeight\ is even steeper than for the power spectrum, with number counts 
going roughly as $\sigeight^{10}$ \citep[e.g.,][]{dudley12}.  Constraints based 
on number counts 
are nearly independent of those using the 
tSZ power spectrum, making the two probes nicely complementary.

The thermal SZ bispectrum offers another approach that complements both the power spectrum and 
cluster-detection methods.  As shown in \citet[][hereafter B12]{bhattacharya12}, the 
$l \sim 3000$
tSZ bispectrum is dominated by more massive, lower-redshift clusters than the tSZ power 
spectrum at similar angular scales. 
This population of clusters is subject to less modeling 
uncertainty than the higher-redshift, lower-mass clusters that dominate the 
tSZ power spectrum.  
Furthermore, the tSZ bispectrum is yet more sensitive to \sigeight\ 
than the tSZ power spectrum and cluster number counts. 
B12 demonstrated that the amplitude of the tSZ bispectrum at $l=3000$ scales 
as $\btsz \propto
\sigeight^{11-12}$.  \citet{hill13} demonstrated a similar scaling for
the real-space tSZ skewness $\langle \ttsz^3 \rangle$, and
\citet{wilson12} used this model prediction and a measurement of
the tSZ skewness in Atacama Cosmology Telescope (ACT) data to place a
$\simleq 5\%$ constraint on \sigeight. 

B12 also showed that by
simultaneously constraining cosmology and cluster physics with the tSZ
bispectrum, one could make a precise prediction for the amplitude of the
tSZ power spectrum. A measurement of the tSZ bispectrum provides new
constraints on intra-cluster gas physics and therefore acts as a
bridge between the very low-redshift, very massive clusters that
currently constrain gas models (mostly through X-ray observations) and
the very high-redshift, low-mass clusters that dominate the tSZ power
spectrum.

The tSZ 
contribution to the CMB power spectrum is difficult to separate from the contribution of the 
kinematic SZ (kSZ) effect \citep{sunyaev80} 
in current data \citep{reichardt12b}.
However, under the assumption that the contribution to the bispectrum 
from the kSZ is negligible (see
Section \ref{sec:sigmodels} for details), the bispectrum-based prediction 
for the tSZ power spectrum can be used
to sharpen the measurement of the kSZ power spectrum.
The resulting constraints on the kSZ effect---which are interesting for
cosmology and for models of reionization \citep[e.g.,][]{knox98,gruzinov98}---are potentially much
stronger than from the power spectrum alone.

Finally, the non-Gaussian signals from extragalactic emissive sources are also
potentially interesting.  Two populations of sources contribute
significantly to measurements at the frequencies at which CMB
experiments typically operate (roughly a few GHz to hundreds of GHz):
synchrotron-dominated ``radio sources''---primarily active galactic 
nuclei---and dusty star-forming galaxies
(DSFGs), the integrated light from which produces the cosmic infrared
background (CIB). Measurements of the bispectrum contribution from radio sources and
DSFGs can constrain source count models beyond the threshold for 
individually detecting sources, and with different flux weighting than measurements 
using the power spectrum.
Perhaps more intriguingly, measurements of the
bispectrum of the clustered CIB (fluctuations in the mean CIB emission
due to large-scale structure) have the potential to constrain models of
galaxy and star formation beyond what can be done with CIB power
spectrum measurements.

The bispectra of the secondary anisotropies and foregrounds are 
characterized by different angular scale dependence as well as different spectral 
signatures.  Multi-band measurements across a wide range of angular scales
can therefore be used to isolate the 
different contributions.  The purpose of this work is to present bispectrum measurements 
in three frequency bands centered at roughly 95, 150, and 220~GHz, 
(corresponding to wavelengths
of $\sim$3.2, $\sim$2.0, and $\sim$1.4~mm), using  
approximately 800~\degs\ of sky from the South Pole Telescope SZ (SPT-SZ) 
survey.  We concentrate on the range of angular scales (or multipole number) 
at which the secondary and foreground sources of non-Gaussianity are expected to 
dominate, namely $\theta \simleq 10^\prime$ ($l \simgeq 1000$), and 
use this information to simultaneously fit for the contributions from tSZ and from
the clustered and spatially uncorrelated contributions from emissive sources.

The paper is organized as follows: we briefly describe the data products used 
in this analysis in Section \ref{sec:data}; we present the method for estimating
the bispectrum in Section \ref{sec:analysis}; we describe the model used to fit
the resulting bispectrum measurements in Section \ref{sec:modeling}; we present
measured bispectra and model fit results in Section \ref{sec:results}; we discuss
the implications of these results for cosmology and models of source emission 
in Section \ref{sec:discussion}; and we conclude in Section \ref{sec:conclusions}.

\section{Data}
\label{sec:data}
The SPT \citep{carlstrom11} is a 10-meter telescope located at the 
National Science Foundation's
Amundsen-Scott South 
Pole station in Antarctica.  The 2500-\degs \ SPT-SZ survey was completed in November 2011.  This 
survey produced maps in three frequency bands (95, 150, and 220~GHz) 
to a depth such that the rms 150~GHz noise level in any of the maps is 
$<18\muk$ per 1-arcminute pixel.
Scientific results from partial or full SPT-SZ survey data include
catalogs of clusters discovered via the SZ effect 
\citep{vanderlinde10,williamson11,reichardt13}, catalogs of emissive sources 
(including a new population of strongly lensed, dusty, high-redshift galaxies, 
\citealt{vieira10,vieira13}), and measurements of the primary CMB power spectrum 
\citep{keisler11,story13} and of the
secondary CMB and foreground 
power spectra \citep{lueker10,shirokoff11,reichardt12b}.

\citet[][hereafter
R12]{reichardt12b} used $\sim$800~\degs\ of SPT-SZ survey data to measure the
small-angular-scale CMB power spectrum and place unprecedentedly tight
constraints on the fluctuation power of the tSZ, kSZ, and CIB at SPT  
observing frequencies.
In this work, we measure the bispectrum over the same area of sky as 
was used in R12 to measure the power spectrum.  

The precision expected for a bispectrum 
measurement using 800 square degrees of SPT data should be high enough
to provide useful new information.
The B12 modeling uncertainty on \atsz\ (the amplitude of the tSZ power spectrum) 
given a perfect measurement of \btsz\ (the amplitude of the tSZ bispectrum) 
is 7\% for the default B12 assumptions about 
gas physics---in which the spread in gas model parameters is constrained
by the pressure profile measurements of \citet{arnaud10}---and 
$\sim$15\% when the extreme cases of no feedback and maximal 
feedback (well beyond the \citealt{arnaud10} limits) are considered. 
The predicted scaling between the two quantities is 
$\btsz \propto \atsz^{1.4}$ (B12). 
This implies that a $\sim$20\% measurement of \btsz\ will be limited
in its constraint on \atsz\ by modeling uncertainty in the pessimistic case, and a 
$\sim$10\% measurement will be limited by modeling uncertainty in the default case.
B12 showed that a survey with the depth and sky coverage
of the full 2500-\degs\ SPT-SZ survey should be able to make a 
$\sim$6\% measurement
(or, equivalently, a $\sim$16$\sigma$ detection) of 
the tSZ bispectrum in the 150~GHz data alone.  Thus, an 
$\sim$11\% measurement ($\sim$9$\sigma$ detection)
should be achievable with 800~\degs\ of SPT-SZ 150~GHz data, with the 95~GHz
data adding additional detection significance. Depending on the level of modeling 
uncertainty assumed and the additional uncertainty on \btsz\ due to systematics (see
Section \ref{sec:systematics}) and sample variance (see Section \ref{sec:tszsampvar}), 
this level of detection has the potential to provide interesting \atsz\ constraints.

The data analysis up to the point of generating maps from single observations of each 
field is identical to that in R12,
and we refer the reader to that work and other SPT data analysis papers
\citep[e.g.,][]{lueker10,schaffer11} for details of the analysis.  Briefly, 
raw, time-ordered detector data from a single observation of a given sky field are 
relatively calibrated, data selection cuts are applied, high-pass filters are
applied to the data to downweight noise from the atmosphere and the readout,
and the data are binned into a single-observation map, using simple 
inverse-variance weighting.  

R12 used a cross-spectrum analysis to estimate the power spectrum.  This choice, involving 
the cross-correlation of single-observation maps, was made mainly to avoid noise 
bias in the power spectrum.  In the bispectrum analysis, no
noise bias is expected if the instrument/atmospheric noise is Gaussian.  For the 
bispectrum estimation, we therefore use a single coadded map for each field and frequency band,
made by taking inverse-variance-weighted averages 
of all single-observation maps (the total number of single observations ranges
from $\sim$200 to $\sim$1000).  We use simulated observations to characterize 
the effect of instrument beam and timestream filtering and to estimate bispectrum
uncertainties (see Section \ref{sec:analysis} for details).
In this work, we use the single-observation maps and simulation products from
the R12 analysis.

The maps used in R12 and in this work are constructed from data taken in the 2008 
and 2009 SPT observing seasons.  In 2008, only detectors at 220~GHz and 150~GHz produced
science-quality data; in 2009, science-quality data was produced in all three frequency
bands.  The two 2008 fields are $\sim$100~\degs\ and roughly square on the sky; 
the three 2009 fields are $\sim$200~\degs\ and extend roughly twice as far (in real degrees on 
the sky) in right ascension as they do in declination.  To simplify the bispectrum calculation, 
we split each of the 2009 fields into a left and a right half, each of which is roughly
the dimensions of the 2008 fields, leaving us with eight $\sim$100~\degs\ fields of
similar shape.  The total sky area analyzed, 
corrected for any overlap between fields and for regions near
bright sources that are interpolated over, 
is 837~\degs.  

\section{Bispectrum estimation method}
\label{sec:analysis}
Previous estimates of non-Gaussianity in CMB data (primordial or otherwise)
have generally made use of an estimator characterizing a single 
amplitude for the non-Gaussian signal.  This amplitude parameter is \fnl\ for 
primordial non-Gaussianity 
\citep[e.g.,][]{yadav08,komatsu11} and $\langle \ttsz^3 \rangle$ for non-Gaussianity 
due to tSZ \citep{wilson12}.  An alternate analysis method is to calculate the 
three-point function or bispectrum in full generality, then extract the best-fit amplitude
of a given non-Gaussian signal template from the full bispectrum.  This more general
approach does not require assumptions about the angular scale dependence of the 
non-Gaussian signal.  It also allows the freedom to simultaneously measure different sources
of non-Gaussianity, such that signals that are not of interest (for example the bispectrum due to
emissive sources in an analysis targeting the primordial CMB bispectrum)
can be marginalized over. 

Historically a calculation of the full bispectrum has been avoided because it is 
computationally unfeasible for full-sky datasets \citep[e.g.][]{yadav10}.  However, 
over a small patch of sky, one can take advantage of the flat-sky approximation, allowing
spherical harmonic transforms to be replaced by fast Fourier transforms (FFTs).
In this work, we use the flat-sky approximation and calculate the full, three-dimensional 
bispectrum over $\sim$800 \degs, or roughly 2\% of the full sky.
\citet{fergusson09} find that the bispectrum estimated using the flat-sky approximation 
agrees with the full, curved sky analysis to $\lesssim 1\%$ if all $l$ values are 
greater than $150$.
In this work, we 
only consider multipole values of 
$l \gtrsim 1000$, so the flat-sky approximation
is a very good one for this analysis.

\subsection{Defining the estimator}
\label{sec:estimator}
Following \citet{hu00}, we define the full-sky (angle-averaged) bispectrum through
the relation
\beq
\left \langle a_{l_1 m_1} a_{l_2 m_2} a_{l_3 m_3} \right \rangle = 
\left (   
  \begin{array}{ccc}
  l_1& l_2& l_3\\
  m_1&m_2&m_3
  \end{array}
\right ) 
B_{l_1 l_2 l_3},
\label{eqn:fullbs}
\eeq
where $a_{l m}$ are the coefficients of the spherical harmonic expansion
of the full-sky temperature field, and the Wigner~3-j symbol enforces  
selection rules on the triplets of angular modes.  
In the flat-sky limit, 
the equivalent relation is
\beq
\left \langle a(\bl_1) a(\bl_2) a(\bl_3) \right \rangle = (2 \pi)^2 \ \delta^2(\bl_1 + \bl_2 + \bl_3) \ \bispv,
\label{eqn:flatbsvec}
\eeq
where $a(\bl)$ are the coefficients of the Fourier transform of the partial-sky 
temperature field, defined by 
\beq
\Delta T (\bx) = \int \frac{d^2 l}{(2\pi)^2} a(\bl) e^{i \bl \cdot \bx},
\eeq
and the Wigner 3-j symbol has been replaced by a Dirac $\delta$-function enforcing
that the locations of the three Fourier modes form a triangle in $l$-space.  If the 
signal responsible for the bispectrum has no preferred direction in the sky,  we can
write this as
\beq
\left \langle a(\bl_1) a(\bl_2) a(\bl_3) \right \rangle = (2 \pi)^2 \ \delta^2(\bl_1 + \bl_2 + \bl_3) \ \bisp,
\label{eqn:flatbs}
\eeq
where $l_i=|\bl_i|$.  In this limit, the flat-sky bispectrum \bisp\ is equivalent to the full-sky 
reduced bispectrum $b_{l_1 l_2 l_3}$, defined through the relation
\beq
B_{l_1 l_2 l_3} = 
\sqrt{\frac{(2 l_1 + 1) (2 l_2 + 1) (2 l_3 + 1)}{4 \pi}}
\left (   
  \begin{array}{ccc}
  l_1 & l_2 & l_3\\
  0 & 0 & 0 
  \end{array}
\right ) 
b_{l_1 l_2 l_3}
\label{eqn:redbs}
\eeq
\citep{komatsu01}.

For a finite-sized map of angular extent $\Delta \theta$, 
Fourier modes separated by $\Delta l \le 2 \pi / \Delta \theta$
are indistinguishable from one another. To avoid significant 
correlations between mode triplets, the bispectrum of such a map should be 
calculated in bins of size $\Delta l > 2 \pi / \Delta \theta$.
\citet{santos02} define an estimator of the bispectrum in
bins of width $\Delta l$:
\begin{eqnarray}
\label{eqn:estimator}
\bispe &=& \nlinv 
\times \\
\nonumber && 
\sum_{l_i - \Delta l/2 \le |\bl_i| \le l_i + \Delta l/2} 
\mathrm{Re} \left [ a(\bl_1) a(\bl_2) a(\bl_3) \right ],
\end{eqnarray}
subject to the constraint $\bl_1 + \bl_2 + \bl_3  = 0$, where 
where $\mathrm{Re}[x]$ is the real part of the complex number $x$,
and \nl \ is the number of mode triplets that satisfy the $l$ bounds
and the triangle condition. We add the option of assigning different weights to different 
mode triplets in a bin by redefining the estimator to be
\begin{eqnarray}
\label{eqn:estimatorwt}
\bispe &=& \frac{1}{\sum W(\bl_1, \bl_2, \bl_3)}
\times \\
\nonumber && 
\sum W(\bl_1, \bl_2, \bl_3) \
\mathrm{Re} \left [ a(\bl_1) a(\bl_2) a(\bl_3) \right ],
\end{eqnarray}
where the sum is over the same mode triplets as in Equation \ref{eqn:estimator}, 
and the same triangle condition applies. The weighting scheme used in this analysis is
described in Section \ref{sec:noise}.
To compare the partial-sky bispectrum estimate to 
predictions for the full-sky reduced bispectrum, we divide by the map area \dommap.

We only calculate the auto-bispectrum $\hat{B}(l_1, \nu_i; l_2, \nu_i ; l_3, \nu_i)$
for each of our three frequency bands. We are not exploiting the full information in the 
bispectrum, since there are also seven unique cross-bispectra
$\hat{B}(l_1, \nu_i; l_2, \nu_j ; l_3, \nu_k)$, where $\nu_i$, $\nu_j$, and 
$\nu_k$ are not all the same.  To keep the computation and interpretation
as simple as possible for this first result, we postpone investigation of the
cross-bispectra to a future publication.

\subsection{Treatment of the instrument beam and filter transfer function}
\label{sec:beamxfer}
The maps from which we estimate the bispectrum in this work do not contain
unbiased estimates of the true sky temperature at all angular scales, due to the 
filtering applied to the detector time-ordered data and due to the instrument beam.
However, in the limit that the filtering is a purely linear operation that is uniform 
over the map, we can define a
single Fourier-domain function $F(\bl)$ that describes the combined effects of 
beam and filtering on the coefficients $a(\bl)$.  We obtain an unbiased estimate
of the true $a(\bl)$ by dividing the raw, biased
$\hat{a}(\bl)$ (estimated by directly Fourier transforming the map) 
by $F(\bl)$.  We estimate $F(\bl)$ by taking
realistic mock skies (described in detail in Section \ref{sec:errorbars}), 
convolving them with the measured beam, and running them 
through the full pipeline up to the coadded 
map stage. We calculate the two-dimensional Fourier-domain ratio 
of output to input maps and use this as our estimate of $F(\bl)$.

Due to the finite size of the detector array and the sky fields measured, the
timestream filtering process cannot truly be represented by a purely linear
map-domain filter that is uniform across the map. 
However, we expect any errors in the measured transfer
function due to this non-linearity or non-uniformity to be very small for two reasons.  
First, the primary effect of departures from our idealization of the transfer function
is to alias power at low spatial frequencies to high spatial frequencies
\citep[e.g.,][]{schaffer11}.  This
is a potentially significant problem when the signal spectrum is very red (as 
in measurements of the primary CMB power spectrum), but the expected 
bispectrum signal in the $l$ range treated here is much closer to flat in $l$.  Second, 
the input signal we use in the simulations is expected to be a reasonable 
approximation to the true input signal, which minimizes the impact 
of non-linearity.  Furthermore, even if the assumed input 
signal is significantly wrong, the expected errors on the transfer function are small.
For example, \citet{lueker10} tested the filter transfer
function for SPT power spectrum analysis and found that changing the input 
power from extragalactic sources by a factor of two made $<1 \%$ changes in the inferred 
transfer function.

\subsection{Apodization and compact-source treatment}
\label{sec:masks}
We use FFTs to calculate $a(\bl)$ from our maps, and FFT algorithms assume 
periodic boundary conditions.
To avoid injecting false signals from discontinuities at the edges of the map,
we enforce periodic boundary conditions by apodizing each map after embedding it in a larger grid and zero-padding.
For a given sky field, we create the apodization mask as follows: we start with
a map representing the total inverse-variance weights used in creating the final 
coadded map for a given field. We smooth the weight map with a Gaussian 
kernel with $\mathrm{FWHM}=4^\prime$, divide by a fiducial 
weight value (equal to 80\% of the median weights, a value empirically determined
to produce well-behaved masks), and set all values above $1.0$ equal to 1.0.
The resulting apodization mask also downweights 
the edges of the map, which are noisier than the nearly-uniform-noise main 
map region.  

The apodization in real space is a convolution in Fourier space, with the convolution
kernel being the Fourier transform of the apodization mask.  This operation will 
correlate otherwise uncorrelated Fourier coefficients.  However, if the mask has a smooth taper and 
no features on small angular scales, the Fourier-domain convolution 
kernel will be compact and of width $\Delta l \simeq 2 \pi / \Delta \theta$; i.e., the 
mode correlation induced will be approximately the same as the correlation that arises 
just from the finite size 
of the map.  If the $l$ bins used in the bispectrum analysis are sufficiently wide, 
this correlation can be ignored.  We choose a bin size of $\Delta l = 200$; the amplitude
of the 2-d Fourier transform of the apodization mask for a typical field at $l=200$ in either 
dimension is $\simleq 0.01$ times the $l=0$ value.  We have verified through simulated
observations that no detectable correlation between bispectrum values in bins of this 
size is induced from our apodization masks, and we ignore any effects of the 
mask---beyond correcting for its effect on \dommap, the area over which modes are 
measured---in subsequent analyses.

It is common practice in CMB analyses to also mask compact sources in the maps
\citep[e.g.,][]{hinshaw03,lueker10,fowler10}. Although we are interested
in the bispectrum from compact sources such as galaxy clusters, radio sources, and 
DSFGs, we do want to remove the signal from the very brightest of these sources. 
Masking the brightest sources reduces sample variance---and, in some cases, uncertainty
in modeling their bispectrum---and it allows us to estimate how much of the bispectrum signal is coming 
from sources that have not been detected and characterized in other analyses of the 
same data (see Section \ref{sec:tszmodel} for details).  However, if we were to multiply 
the maps by a mask that had holes at source locations, we could no longer ignore the 
effects of the mask on the measured bispectrum because the mask would now have 
small-scale features.  

To avoid having to calculate the bispectrum equivalent of
the \citet{hivon02} pseudo-$C_l$ mode-mixing kernel, we instead choose to remove 
compact-source signals from our maps via harmonic inpainting.  We use the procedure 
described in \citet{vanengelen12}; briefly, a square region around a bright source in the 
map is interpolated over using the correlation properties measured in the rest of the map 
to create the interpolates.  In all maps, we interpolate over sources detected at $5 \sigma$
at 150~GHz (using the catalogs of \citealt{vieira10} and \citealt{mocanu13}).
Because the different
sky fields used in this analysis were observed to slightly different depths, the 150~GHz flux cut to which
this significance is equivalent varies from 5.7~mJy to 6.6~mJy.  In some versions of 
the bispectrum analysis, we also interpolate over galaxy clusters above a given mass from the 
\citet{reichardt13} catalog (see Section \ref{sec:tszmodel} for details).  In all cases, 
the inpainting is done over only $\sim$1\% of the total map area. We test for any effects of 
this inpainting on bispectrum measurements by studying simulated data.  We see no effect 
from inpainting---beyond the obvious effect of eliminating the contribution to the bispectrum
from the painted-over sources---at the 
sensitivity of our
tests, which probe down to roughly $1\%$ of the expected secondary/foreground 
bispectrum signal level.

\subsection{Noise and bispectrum weighting} \label{sec:noise}

The signals we are interested in for this work are non-Gaussian
contributions to the sky temperatures recorded in our maps.  Any purely
Gaussian contributions will, by definition, produce no average
bispectrum. However, Gaussian components of the maps will contribute to
the variance on the bispectrum measurement.  Therefore, in addition to
instrumental and atmospheric sources of noise, astrophysical and cosmological sources of
Gaussian power (such as the primary CMB and populations of emissive
sources) will also contribute noise to this analysis.

When constructing the binned bispectrum, we average together the
products of many mode triplets to estimate the bispectrum in each bin. 
Considering a single mode triplet, the variance on the product of three
Fourier coefficients $a(\bl_1) a(\bl_2) a(\bl_3)$ has contributions from the Gaussian
components of the map (including the Gaussian part of intrinsically
non-Gaussian sources of power such as emissive sources) as well as from
the non-Gaussian components.  In the limit of very small non-Gaussian
signatures in the maps, the Gaussian components dominate this variance.

The three most significant sources of fluctuation power for the maps
used in this analysis are noise (instrumental and atmospheric), primary
CMB fluctuations, and power from extragalactic sources (mainly DSFGs)
below the SPT detection threshold. We measure the SPT noise to be
Gaussian at the level necessary for this analysis by calculating the
bispectrum of noise-only maps (described in more detail in Section
\ref{sec:weights} below) using the bispectrum estimator and obtaining 
the expected null result. 
The non-Gaussianity of the primary CMB is constrained through estimates
of \fnl \ from \wmap\ data. The signal from extragalactic sources is
intrinsically Poisson-distributed, but at flux levels at which we expect
many sources per SPT beam, the distribution of fluxes
in a single SPT beam or pixel will approach a Gaussian. Hence the
extragalactic source signal will have a non-Gaussian part, which we consider as
a potential bispectrum signal, and a Gaussian part, which will
contribute to the noise of the bispectrum measurement.
The processes used to estimate weights and error bars are
described in more detail in Sections 
\ref{sec:weights} and \ref{sec:errorbars}.

When averaging many mode triplets together, we use weights derived from
estimates of the Gaussian variance from noise, primary CMB, and point
sources (estimated as described in the next section).
Error bars on the binned bispectrum values are also constructed from
estimates of the Gaussian variance in the maps.  
These weights and error bars do not
take into account non-Gaussian signal variance.  Signal variance for the
binned bispectrum can be significant: individual strong sources in the
maps produce bispectrum signals that are highly correlated across mode
triplets and will vary from field to field across the sky.  We take
this signal variance into account when interpreting our model fits in a
cosmological context, as described in Section \ref{sec:tszsampvar}.

\subsubsection{Weights} 
\label{sec:weights}
In this work, the weights used in the bispectrum estimator described by Equation \ref{eqn:estimatorwt} are constructed from
estimates of the Gaussian variance for each mode triplet.  A purely
Gaussian component with angular power spectrum $C(\bl)$ will contribute variance
to the bispectrum measurement equal to 
\beq \left \langle \left |
\bispv \right | ^2 \right \rangle = C(\bl_1) C(\bl_2) C(\bl_3) 
\eeq 
for $\bl_1 \neq \bl_2 \neq \bl_3$, so we use as our bispectrum weights 
\beq
W(\bl_1,\bl_2,\bl_3) = \frac{1}{C(\bl_1) C(\bl_2) C(\bl_3)}.
\eeq
Our estimate of the total $C(\bl)$ contributing to bispectrum variance
is the sum of the contributions from 
primary CMB fluctuations, emissive sources below the SPT 
detection threshold, and instrumental/atmospheric noise.

The input CMB and point-source spectra are identical to those 
used in the simulated skies in R12, and we refer to that
work for details on the input spectrum. Briefly, 
the contribution from the primary CMB is a $\Lambda$CDM 
model of the
primary CMB power spectrum from \citet{keisler11}, and the contribution
from emissive sources is based on measurements in \citet{shirokoff11}.  The contribution
from instrumental and atmospheric noise is estimated using the
two-dimensional $l$-space noise  power spectrum associated with the map.
 The noise power spectrum for a given map is calculated via a jackknife
procedure, in which many combinations of the individual-observation maps
are created, each one with half the maps multiplied by $-1$ so that the
resulting combination has no astronomical signal. The power spectrum of
each of these combinations is computed, and the results are averaged to
produce the final estimate of the noise power spectrum
$C_\mathrm{noise}(\bl)$, which is divided by the square of the beam and
transfer function estimate $F(\bl)$ (see Section \ref{sec:beamxfer})
before being included in the variance calculation.

The expected variance of individual Fourier modes due to instrumental/atmospheric noise 
and the primary CMB 
varies across the Fourier plane.  The primary CMB variance depends on
$l$, and the noise power spectrum depends on $\bl$ (see
\citealt{schaffer11} for details).  Our weights take these variations
into account.

We address some mode triplets as special cases.  For
mode triplets in which two of the $\bl$ values are the same, the
bispectrum variance will be elevated by a factor of three (because
$\langle a(\bl_1) a(\bl_2) a(\bl_3) \rangle \rightarrow \langle a^2(\bl_1) a(\bl_3) \rangle$).  
Additionally,
because the estimator we are using takes the real part of $a(\bl_1)
a(\bl_2) a(\bl_3)$, the same noise elevation occurs for mode triplets in
which $\bl_1 = -\bl_2$ or $\bl_1=-\bl_3$, etc.  Because the fraction of
such mode triplets is small (fewer than $0.01 \%$ of the total number 
of triplets over the $l$ range considered here), 
and they would be significantly
downweighted in our weighting scheme, we choose to simply give these
triplets zero weight in the estimator.

\subsubsection{Binned Bispectrum Error Bars}
\label{sec:errorbars}

The total inverse-variance weight in a given $l$ bin is also directly
related to the uncertainty on the estimate of the bispectrum in that
bin: $\sigma^2(\bispe)$ should be equal to $1/W_\mathrm{tot}$, where
$W_\mathrm{tot}$ is the total weight in that bin 
\beq W_\mathrm{tot} =
\sum_{l_i - \Delta l/2 \le |\bl_i| \le l_i + \Delta l/2}
W(\bl_1,\bl_2,\bl_3). 
\eeq 
In practice, we estimate the bispectrum
variance $\sigma^2(\bispe)$ from the scatter in the bispectrum measured
from 100 simulated observations. 
As a cross-check, we have compared the binned bispectrum
error bars estimated from the weights to those estimated from the
simulations.  The two are the same up to an overall scaling factor (of
order unity) related to the ratio of the area under the apodization mask
to the total area of the field, which affects the number of truly
independent mode triplets in a bispectrum bin.  This decrease in
independent modes is due to mode correlation from the mask.  This
correlation also increases the number of mode triplets with elevated
noise (as described in the previous section). However, for masks that
are smoothly tapered and cover nearly the full field (such that they
are strongly localized in the Fourier domain), the fractional increase 
in noisy triplets---which were $<0.01 \%$ of total triplets to begin with
(see Section \ref{sec:weights})---is small.
Therefore, we ignore this effect in this analysis (i.e., we include these
triplets in the estimator and give them the weight they would have in
the absence of masking).

We create our final estimate of bispectrum error bars by running 100
sets of mock observations of our eight fields through the bispectrum
estimator and calculating the scatter in the measured bispectrum in each
$l$ bin across the 100 sets.  The input skies are composed of Gaussian
realizations of the same sky power spectrum used for the weights 
described in the previous section, 
convolved with the measured
beam-and-filter transfer function $F(\bl)$, with a realization of the
instrumental/atmospheric noise added. 
For the simulated noise in a given
field, we use one of the signal-free combinations of individual
observations used in the noise power spectrum estimation described
above.  The simulated observations are apodized in the same manner as
the real data, so  any effects of the apodization are taken into account
in the uncertainty estimation. The simulations do not contain correlated
signal between fields, so overlap between fields is not taken into
account; however, the overlap is $\sim$2\% of the total area, and any
error caused by neglecting it is small compared to the statistical
precision of our final results.

We have examined the $l$-bin-to-$l$-bin covariance over the 100
simulated observations and see no bin-to-bin correlation above the level expected
from this number of independent measurements.
We treat the noise in each
$l$ bin as independent in all subsequent analyses. We do see
correlations within an $l$ bin among the three observing bands, as
expected given the contribution to the bispectrum variance from Gaussian
sky signal (particularly the primary CMB, which is perfectly correlated
among the three bands). We account for these correlations by expressing
the bispectrum covariance as a $3 \times 3$ matrix in each $l$ bin. We
also note that we detect no mean bispectrum in these simulated
observations, which include actual 
instrumental/atmospheric noise, demonstrating that the noise is Gaussian
to a very good approximation.
The bispectrum covariance matrix used in further analysis is thus 
\beq
\label{eqn:bispcov} C_{ij}(l_1,l_2,l_3) = \left \langle
\hat{B}_\mathrm{sim}(l_1,l_2,l_3,\nu_i) \
\hat{B}_\mathrm{sim}(l_1,l_2,l_3,\nu_j) \right \rangle, 
\eeq 
where
$\hat{B}_\mathrm{sim}$ is the estimated bispectrum from a single
simulation, and the expectation value is over all simulations. 

\section{Bispectrum modeling and model fitting}
\label{sec:modeling}
To interpret any detection of the secondary/foreground bispectrum in an 
astrophysical or cosmological context, we need a model of the expected signal and 
a model-fitting procedure.  In this section, we describe the signal models we adopt 
and the procedure we use for fitting the multi-band data to these models.  We also describe 
how we account for instrument-related systematic effects such as uncertainties in beams, 
spectral bandpasses, and calibration.

\subsection{Signal models}
\label{sec:sigmodels}
We include three types of non-Gaussian signal in our modeling: tSZ from  
galaxy clusters, the spatially uncorrelated signal from extragalactic sources
(hereafter ``point sources," since the vast majority of such sources will 
appear point-like at the $\sim$1$^\prime$ resolution of the SPT), and the 
expected clustered emission from one source population (DSFGs).  
We describe our modeling choices for each of these in turn, but first we note
that we do not include other potential sources of millimeter-wave signal---particularly 
the kSZ effect, clustered radio sources, and galactic foregrounds---in the bispectrum
model.  

We do not include the kSZ primarily because none of the predicted kSZ-generating
mechanisms---including the peculiar velocity of free electrons in galaxies 
\citep{ostriker86} or galaxy clusters \citep{sunyaev80}, and patchy reionization 
\citep{knox98,gruzinov98}---is 
expected to impart a net skewness on the CMB temperature 
distribution. This is because the velocities of ionized gas in the universe should be random with
respect to the observer, meaning that the induced kSZ signal should be symmetric 
around the mean CMB temperature.  We expect the signal from the clustered radio
background to be much smaller than the clustered DSFG signal (see Section \ref{sec:cibmodel}
for details), and we do not include such a term in our signal model.  We do not 
include any expected signal from our own galaxy, both because the sky fields used here 
are at high galactic latitude, and because such signals are 
expected to fall steeply with $l$ \citep[e.g.,][]{finkbeiner99} and thus be negligible at 
the angular scales or $l$ values of interest to this work.

\subsubsection{Spatially uncorrelated (``Poisson'') point-source contribution}
\label{sec:poissmodel}

We introduce the model for spatially uncorrelated point sources first to illustrate the basic
properties of the bispectrum arising from any population of discrete, spatially uncorrelated 
sources with a given angular profile.  Following, e.g., \citet{hall10}, we will use ``Poisson'' 
as shorthand for the component of the point-source contribution to the power spectrum
or bispectrum that arises from spatially uncorrelated sources.

For a CMB map made at observing frequency $\nu$ with pixels of size 
$\dom_p$ (where, for this toy example, the pixel size is much larger than the beam size), 
containing only a point source of flux $S$, 
we can write the signal in the map as
\beq
T_\mathrm{source} (\bx) = 
    \begin{cases}
      \tpeak, & \text{if}\ \bx \in \text{source pixel} \\
      0, & \text{otherwise}
    \end{cases}
\eeq
We know that the total flux in the map must equal $S$, which means 
\beq
\tpeak(\nu) = g(x_\nu) \times \frac{S}{\dom_p},
\eeq
where $g(x_\nu)$\  is the conversion factor between CMB fluctuation temperature and flux per solid angle 
(in units of $\mathrm{Jy} \ \mathrm{sr}^{-1}$) at observing frequency $\nu$:
\beq
\label{eqn:conversion}
g(x_\nu) = 10^{-26} \times \left [ \frac{2k_B}{c^2} \left ( \frac{k_B \tcmb}{h} \right )^2 \frac{x_\nu^4 e^{x_\nu}}{(e^{x_\nu}-1)^2} \right ]^{-1},
\eeq
and $x_\nu=h\nu/(k_BT_{\rm CMB})$.
For angular frequencies well below the cutoff of the pixel window function
($l \ll 2 \pi/\sqrt{\dom_p}$), the Fourier transform of this map is 
\begin{eqnarray}
a_\mathrm{source}(\bl) &=& \int d^2x \ T_\mathrm{source} (\bx) \ e^{-i \bl \cdot \bx} \\
\nonumber &\simeq& \dom_p \ \tpeak \ e^{-i \bl \cdot \bx_\mathrm{source}} \\
\nonumber &=& g(x_\nu) \ S \ e^{-i \bl \cdot \bx_\mathrm{source}}.
\end{eqnarray}

The estimated bispectrum due to this single source of flux $S$ is then
\begin{eqnarray}
\nonumber \bispenu &=& \nlinvsh \sumsh \mathrm{Re} \left [ a(\bl_1) a(\bl_2) a(- (\bl_1 + \bl_2)) \right ] \\
\nonumber &=& \frac{g^3(x_\nu) \ S^3}{N_l}  \sumsh 
\mathrm{Re}  [ e^{-i \bl_1 \cdot \bx_\mathrm{source}} e^{-i \bl_2 \cdot \bx_\mathrm{source}}  \\
\nonumber & \ & \times e^{i (\bl_1 + \bl_2) \cdot \bx_\mathrm{source}}   ] \\
&=&  g^3(x_\nu) \ S^3,
\label{eqn:bispone}
\end{eqnarray}
where we have used the triangle condition $\bl_1 + \bl_2 + \bl_3 = 0$ to redefine $\bl_3$.
When there are two or more sources in the map, 
Equation \ref{eqn:bispone} becomes
\begin{eqnarray}
\bispenu &=& g^3(x_\nu) \ \big ( S_1^3 + S_2^3 + ... + S_N^3 \\
\nonumber &+& \text{cross terms} \big ),
\end{eqnarray}
where the cross terms are of the form 
$S_1^2 S_2 e^{-i (\bl_1 + \bl_2) \cdot (\bx_1 - \bx_2)}$. 
If the sources are spatially uncorrelated, the phase of the cross terms 
is random, and these terms will on average be zero, leaving 
\beq
\label{eqn:bispmulti}
\bispenu = g^3(x_\nu) \left ( S_1^3 + S_2^3 + ... + S_N^3 \right )
\eeq
as the only nonzero average bispectrum contribution.  

For a population of sources with number 
density per unit solid angle per unit flux $dN/dS/d\Omega$, 
Equation \ref{eqn:bispmulti} is easily generalized to
\beq
\bispenu = g^3(x_\nu) \ \dommap \int_0^{\infty} S^3 \frac{dN}{dS d\Omega} dS,
\eeq
or, if sources have been cleaned down to some threshold $S_\mathrm{max}$,
\beq
\bispenu = g^3(x_\nu) \ \dommap \int_0^{S_\mathrm{max}} S^3 \frac{dN}{dS d\Omega} dS,
\label{eqn:dndseq}
\eeq
which, after we apply the \dommap\ correction, is identical to the familiar result of, e.g., \citet{komatsu01}.

We have chosen to only use bispectrum shape information to fit for the 
Poisson contribution to our multi-band bispectrum (i.e., we fit for a contribution
that is flat in $l$).  We include a single 
free parameter in the fit for the amplitude of the Poisson source bispectrum in 
each observing band.  

\subsubsection{Thermal SZ model}
\label{sec:tszmodel}
We use the model described in B12 for the bispectrum due to the tSZ effect in 
galaxy clusters.  We describe the model briefly here and refer the reader to B12
for details.  The tSZ bispectrum is calculated assuming the signal arises from spatially 
uncorrelated galaxy clusters and so is conceptually 
identical to the result from the previous section, with the modification that
the intrinsic angular profile of the clusters modifies the bispectrum shape.
For a family of astrophysical sources with angular profile $F(\bx)$ or 
Fourier-domain profile $F(\bl)$, $a_\mathrm{source}(\bl) \rightarrow F(\bl) \ 
a_\mathrm{source}(\bl)$, and the bispectrum $\bisp \rightarrow F(l_1) F(l_2) F(l_3) \bisp$.

The tSZ bispectrum at multipole numbers $l_1$, $l_2$, and $l_3$ and
observing frequency $\nu$ is calculated as the integral over cosmological volume 
of the product of the Fourier-domain cluster pressure profile at the three $l$ values, weighted by 
the halo mass function:
\begin{eqnarray}\nonumber
\btsz&(l_1,l_2,l_3,\nu)&= f^3(x_\nu)\int dz\,\frac{dV}{dz}\int d \ln M\, \frac{dn(M,z)}{d \ln M} \\
&\times&y (M,z,l_1)y (M,z,l_2)y (M,z,l_3),
\label{eq:tsz}
\end{eqnarray}
where $f(x_\nu)$ is the dimensionless function specifying the  
dependence of the tSZ on observing frequency \citep{sunyaev80}. 
We do not include relativistic corrections to $f(x_\nu)$ (see discussion below). 
The Fourier-domain pressure profile $y(M,z,l)$ is calculated from 
the analytic model of \citet{shaw10}, using their fiducial values of the intracluster medium
(ICM) parameters.  The halo mass function $dn(M,z) / d \ln M$ is from \citet{tinker08}.
A $\Lambda$CDM cosmology is assumed in calculating the halo mass function, with
fiducial parameters as in B12, namely
$\Omega_b=0.045$,
$\Omega_m=0.27$,
$h=0.71$,
$n_s=0.97$, and
$\sigeight = 0.8$.

The model is calculated at the center of each $l$ bin in which the bispectrum is 
estimated from the data.   The signal is sufficiently flat in $l$ that this is within 2\%
of the value that would be obtained by calculating the model at higher resolution in $l$
and averaging over the bin.  The tSZ frequency factor $f(x_\nu)$ is evaluated for each
observing band at the effective center frequency of the band assuming a 
non-relativistic tSZ spectrum.
R12 calculated these frequencies to be 97.6, 152.9, and 218.1~GHz.
(This value is an average over the 2008 and 2009  
observing seasons for the 150 and 220~GHz bands, see R12 for details.)

There is some uncertainty in how well mass function fits to simulation output 
capture the high-mass end, with potential 5-10\% uncertainties at halo masses 
above $\sim$10$^{15} M_\odot$ \citep{tinker08, bhattacharya10}.  For this reason, we also 
calculate a version of the tSZ model with the mass function truncated above
$\mtwoh=\bigmass$, where \mtwoh\ is the mass enclosed inside 
$R_{200} (\rho_\mathrm{crit})$, defined as the radius within which the average density is 
200 times the critical density.  To compare to this prediction, we construct a 
bispectrum estimate in all three SPT bands with clusters above this same mass
removed from the maps (using the same inpainting procedure used for the point 
sources, see Section \ref{sec:masks}).  The cluster masses are taken from the 
\citet{reichardt13} catalog and converted from $M_{500}$ to $M_{200}$ assuming 
a \citet{navarro96} profile and the \citet{duffy08} mass-concentration relation.
A total of four clusters above this threshold are masked in the full $\sim$800~\degs\ dataset.

This level of cluster masking also reduces potential systematic errors caused by 
ignoring relativistic corrections to the predicted tSZ bispectrum amplitude. The most massive,
hottest clusters have gas temperatures of $\gtrsim$10~keV \citep[e.g.,][]{allen08}.
At these temperatures, the relativistic correction to the tSZ temperature decrement is 
$\sim$6\% at 150~GHz \citep{nozawa00}. Limiting the cluster sample to $\mtwoh < \bigmass$
is roughly equivalent to a temperature limit of $T < 5 \ \mathrm{keV}$ \citep{stanek10}.
At these temperatures, the maximum error in $f(x_\nu)$ from ignoring relativistic 
corrections is $\sim$3\%.

We also construct tSZ models and bispectrum estimates with the mass function
truncated and clusters masked above $\mtwoh=\threshmass$.  This mass threshold
closely approximates the selection of clusters used to constrain cosmology
in \citet{reichardt13}, namely signal-to-noise ratio greater than five and 
$z \geq 0.3.$  This allows us to estimate the amount of information we 
are extracting from the tSZ bispectrum above and beyond what has already 
been extracted using cluster counts.
A total of 117 clusters above this threshold are masked in the full $\sim$800~\degs\ dataset.
For comparison, the total number of clusters used in the \citet{reichardt13}
cosmological results was 100.

Masking clusters in the data will lead to a smaller absolute amplitude of the tSZ 
bispectrum but will also lead to smaller sample variance, since the sample variance
is dominated by the presence or absence in a map of the rarest, most massive 
clusters. On the other hand, any systematic uncertainty in the method used to 
estimate cluster masses will lead to an uncertainty in the true mass threshold 
used for masking, resulting in a systematic uncertainty when comparing the 
masked data to a tSZ bispectrum model (in which the mass threshold for masking
is known perfectly).
The different scalings with cluster mass of these various contributions to the 
tSZ bispectrum error budget
imply that there may be some optimal mass cut that reduces the 
combined statistical-plus-systematic-plus-sample-variance uncertainty on the tSZ bispectrum, similar 
to the results in \citet{shaw09} for the tSZ power spectrum. We investigate
this further in Section \ref{sec:tszsampvar}.

\subsubsection{Clustered CIB model}
\label{sec:cibmodel}
Not only will emissive sources contribute
to the Poisson bispectrum, they can also be spatially clustered,
possibly leading to a detectable bispectrum 
signal with a different shape. Because this signal arises from a spatial modulation of the mean intensity, 
and because the CIB is much brighter than the radio background at SPT observing
frequencies \citep[e.g.][]{hauser01}, we expect the clustering signal from DSFGs to be much 
stronger than that from radio sources, as has been found in power spectrum 
measurements \citep[e.g.,][]{hall10,holder12}. We do not include the clustered 
radio background in our modeling and concentrate on the potentially measurable
signal from the clustered CIB.

As pointed out in \citet{lacasa12}, a single population of sources with clustering 
properties described by a single correlation function or 
angular power spectrum will have a bispectrum 
equal to 
\beq
B_\mathrm{tot}(l_1,l_2,l_3) = \alpha \sqrt{C_\mathrm{tot}(l_1) C_\mathrm{tot}(l_2) C_\mathrm{tot}(l_3)},
\label{eqn:cibone}
\eeq
where $\alpha$ is a constant, and $C_\mathrm{tot}$ is the total Poisson-plus-clustering 
angular power spectrum: $C_\mathrm{tot} = C_\mathrm{clust} + C_\mathrm{poiss}$.
\citet{lacasa12} further showed that this formulation provides an 
accurate characterization of the CIB bispectrum in the simulated sky maps of 
\citet{sehgal10}.  

In this formulation, for $l$ triplet bins in which the clustering signal dominates 
over the Poisson
contribution, we can write the clustering signal as
\beq
\bclust(l_1,l_2,l_3) \propto \sqrt{C_\mathrm{clust}(l_1) C_\mathrm{clust}(l_2) C_\mathrm{clust}(l_3)}.
\label{eqn:cibtwo}
\eeq
We use this as the $l$-space template for the clustered CIB bispectrum in our model fits, with a simple
power-law model for the clustered power spectrum
$C_\mathrm{clust}(l) \propto l^{-n}.$  Both \citet{addison12a} and R12 have
found this to be a good description of the clustered CIB power spectrum over a large range
of angular scales, including the entire range considered in this work.
We choose $n=1.2$, which is consistent with the best-fit values in \citet{addison12a} 
and R12. 

The spectral behavior of the clustered CIB at millimeter wavelengths is fairly well constrained
from power spectrum measurements, and we use existing results to inform our model fitting.
We assume a single spectral index $\acib$ over our three observing bands and model the clustered
CIB bispectrum at observing frequency $\nu$ as
\begin{eqnarray}
\label{eqn:cibthree}
\bclust(l_1,l_2,l_3,\nu) &=& \bclust(l_1,l_2,l_3,\nu_0) 
\left ( \frac{\nu}{\nu_0} \right )^{3 \acib} \\
\nonumber &=&  \bclust^{2000}(\nu_0) 
\left ( \frac{\nu}{\nu_0} \right )^{3 \acib} \\
\nonumber & & \times \left ( \frac{l_1}{2000} \frac{l_2}{2000} \frac{l_3}{2000} \right)^{-n/2},
\end{eqnarray}
where $ \bclust^{2000}(\nu_0) = \bclust(l_1=2000,l_2=2000,l_3=2000,\nu_0)$.  
We use $\nu_0= 220$~GHz in our CIB model.

Using the results of R12, we adopt a nominal value and $1 \sigma$ uncertainty for
the clustered CIB spectral index of $\acib=3.72 \pm 0.12$ (see Section 
\ref{sec:systematics} for how this uncertainty is included in the fit).  
In Equation \ref{eqn:cibthree}, $\nu$ for each 
observing band is the effective center frequency of the band assuming a $\nu^{\acib}$ 
spectrum. R12 calculated these frequencies to be 97.9, 153.8, and 219.6~GHz for 
$\acib=3.5$; the difference in effective frequencies for $\acib=3.5$ and 
$\acib=3.7$ is negligible.

We have chosen not to explore more complicated CIB modeling, involving 
(for instance) spatial correlations between the sources of the tSZ and CIB bispectra, 
different CIB bispectrum shapes, and spectral behavior beyond a single spectral index.
As will be shown in Section \ref{sec:cibinterp}, the simple model adopted here is
adequate to describe the data, and extensions to this model would not be strongly
constrained using the data in this work. 

For the particular case of tSZ-CIB
correlation, we note that this effect should be a far less significant bias to the measurement
of the tSZ bispectrum than it is to the tSZ power spectrum. Galaxies in the high-mass, low-redshift
clusters that make up the bulk of the tSZ bispectrum signal are measured to have
significantly less star formation per unit mass than galaxies in lower-mass clusters or 
low-redshift field galaxies \citep[e.g.,][]{hashimoto98}. Furthermore, the star-forming fraction is also seen to 
increase with redshift \citep[e.g.,][]{butcher84}. This evidence all indicates that, even if
tSZ-CIB correlation has a significant effect on the tSZ power spectrum---which is sourced by lower-mass, 
higher-redshift clusters than the bispectrum---the tSZ bispectrum is unlikely to be signifcantly
affected. When R12 allow tSZ-CIB correlation as a free parameter, the best-fit
tSZ power spectrum amplitude shifts by $<20 \%$, so we assume that the effect of tSZ-CIB
correlation on our measurement of the tSZ bispectrum will be $\ll 20 \%$ and hence subdominant
to other uncertainties.

\subsection{Fitting procedure}
\label{sec:fitting}
We use a simple linear least-squares procedure \citep[e.g.,][]{press86}
to fit the measured bispectrum with the three-component model described above.  
Least-squares fitting results in the maximum-likelihood estimate of model parameters
only if the measurement uncertainties are Gaussian-distributed. While the distribution 
of the individual $l$-space mode triplets is highly non-Gaussian, each $\Delta l = 200$ bin
contains $>10^4$ of these triplets, so we expect the distribution of
binned bispectrum uncertainties to be very nearly Gaussian. We confirm this through
simulations.

We fit all three bands' bispectrum data simultaneously.
The data vector has $3 \times N_\mathrm{bin}$ elements, where
$N_\mathrm{bin} = (l_\mathrm{max}/\Delta l)^3$, 
$l_\mathrm{max}$ is the maximum angular frequency used in the fit and $\Delta l$
is the size of the bins in $l$-space, in this case 200. 
None of the signal 
models described in the previous section have features on the scale of $\Delta l = 200$, 
so this resolution should be adequate to characterize the measured bispectrum.
The maximum $l$ used in this analysis is $\num{11000}$, which was chosen by investigating
the factor by which the bispectrum variance is inflated by deconvolving  the beam and transfer
function $F(\bl)$ from the maps.  The raw SPT map noise at high $l$ has a nearly white spectrum. 
After deconvolving $F(\bl)$, the noise power spectrum of the maps will be 
proportional to $1/F^2(\bl)$ at high $l$. The bispectrum variance from this map will thus
be proportional to $1/F^6(\bl)$ at high $l$.  This factor $F^6(\bl)$ is $>500$ for all SPT bands at $l>\num{11000}$.

We write the data vector as $d_\mu$ where the index $\mu$ is the product of an $l$-bin
index $a$ and an observing-frequency index $i$ (such that $\mu$ takes on a unique 
value for each bin $l_\alpha, l_\beta, l_\gamma$ and observing frequency $\nu_i$):
\begin{eqnarray}
d_\mu &=& d_{[ia]} = d_{[i \alpha \beta \gamma]} = \hat{B}(l_\alpha,l_\beta,l_\gamma,\nu_i), \\
\nonumber a &=& \alpha N^2_\mathrm{bin} + \beta N_\mathrm{bin} + \gamma, \\
\nonumber \mu &=& 3 \times a + i.
\end{eqnarray}

The weight matrix, which is the inverse of the bin-bin-band-band bispectrum
covariance matrix, is assumed to be block-diagonal in this analysis---i.e., we assume no
bispectrum covariance between $l$ bins due to noise or Gaussian sky components, 
but we do include the band-band covariance of the Gaussian sky
terms (see Section \ref{sec:noise}). 
Under this assumption, each sub-matrix characterizing the   
covariance between bands for a given bispectrum bin is an independent
$3 \times 3$ matrix given by the inverse of the band-band covariance matrix for that bin. 
This covariance matrix is estimated from simulations, as described in 
Section \ref{sec:noise}. Thus, we can write 
the weight and covariance matrices as 
\begin{eqnarray}
W_{\mu \nu} &=& C_{\mu \nu}^{-1} \\
\nonumber C_{\mu \nu} &=& C_{[ia][jb]} = C_{[ia][ja]} \delta_{ab},
\end{eqnarray}
where 
$C_{[ia][ja]} = C_{[i \alpha \beta \gamma][j \alpha \beta \gamma]} = C_{ij}(l_\alpha,l_\beta,l_\gamma)$,
as defined in Equation \ref{eqn:bispcov}, and, again, the indices $i$ and $j$ run over
observing bands and the remaining indices over $l$-space bins.

The model or design matrix $A$ is composed of five 
$3 \times N_\mathrm{bin}$-element vectors, each representing
the unnormalized signal shape for one of the signal components in all observing bands.
The tSZ and clustered CIB vectors have non-zero values in all observing bands (although
the model amplitude for the tSZ in the 220~GHz band is very small, since that band is 
very near the tSZ null), while the three vectors representing the Poisson point-source
power in each band (assumed to be independent in this fit) are non-zero only in the 
$N_\mathrm{bin}$ elements corresponding to that band.

The five free parameters of the model $\lambda$ are the amplitudes for each model 
component: tSZ, clustered CIB, and the Poisson point-source component in each of three bands.
The best-fit values of these parameters are estimated from the data as
\beq
\label{eqn:bestfitparams}
\overline{\lambda_\psi} = \left ( A_{\psi \mu}^T W_{\mu \nu} A_{\nu \omega} \right )^{-1} A_{\omega \pi}^T W_{\pi \rho} d_\rho,
\eeq
where sums over repeated indices are assumed.
This estimate of parameters has a covariance matrix equal to 
\beq
C^\mathrm{param}_{\psi \omega} = \left \langle \delta \lambda_\psi \ \delta \lambda_\omega \right \rangle 
= \left ( A_{\psi \mu}^T W_{\mu \nu} A_{\nu \omega} \right )^{-1} .
\eeq

\subsubsection{Incorporating systematic uncertainties}
\label{sec:systematics}
The disadvantage of a simple linear least-squares fit (in comparison to a more 
general parameter-space search such as a Markov chain Monte Carlo) is that there is
no way to trivially incorporate systematic uncertainties in such quantities as the 
instrument beam measurement or the spectral index of clustered CIB fluctuations
without introducing strong covariance among all $l$ bins and significantly complicating
the inversion of the covariance matrix.
To retain the advantages of the linear fit (speed, simplicity, and robust parameter
covariance estimation), 
we account for such systematics by running the linear fit many times, 
each time using a different realization of each systematic effect.  We then calculate
a systematic parameter covariance matrix $C_{\psi \omega}^\mathrm{syst}$ by directly 
computing the outer product $\delta \lambda_\psi \ \delta \lambda_\omega$
in each realization and averaging over all realizations.  

We account for four independent sources of systematic uncertainty: 
1) instrument spectral bandpasses; 2) the spectral index of
CIB fluctuations \acib; 3) instrument calibration; 4) instrument beams.  
Based on measurements described in \citet{schaffer11}
and similar measurements in 2009, the band centers for SPT are estimated
to be accurate to 0.3~GHz.  The major source of this uncertainty is the frequency
calibration of the Fourier transform spectrometer used to measure the bandpasses,
implying that the uncertainty should be highly correlated between the three bands.
For each systematic realization, we draw a bandpass error from a Gaussian of 
width $\sigma_\mathrm{band} = 0.3 \ \mathrm{GHz}$ and calculate the signal models 
using band centers shifted by this error. To account for uncertainty in the spectral 
behavior of the CIB, in each systematic realization, we draw
a value for \acib\ from the R12 distribution $\acib = 3.72 \pm 0.12$ and use that value in 
calculating the clustered CIB model.

R12 estimated the calibration uncertainty in the three bands 
to be 0.035, 0.032, and 0.048 in power, or 0.018, 0.016, and 0.024 in temperature.  
These uncertainties are
highly correlated, because a primary source of uncertainty in each band's calibration
is the noise in the WMAP
power spectrum in the range $l \in \left [650,1000 \right]$. We approximate the 
calibration covariance matrix by assigning the fractional uncertainty at 150~GHz
to all bands as a fully correlated component and augmenting that with uncorrelated
components at 95 and 220~GHz to make the on-diagonal elements equal to the 
square of the measured uncertainties in each band. For each systematic realization 
we create a three-element vector $\sigma_\mathrm{cal}(\nu)$
with the appropriate covariance and
multiply the elements of the data vector $d$ 
corresponding to band $\nu$ by 
$\left [ 1 + \sigma_\mathrm{cal}(\nu) \right ]^3$. The mean and $1 \sigma$ width
of the systematic distributions for bandpass, CIB spectral index, and calibration
errors are summarized in Table~\ref{tab:syst}.

Uncertainties in the measurement of the instrument beam are incorporated by 
creating realizations of the beams using the full beam covariance matrix described
in \citet{keisler11}.  For each systematic realization, a beam realization is created 
for each observing band and observing season, including the correlations in the 
uncertainties between bands and seasons.  The bispectrum estimate from each 
100~\degs\ field and each band is multiplied by the cube of the ratio of the appropriate 
beam realization (for the year the field was observed) to the nominal beam.  The 
data from all fields are then combined using the nominal weights, and this combined 
beam-error-multiplied bispectrum is used to construct the data vector $d$.

\begin{table}[]
\caption{Systematic error accounting}
\centering
\begin{tabular}{ | l c c | }
\hline
Parameter & nominal value & $1 \sigma$ uncertainty \\ 
\hline
tSZ band center, 95 GHz & 97.6 GHz & 0.3 GHz \\ 
tSZ band center, 150 GHz & 152.8 GHz & 0.3 GHz \\ 
tSZ band center, 220 GHz & 219.1 GHz & 0.3 GHz \\ 
CIB band center, 95 GHz & 97.9 GHz & 0.3 GHz \\ 
CIB band center, 150 GHz & 153.8 GHz & 0.3 GHz \\ 
CIB band center, 220 GHz & 219.6 GHz & 0.3 GHz \\ 
CIB spectral index & 3.72 & 0.12 \\ 
calibration, 95 GHz & 1.00 & 0.018 \\ 
calibration, 150 GHz & 1.00 & 0.016 \\ 
calibration, 220 GHz & 1.00 & 0.024 \\ 
\hline
\end{tabular}
\label{tab:syst}
\begin{tablenotes}
\item{Distributions from which the systematic error realizations
described in Sections \ref{sec:systematics} and \ref{sec:fitresults} are applied.
The band center and calibration uncertainties are highly
correlated between bands.
Uncertainty in the instrument beam in each frequency band
is also taken into account using realizations of ``error beams" as
described in the text.}
\end{tablenotes}
\end{table}

\begin{figure*}
\begin{center}
\subfigure[95~GHz 1d bispectrum, with best-fit model overplotted]{\includegraphics[width=0.48\textwidth]{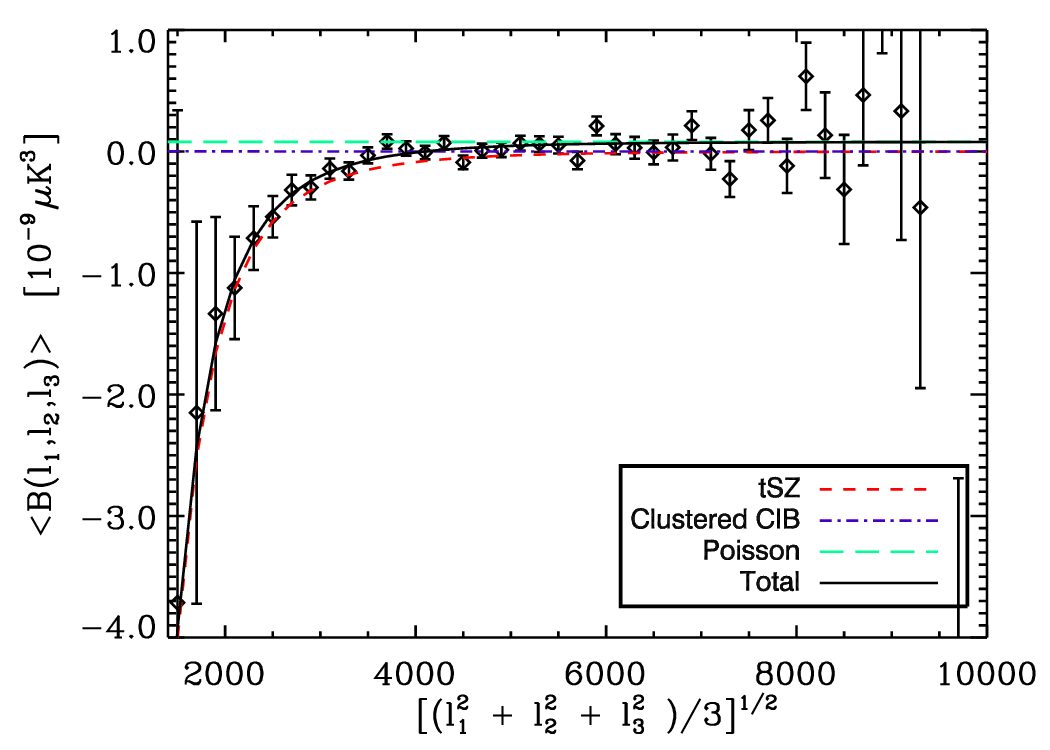}}
\subfigure[150~GHz 1d bispectrum, with best-fit model overplotted]{\includegraphics[width=0.48\textwidth]{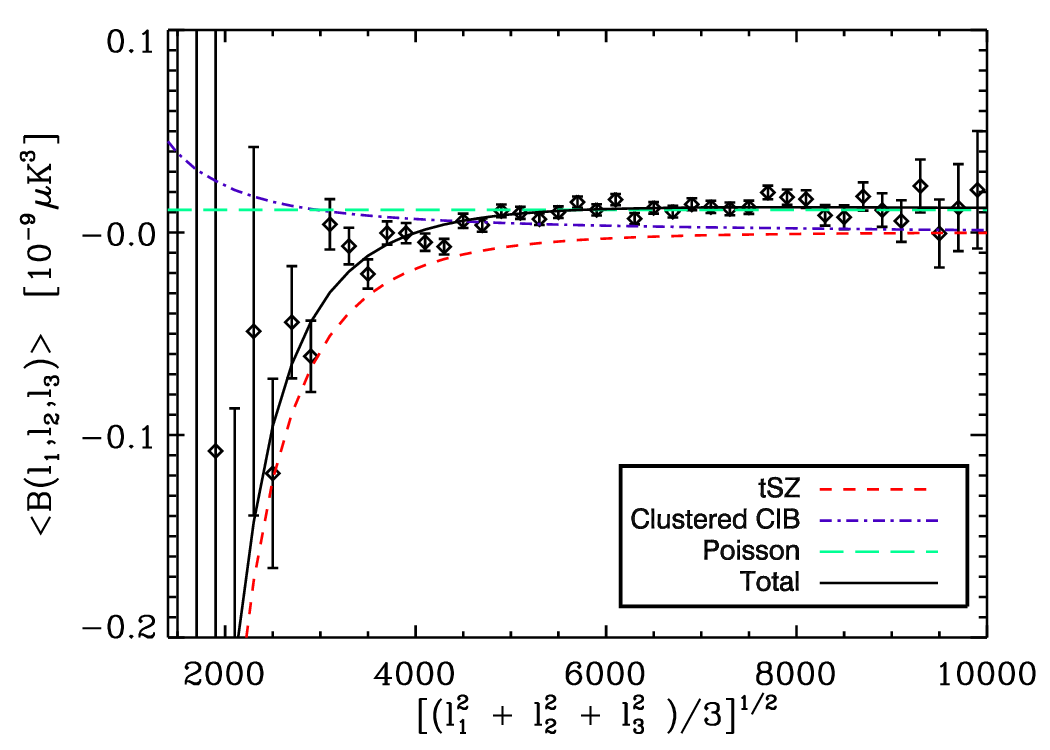}}
\subfigure[220~GHz 1d bispectrum, with best-fit model overplotted]{\includegraphics[width=0.48\textwidth]{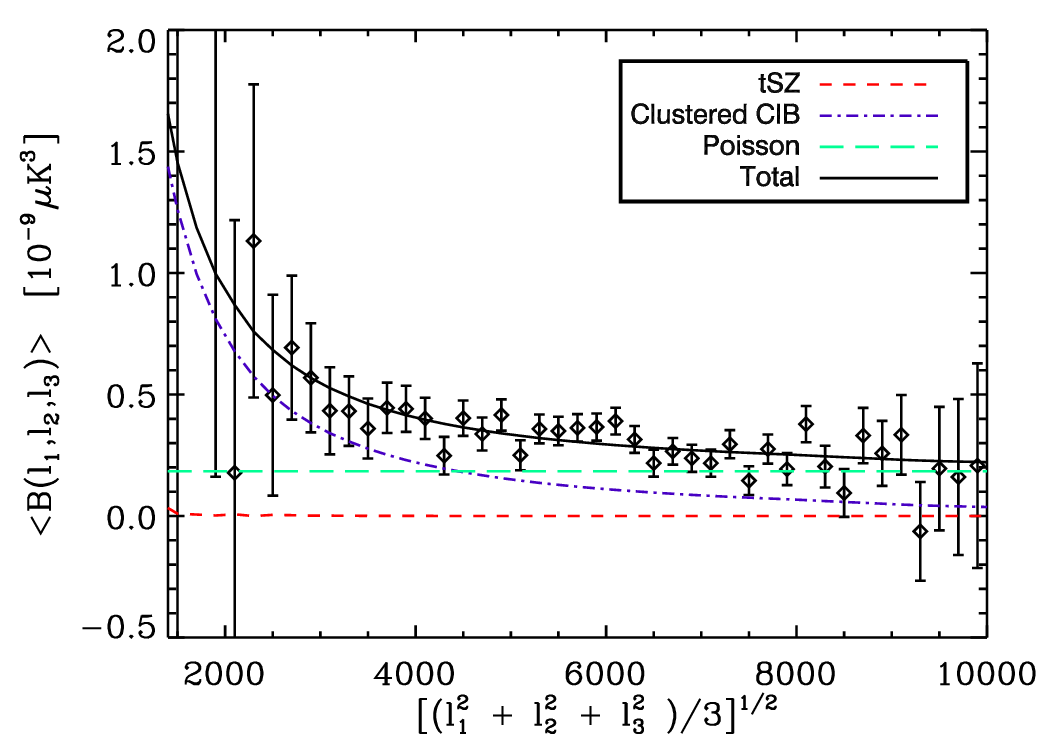}}
\end{center}
\caption{
Measured bispectrum (with no clusters masked) in each of the three SPT frequency bands.  Bispectra 
have been collapsed from three dimensions to one dimension as described 
in the text.  The solid black line shows the best-fit model estimated from the 
full 3d bispectrum (collapsed to 1d using the same procedure and weighting as used for the data).  The three individual components
of the best-fit model are also plotted:  tSZ 
(short-dashed red line), clustered CIB (dot-dashed purple line), and the Poisson
point-source component (long-dashed green line). See Section \ref{sec:modeling} for 
more details on the model.
The bispectrum error bars
shown are statistical only. The data and best-fit models shown are for 
the nominal values of the systematic parameters and with no cluster
masking (see text for details).
}
\label{fig:measbispecmodel}
\end{figure*}

\section{Results}
\label{sec:results}

\subsection{Measured bispectra and single-band detection significance}
\label{sec:measbispec}
The bispectrum in each frequency band (with no cluster masking), as estimated using the 
analysis procedures detailed in Section \ref{sec:analysis} and the bispectrum estimator 
discussed in Section \ref{sec:estimator}, is plotted in Figure \ref{fig:measbispecmodel}.  
Values of the bispectrum and inverse-variance weight in each band and each 
$l_1,l_2,l_3$ bin are available for download from the SPT 
website.\footnote{http://pole.uchicago.edu/public/data/crawford13}

We note that displaying this inherently three-dimensional data product in one dimension
requires the data to be contracted along two axes.  There is no ``industry 
standard" for displaying bispectra, particularly real measurements with noise.  B12
used the ``skewness spectrum" 
$\Lambda(l) = \sqrt{\sum_{l_1,l_2} b^2 (l, l_1, l_2)}$; however, 
this quantity will have a positive expectation value for a bispectrum estimated from
data with zero non-Gaussianity but finite noise and Gaussian sky power.  We choose
to define an $l$-space radius
\beq
\label{eqn:lrad}
\lrad = \sqrt{\frac{l_1^2 + l_2^2 + l_3^2}{3}}
\eeq
and to plot 
\beq
\label{eqn:blrad}
\blrad = \frac{\sum_{l_1,l_2,l_3 \in \lrad} \wisp \bispe}{\sum_{l_1,l_2,l_3 \in \lrad} \wisp},
\eeq
where \wisp\ are the bispectrum weights in an $l$ bin defined in Section \ref{sec:noise}.
The error bar on this one-dimensional quantity is
\beq
\label{eqn:wlrad}
\sigma(\blrad) = \sqrt{ \frac{1}{\sum_{l_1,l_2,l_3 \in \lrad} \wisp}}.
\eeq
We emphasize that this contraction to one dimension is only for display 
purposes; all model fitting and \chisq\ estimation is performed in the 
full three-dimensional $l$ space. 
However, when we calculate \btsz\ from the B12 model to 
study the cosmological scaling and modeling uncertainties, we do use the 
value of this one-dimensional quantity at $\lrad=3000$ as a convenient proxy, 
rather than performing the full three-dimensional fit---see Section \ref{sec:tszsigeight}
for details.

Three features of the bispectrum data are immediately clear from 
Figure \ref{fig:measbispecmodel}: 1) the data are highly inconsistent with 
zero bispectrum in all bands; 2) all bands show evidence of two 
signal components, namely a component
that is larger at low \lrad\ than high \lrad\ and is roughly consistent with a power law in \lrad, 
and a flat-in-\lrad\ component consistent with a Poisson
point-source component 
(note that a signal that is flat in $l$ will also be flat in \lrad); 
3) the power-law component is negative 
at 95 and 150~GHz but positive at 220~GHz, as would be expected 
from a bispectrum dominated by tSZ at 95 and 150~GHz and by clustered
CIB at 220~GHz. 

The results of fitting this data using the multi-frequency model from Section \ref{sec:modeling} are 
discussed below. However, 
to make a reasonably model-independent statement
about the preference for these two components in the data, we first fit each band's 
data individually to a toy model that includes a
flat component and a simple power law in $l$, with a power-law index chosen to 
match the observed signal in all bands.  This turns out to be roughly
$\bisp \propto (l_1 l_2 l_3)^{2/3}$ or $B \propto l^2$ in the equilateral configuration 
($l_1=l_2=l_3$).
Both components 
are detected strongly in all three bands, with the significance of the 
power-law component ranging from $5 \sigma$ at 220~GHz to $9 \sigma$ at 150~GHz.

To assess whether the 
data still prefer a power-law bispectrum component with the most significantly
detected clusters masked, we estimate the bispectrum in each band while masking 
all clusters with $\mtwoh > \threshmass$, which is
very close to a cut at signal-to-noise of five in the \citet{reichardt13} catalog. 
With this level of masking, the detection significance of the power-law 
component at 95 and 150~GHz data is much reduced but still 
1-2$\sigma$ in each band.

Perhaps most intriguing is the detection of a power-law
component in the 220~GHz data, which
is near the tSZ null and should not be measuring a tSZ bispectrum.  We 
interpret this signal as the bispectrum of the clustered CIB, and 
we discuss the implications of this signal in Section \ref{sec:cibinterp}.

\subsection{Results of model fits}
\label{sec:fitresults}
Having established that the bispectrum data in each band contain significant
detections of a power-law component and a flat-in-$l$ component, we move
on to fitting these data to the model described in Section \ref{sec:sigmodels},
using the fitting procedure described in Section \ref{sec:fitting}.  As described 
in Section \ref{sec:systematics}, the linear least-squares fit is repeated many 
times with different realizations of systematic uncertainties, drawn from 
distributions summarized in Table \ref{tab:syst}, or, in the case of the instrument
beam uncertainties, using beam realizations described in 
Section \ref{sec:systematics}.  To minimize uncertainty in interpreting the 
tSZ result due to uncertainties in the assumed halo mass function, we 
repeat the fit using data in which all clusters above $\mtwoh=\bigmass$
masked and a tSZ model template with the mass function truncated 
at that value.  To determine how much of the tSZ bispectrum is coming from 
clusters not already used for cosmological constraints from cluster count 
analyses, we repeat the fit using data and model templates with no 
clusters above $\mtwoh=\threshmass$ (see Section \ref{sec:tszmodel} 
for details). 

The results of the fit with no clusters masked are shown in 
Figure \ref{fig:measbispecmodel} and summarized in Table \ref{tab:fitresults}.
The results of the fit with the two levels of cluster masking and mass function
truncation are summarized in
Tables \ref{tab:fitresults_mask1} and \ref{tab:fitresults_mask2}.
The best-fit parameter values using the nominal values of the beam and 
other sources of systematic uncertainty (see Section \ref{sec:systematics}) 
are shown in the tables, 
along with $1 \sigma$ statistical uncertainties (from the covariance matrix 
in the linear least-squares fit), $1 \sigma$ systematic uncertainties (from 
the scatter in best-fit parameter values over 1000 realizations of systematic uncertainties), 
and the quadrature sum of the two $1 \sigma$ uncertainties.  
The uncertainties on each parameter are the 
square root of the diagonal of the covariance matrix, i.e., the  
uncertainty of each individual parameter marginalized over the others.
The parameter correlation matrix (statistical-only and statistical-plus-systematic)
for the fit results with no cluster masking is shown in Table \ref{tab:fitresults_corr}.
Full statistical and
systematic error covariance matrices are downloadable from the SPT 
website, as are the tSZ bispectrum templates and 
the 1000 beam realizations used in the fits.

\input{tab_fitresults_all.tex}

\input{tab_fitresults_8e14.tex}

\input{tab_fitresults_3e14.tex}

\input{tab_fitresults_corrmat.tex}

\input{tab_fitresults_chisqall.tex}

\subsubsection{Best-fit thermal SZ amplitudes}
\label{sec:tszampl}
We discuss the cosmological implications of our tSZ bispectrum measurement in 
Section \ref{sec:tszcosmo}; here we briefly discuss the best-fit amplitudes 
at the three different masking levels, and we compare
the best-fit amplitude with no masking to the measurement of the tSZ real-space
three-point function (skewness) in ACT data from \citet{wilson12}. 

First, we note that
the best-fit amplitudes at each masking level (no clusters masked, clusters above
\bigmass\ masked, clusters above \threshmass\ masked) relative to the model
prediction for that level of masking are statistically consistent with one another
and indicate a lower tSZ bispectrum amplitude than predicted by the fiducial
model. The implications of this result are discussed in Section \ref{sec:tszcosmo}.
The model predicts a tSZ bispectrum amplitude at 
$l_1=l_2=l_3=3000$ and 
152.8~GHz (the tSZ-weighted band center of the SPT 150~GHz band) of
$\tsztemplnomask$, $\tsztemplhugemask$, and 
$\tsztempldetmask \times 10^{-11} \muk^3$ for the three masking levels.
The model prediction for $\btsz(\lrad=3000)$ and 152.8~GHz is 
$\tszlradtemplnomask$, $\tszlradtemplhugemask$, and 
$\tszlradtempldetmask \times 10^{-11} \muk^3$ for the three masking levels.
The best-fit results from Tables \ref{tab:fitresults}-\ref{tab:fitresults_mask2} 
therefore translate to 152.8~GHz tSZ amplitudes of  $\tszmeasnomask$, $\tszmeashugemask$, and 
$\tszmeasdetmask \times 10^{-11} \muk^3$ 
at $l_1=l_2=l_3=3000$ and values of 
$\tszlradmeasnomask$, $\tszlradmeashugemask$, and 
$\tszlradmeasdetmask \times 10^{-11} \muk^3$ 
for $\btsz(\lrad=3000)$
for the three masking levels.

Roughly 1/3 of the total tSZ bispectrum is coming from clusters 
below the mass threshold used for the cosmological constraints in 
\citet{reichardt13}, implying that cosmological constraints 
from the tSZ bispectrum do contain information beyond what is already 
measured using cluster counts. This would appear to be somewhat inconsistent with Figure 3
in B12, which shows that less than 10\% of the tSZ skewness spectrum 
at $l=3000$ is predicted to come from clusters with
$M_{500}(\rho_\mathrm{crit}) < 2 \times 10^{14} M_\odot / h$.
(roughly equivalent to the \citealt{reichardt13} mass threshold of 
$\mtwoh = \threshmass$). However, the contribution in mass and redshift 
to our measurement of the full three-dimensional bispectrum is weighted
slightly differently than the contribution to the skewness spectrum at 
$\l=3000.$ When we calculate $B(<z, >\mtwoh)$ using $\btsz(\lrad=3000)$
(which tracks the full three-dimensional measured bispectrum very closely,
see Section \ref{sec:tszsigeight} for details), we find that the prediction is that 
roughly 25\% of our measured signal should come from clusters below
the \citet{reichardt13} mass threshold, consistent with what we observe.
The general statement that the bispectrum is dominated by massive, 
low-redshift clusters still holds when $\btsz(\lrad=3000)$ is used as the
observable: in our model, 75\% of the $\btsz(\lrad=3000)$ signal is predicted 
to come from clusters with $\mtwoh > \threshmass$ and $z < 0.6$.

To compare our Fourier-domain three-point function (i.e., bispectrum) tSZ amplitude 
to the real-space three-point function (skewness) of the tSZ measured in ACT
data by \citet{wilson12}, we first take the $l$-space filter shown in Figure 1 of
\citet{wilson12} and calculate the real-space skewness that should be observed
in a map convolved with this filter, if our tSZ bispectrum template is correct.
To calculate this, we create a three-dimensional bispectrum filter from the one-dimensional 
\citet{wilson12} filter
(by defining $F_\mathrm{bisp}(l_1,l_2,l_3)=F_{1d}(l_1) F_{1d}(l_2) F_{1d}(l_3)$), 
multiply the predicted tSZ bispectrum by this three-dimensional filter, and calculate
$\langle T^3 \rangle$ 
following \citet{komatsu01}:
\begin{eqnarray}
\left \langle T^3_\mathrm{tSZ,filt} \right \rangle &=& \frac{1}{2 \pi^2}
\sum_{l_1 l_2 l_3}
\left (l_1 l_2 l_3 \right )
\left (   
  \begin{array}{ccc}
  l_1 & l_2 & l_3\\
  0 & 0 & 0 
  \end{array}
\right ) \times \\
\nonumber && F_\mathrm{bisp}(l_1,l_2,l_3)
B(l_1, l_2, l_3).
\end{eqnarray}
To evaluate the Wigner 3-$j$ symbol, we use the high-$l$ approximation in Equation 8 of B12.

The predicted skewness from our no-masking tSZ bispectrum template multiplied 
by the bispectrum version of the \citet{wilson12} filter is $\tszskewtemplspt \ \muk^3$ at 152.8~GHz.
At the ACT band center of 148~GHz, the template and filter predict $\tszskewtemplact \ \muk^3.$ 
Given the amplitude we measure (in no-cluster-masked data) relative to the model prediction, 
and assuming the shape of the bispectrum model template is correct, our bispectrum
measurement corresponds to a real-space tSZ skewness at 148~GHz of 
$\tszskewmeasact \pm \dtszskewmeasact \ \muk^3$, or 
$\tszskewmeasact \pm \dtszskewmeasactsv \ \muk^3$ when we add the 20\% sample
variance uncertainty estimated in Section \ref{sec:tszsampvar}.
This is consistent with the value of $-31 \pm 14 \ \muk^3$ (also including
sample variance uncertainty) reported in \citet{wilson12}. 

Given the detection in this work
of a significant clustered CIB bispectrum, it is likely that the \citet{wilson12} tSZ skewness
measurement is biased low by $\sim$15\% (see Section \ref{sec:tszcib} for details). 
\citet{wilson12} recognized this potential bias.
Their approach was to correct for it
by measuring the CIB bispectrum in the \citet{sehgal10} simulations
(scaled down by a factor of 1.7 in temperature) and subtracting that value from the tSZ skewness
before using that number in cosmological fits. The correction was $-3.9 \pm 0.1 \ \muk^3$, 
or 11\% of the corrected tSZ skewness value of $-35 \pm 14 \ \muk^3$. This bias estimate
is roughly consistent with the prediction from the CIB bispectrum measured here, and the
corrected \citet{wilson12} tSZ skewness is even more consistent than the uncorrected 
one with the tSZ bispectrum we measure.

\subsubsection{\chisq\ and goodness-of-fit values}
\label{sec:chisq}
The \chisq\ values from the fits using the three levels of cluster masking
and the nominal values of beams and other sources of systematic 
uncertainty are summarized in Table \ref{tab:fitresults_chi}.  The table includes values of
absolute \chisq, reduced \chisq, and 
\dchisq\ between the best fit and the null hypothesis.  The formal probabilities to exceed
(PTEs) associated with the reduced \chisq\ for all three levels of cluster
masking are vanishingly small, but a small 
underestimate of the bispectrum variance would cause an otherwise 
good fit to have such a PTE.  Since the \chisq\ of the bispectrum fit will scale as
the amplitude of the 
map noise and simulated Gaussian sky signal to the $-6$ power, 
the observed \chisq\ excess is consistent with 
a percent-level misestimate in the noise or the Gaussian sky amplitude.

The map noise used to estimate the bispectrum variance is 
taken from the same data used to construct the real maps used to measure
the bispectrum (see Section \ref{sec:noise} for details). Thus, 
we can calculate a \chisq\ from the ``measured"
bispectrum of every simulated map and use the scatter in \chisq\ across the 
simulations as a measure of how closely the estimated bispectrum variance 
from map noise should match the map-noise variance contribution to the real
data. None of the 100 simulations had \chisq\ as high as the data, so it is unlikely
that the excess \chisq\ is due to a map noise misestimate. On the other hand, it
is plausible that the Gaussian sky amplitudes could be mismatched between the
simulations and the data at the 1\% level. Our estimate for the amplitude of CMB  
fluctuations in SPT maps of
this 800 \degs\ region is limited by cosmic variance and the uncertainty on
our absolute calibration (which is 1-2\% in temperature, see Section \ref{sec:systematics}), 
while our estimate for the Poisson point-source power
is limited by calibration and beam uncertainties and by the lack of high-precision measurements of the 
Poisson amplitude at millimeter wavelengths.

Alternatively, the excess \chisq\ could be evidence of departures 
from our models for either tSZ or clustered CIB. However, 
the total $\Delta \chisq$ between the null hypothesis of zero bispectrum
and the best-fit model is smaller than the difference between the \chisq\ of
the best-fit model and a \chisq\ that would reduce to 1.0.
Misestimates of the beam or 
filter transfer function could also be responsible.
We can test this hypothesis by examining the \chisq\ values for each systematic
realization, and we do not find any trend of \chisq\ with beam realizations; in fact, the 
total spread in \chisq\ across all systematic realizations is roughly $\pm 10$, indicating that 
none of the identified sources of systematic uncertainty are responsible for 
the excess \chisq.  Finally, the excess \chisq\ is not strongly concentrated in one frequency 
band or region of $l$-space. This points to a slight underestimate in the simulated Gaussian 
sky signal as the source of the excess \chisq.

\section{Discussion}
\label{sec:discussion}
In this section, we discuss the implications of each bispectrum component 
measured in the 
multi-band fit. We begin by comparing the amplitude of the Poisson point-source bispectrum in 
each band to model predictions; we then discuss the clustered CIB 
amplitude, both as an interesting signal in its own right and as a possible
contaminant to the tSZ amplitude; finally, we implement the analysis 
introduced in B12 and use the tSZ bispectrum amplitude constraint 
to measure \sigeight\ and to sharpen the kSZ amplitude measurement
from R12.

\input{tab_poissresults.tex}

\subsection{Poisson point-source component amplitudes vs.~model predictions}
\label{sec:poissresults}
Given a model for the number of sources in a given flux interval per 
unit solid angle $dN/dS/d\Omega$, we can predict the contribution 
to the bispectrum from the Poisson component of
those sources.  We can then compare these predictions to the results 
in Tables \ref{tab:fitresults}-\ref{tab:fitresults_mask2} as a test of the source 
models.  Because the Poisson contribution to the bispectrum is weighted by 
the individual source fluxes cubed---compared
to the source-flux-squared weighting in the power spectrum---this 
test is largely independent of power-spectrum-based tests of source 
models.  And, because the bispectrum in this work is calculated with all sources 
detected above $5 \sigma$ masked, the bispectrum constraints on models are 
nearly independent of 
source-count constraints from the same data (\citealt{vieira10,mocanu13}).

In Table \ref{tab:poiss}, we show the predicted bispectrum amplitude
in all three SPT bands from two models of radio-loud sources (\citealt{dezotti10} and 
\citealt{tucci11}), two models of radio-quiet dusty sources 
(\citealt{bethermin11} and \citealt{bethermin12}), and the four possible combinations
of these models.  We also repeat the measured values of the Poisson 
bispectrum component from Tables \ref{tab:fitresults}-\ref{tab:fitresults_mask2} 
for comparison.  In some cases, the model predictions are at the nominal SPT
bands, in others, the predictions are for the analogous {\it Planck} bands; in 
either case, we transform the models to the appropriate effective SPT band center
for that source family, using assumed spectral indices of $\alpha_\mathrm{radio}=-0.5$
and $\alpha_\mathrm{dusty}=3.5$, consistent with the results of \citet{vieira10} and
R12.  We also use this assumed spectral behavior to transform the 150~GHz 
flux cut to the other two bands.

Two things are immediately clear from Table \ref{tab:poiss}.  The first is that 
only in the 150~GHz band is the bispectrum expected to contain a significant contribution
from both families of sources: at 95~GHz, the dusty sources are expected to contribute $<5\%$
of the total amplitude, while at 220~GHz, the radio-loud sources are expected to contribute 
$<1\%$ of the total amplitude.  The second is that, while the 
\citet{dezotti10} model prediction is within $1 \sigma$ of the measured
95~GHz measurement,
there are significant differences between the model predictions and the measured 
amplitudes in all other cases.

We first investigate whether this discrepancy between measured and predicted 
Poisson bispectrum amplitudes could be due to effects that are not included in the
measured uncertainty, in particular sample variance and the effect of a varying flux
cut.  
Near the $\sim$6 mJy (at 150~GHz) flux cut used in this work, the dependence of radio source counts on flux is shallow (\citealt{dezotti10,vieira10,tucci11,mocanu13}).  This means that
the radio-source bispectrum will be dominated by the brightest (and rarest) sources
just below the flux cut.  This will tend to make the radio source bispectrum more sensitive
to sample variance and to the fact that, while the real flux cut used in this work varies
from field to field by $\sim$10\%, we calculate the predicted bispectrum from source models
using a single flux cut.  We estimate the magnitude of both of these effects by simulated
observations of many 800-\degs\ patches of sky containing sources drawn from the 
source count models listed in Table \ref{tab:poiss}.  In one set of simulated 
observations, we use the nominal 6~mJy 150~GHz source cut in every field; in another
set, we use the actual 150~GHz source cut used in each individual field in this work; these 
cut levels range from 5.7 to 6.6 mJy.  Both the bispectrum sample variance (calculated 
as the scatter among the best-fit Poisson bispectrum in all simulated observations) and 
the effect of the different flux cuts were largest at 95 GHz---not surprising, given that this
is the band in which the radio source contribution is largest---but even in that band, the
square root of the sample variance was only $2\%$ of the average Poisson bispectrum,
and the difference between using the true flux cut for every field and using the nominal 
flux cut was only $6\%$.  
The effect of sample variance on the bispectrum due to dusty sources
should be significantly smaller than this, because the predicted dusty source bispectrum
is dominated by sources well below the flux cut used here.
We conclude that the discrepancy between model predictions
and the measured Poisson bispectrum cannot be explained by sample variance and 
varying flux cuts.

The simplest modifications to the source models that would bring them in line with 
the bispectrum results in this work would be: 1) to steepen the spectral behavior used to 
extrapolate the dusty source behavior from higher frequencies to the SPT bands, thus 
reducing the dusty-source bispectrum by a small amount at 220~GHz and a larger amount 
at 150~GHz; and 2) to assume a slightly shallower spectral index in extrapolating 
counts at radio frequencies to 95~GHz,
particularly for the \citet{tucci11} model. 
It remains to be seen whether such modifications
would be compatible with constraints from other measures of point-source behavior
such as number counts and the power spectrum. An interesting possibility for future work 
is to combine these probes into a simultaneous constraint on source models.

\subsection{The clustered CIB bispectrum}
\label{sec:cibinterp}
Measurements of the two-point function of CIB clustering (either the real-space
two-point angular correlation function or the angular power spectrum) are currently
providing key constraints on the relationship between star-forming galaxies and 
their dark-matter halos, or, equivalently, on the relationship between luminosity
and mass in star-forming galaxies (see, e.g., \citealt{viero12b} and references 
therein). Equally interesting will be constraints on this relationship from the 
clustered CIB bispectrum. As is the case for tSZ and the Poisson point-source component, the relative
weighting of sources that contribute to the clustered CIB power spectrum and 
bispectrum will be different---with the bispectrum generally sensitive to brighter
sources because of the $S^3$ weighting---implying that the two probes can 
give independent constraints on models of the mass-luminosity relationship.

There are currently no physically motivated predictions for the clustered CIB 
bispectrum in the literature---although \citet{lacasa12} present a heuristic model
based on power spectrum measurements, which we adopt as our fitting template.
However, any model of the mass-luminosity relationship of star-forming galaxies
that can predict the clustered CIB power spectrum should also be able to predict
the bispectrum, and we expect the measurement of the clustered CIB bispectrum
in this work to provide new constraints on such models.

For now, our main conclusions about the clustered CIB bispectrum are that it is 
clearly detectable in 800~\degs\ of 220~GHz data at SPT noise levels, and that
the angular shape of the signal is fit reasonably well by a pure power law.
Our model, based on the ansatz of 
\citet{lacasa12} and described in Section \ref{sec:cibmodel}, 
has $\bclust(l_1,l_2,l_3) \propto (l_1 l_2 l_3)^{-n/2}$,
with $n=1.2$ and scales with observing frequency as $\nu^{3 \alpha}$, with
$\alpha=3.72$. As shown in Figure \ref{fig:measbispecmodel}, this is by eye a reasonable fit 
to the data. Although the PTE associated with the reduced \chisq\ is vanishingly 
small, as discussed in Section \ref{sec:chisq}, this is consistent with a very small
noise misestimate. Neither changing the power-law index of the angular shape of the 
signal nor changing the assumed frequency scaling results in significant
improvements in \chisq\ ($| \Delta \chisq | < 2$ for a $2 \sigma$ shift in $\alpha$
or a $30 \%$ change in the power-law index). 
For the fiducial model, using the results of the no-cluster-masking
multi-band fit, the amplitude of the clustered CIB bispectrum at $l_1=l_2=l_3=3000$ and
219.6~GHz (the CIB-weighted band center of the SPT 220~GHz band) is 
$(\cibmeasnomask \pm \dcibmeasnomask) \times 10^{-10} \ \muk^3.$ Using the $> \bigmass$ masking 
result, the amplitude of the clustered CIB bispectrum at at $l_1=l_2=l_3=3000$ and
219.6~GHz is $(\cibmeashugemask \pm \dcibmeashugemask) \times 10^{-10} \ \muk^3.$

\subsubsection{The clustered CIB as a contaminant to the thermal SZ bispectrum}
\label{sec:tszcib}
The clustered CIB bispectrum that we detect in the SPT 220~GHz band will also 
contribute to the bispectrum at 150~GHz (and, to a much lesser extent, at 95~GHz).
With the assumed frequency scaling of $B \propto \nu^{3 \alpha}$, with $\alpha=3.72,$
the ratio of clustered CIB bispectrum amplitude in the 150 and 95~GHz bands 
compared to that in the 220~GHz band should be 0.031 and 0.001, respectively,
meaning we would expect roughly $1 \times 10^{-11} \ \muk^3$ of clustered CIB
bispectrum at $l=3000$ and 150~GHz (compared to an expected tSZ bispectrum
of $-5.4 \times 10^{-11} \ \muk^3$)
and a negligible contribution of $< 10^{-12} \ \muk^3$
at 95~GHz. This implies that, if we were to neglect the clustered CIB, we would 
underestimate the tSZ bispectrum amplitude by roughly 20\% at 150~GHz (because
the bispectrum shape of the tSZ and clustered CIB are similar, but with different 
polarities). If we fit both the 95 and 150~GHz bispectra individually to a 
two-component model consisting of tSZ and a Poisson point-source term, the results are
as expected. The best-fit tSZ amplitude with no clusters cut from the 95~GHz data alone 
($\btsz = 0.54 \pm 0.07$) is consistent 
with the multi-band fit results ($\btsz = 0.54 \pm 0.04$, cf.~Table \ref{tab:fitresults}), 
but the best-fit tSZ amplitude from the 150~GHz data 
alone ($\btsz = 0.43 \pm 0.05$) is $\sim$20\% lower than the multi-band result.

The multi-band fit properly accounts for this, and if the CIB behavior were very 
different from what we assume in the model, this would manifest in a noticeably
poorer \chisq\ in the multi-band fit relative to single-band fits, which we do not see.
In particular, if there were a significant level of tSZ-CIB correlation in the bispectrum,
we would expect to see a very different best-fit tSZ amplitude from the multi-band fit
from what we obtain with the 95~GHz-only fit; in fact, our cosmological constraints
detailed below would not substantively change if we used only 95~GHz data in
the fit, although the error bars would increase somewhat.
We conclude that, at the current level of statistical precision, we see no evidence
that our model of the CIB is inadequate or that the CIB is significantly biasing our measurement 
of the tSZ bispectrum. More complicated models involving spatial correlations 
between the sources of the tSZ and CIB bispectra, different CIB bispectrum shapes, 
and spectral behavior beyond a single spectral index will be explored in future 
analyses which include measurements of the cross-bispectra among the three
SPT bands (in addition to the auto-bispectra considered here).

As noted in Section \ref{sec:tszampl}, according to our model and fit results, the 148~GHz
tSZ skewness measurement of \citet{wilson12} in ACT data is likely to be biased low 
by roughly 15\% (less than the 20\% we see at 152.6~GHz, the CIB-weighted band center of 
the SPT 150~GHz band, due to the very steep frequency scaling of the CIB). 
This is slightly smaller than the statistical + sample variance uncertainty
in that result---and significantly smaller than the difference in tSZ skewness predicted
by the range of ICM models they consider; we also note that \citet{wilson12} 
included an 11\% correction for CIB contamination in the tSZ skewness value they
used in cosmological fits.

\subsection{Cosmological interpretation of the thermal SZ bispectrum amplitude}
\label{sec:tszcosmo}
In this
section, we use our measurement of \btsz, the tSZ bispectrum amplitude, to
place a constraint on \sigeight\ and to predict the tSZ power spectrum
amplitude, \atsz. We use this \atsz\ prediction to break degeneracies between 
tSZ and kSZ in measurements of the CMB power spectrum.  
First, however, we 
need to estimate two properties of the \atsz\ and \btsz\ distributions, namely the 
sample variance of \btsz\ and the covariance of \atsz\ with \btsz\ over the same
patch of sky. 

\subsubsection{Sample variance in the measurement of \btsz}
\label{sec:tszsampvar}
To estimate the sample variance contribution to our measurement of \btsz, we use
a set of cosmological simulations. These simulations use the same gas physics prescription, gas physics
parameter settings, and cosmological parameter settings that went into the template
predictions used in the model fitting procedure described in Section \ref{sec:fitting}.
The simulations cover an octant of sky, from which we extract 40 independent 100~\degs\ 
fields. We run the bispectrum estimator over these fields with the same
weighting used in running the estimator on the 150~GHz data. We fit each of the resulting 
40 bispectrum measurements to the predicted template, again using the weights from
the 150~GHz data and restricting the fit to $l \le 4000$ to roughly account for the 
effects of the Poisson point-source component in the 
fit to the data. The calculation is performed with ten levels of cluster
masking, ranging from no masking to masking clusters above 
$\mtwoh \ge 2 \times 10^{14} M_\odot / h.$

We then estimate the scatter in 800~\degs\ regions for each cluster masking level
by grouping the 40 amplitudes into five independent groups of eight amplitudes, 
averaging each group, and calculating the scatter among groups. This is a noisy
estimate of the sample variance. In particular, the sample variance as a function of 
mask threshold is affected by the masking of individual clusters at high enough mass 
that only a few such clusters exist in the entire octant simulation. For this reason, we
fit a smooth function to the measured sample variance as a function of masking, and
we report our sample variance as the best-fit value at the three masking levels used 
for the data.
The fractional scatter in \btsz\ due to 
sample variance is estimated to be $0.20$,  $0.15$, and $0.06$ at the three levels of
cluster masking (none, $> \bigmass$, $> \threshmass$) used for the data. We note that the value 
for the no-masking case is consistent with the sample variance of the tSZ skewness
measured by \citet{wilson12}, given the relative sky coverage of the two 
analyses. (\citealt{wilson12} measured 41\% scatter for 239~\degs\ 
as compared to 20\% for 837~\degs\ in this work.)

There is also a potential
systematic uncertainty in \btsz\ introduced when clusters are masked. If the mass estimates
for all clusters are systematically biased, then the mask threshold used in the 
model to which the data is compared is different than the mask threshold actually
used in the data. 
The uncertainty on the overall scaling between the
SPT measure of SZ signal and cluster mass, as estimated in
\citet{benson13}, is 10\% at $z=0$ and 15\% at $z=1$. This includes the
contribution from the uncertainty in the weak-lensing-derived
normalization of the X-ray $Y_X$-mass relation. Although the bispectrum
is dominated by low-redshift clusters, we adopt the 15\% uncertainty to
be conservative. Using our model for the tSZ bispectrum---specifically 
$\btsz(\lrad=3000)$---as a function of mass and redshift, we estimate the
effect of the systematic mass uncertainty on our determination of
\btsz\ by integrating the model prediction over redshift and up to
three different maximum mass values: the nominal value we use, and
that value times 1.15 and 0.85. We find that a 15\% error in mass leads
to a $\sim$10\% error in \btsz\ for our $\mtwoh=\bigmass$ mass cut and a $\sim$30\%
error for our $\mtwoh=\threshmass$ cut. We add this to the total error budget 
on \btsz\ in all calculations that follow.

As discussed in Section \ref{sec:tszmodel}, the sample variance and 
statistical-plus-systematic-plus-mask-threshold
uncertainties on \btsz\ go in opposite directions as more clusters are masked, implying
that there is an optimal mass threshold, at which level the total 
error on \btsz\ is minimized. Table \ref{tab:tszplussampvar}
shows the fractional uncertainty from each of these sources (and the quadrature sum of all 
of them) for the three masking levels used for the data. Among these three masking 
levels, the total uncertainty is smallest when clusters above \bigmass\ are masked.

To investigate whether a different threshold would be optimal, 
we calculate the total uncertainty at the ten mask thresholds used in the 
sample variance calculation. At each mask threshold, we re-calculate the
uncertainty in \btsz\ due to cluster mass systematic error, and we scale the
fractional statistical error by the best-fit \btsz\ in the simulations using that
mask threshold. This calculation implies that the total fractional scatter has a broad
minimum between $6 \times 10^{14} M_\odot / h$ and $9 \times 10^{14} M_\odot / h$. 
We use the \bigmass\ cut values of \btsz\ for our cosmological results.

\subsubsection{Covariance between \atsz\ and \btsz}
\label{sec:abcov}
We estimate the covariance between \atsz\ and \btsz\ using the halo-model 
approach of \citet{kayo13}, together with the gas physics prescription of B12.
We find that the square root of the fractional covariance between
\atsz\ with no clusters cut and \btsz\ with clusters above \bigmass\ cut 
is $\sim$0.06. This is small compared
to the other sources of uncertainty in our prediction of \atsz, and we ignore 
it for this analysis. Details of the \atsz/\btsz\ covariance calculation are given 
in the Appendix.

\input{tab_tszplussampvar.tex}

\begin{figure*}
\begin{center}
{\includegraphics[width=0.48\textwidth]{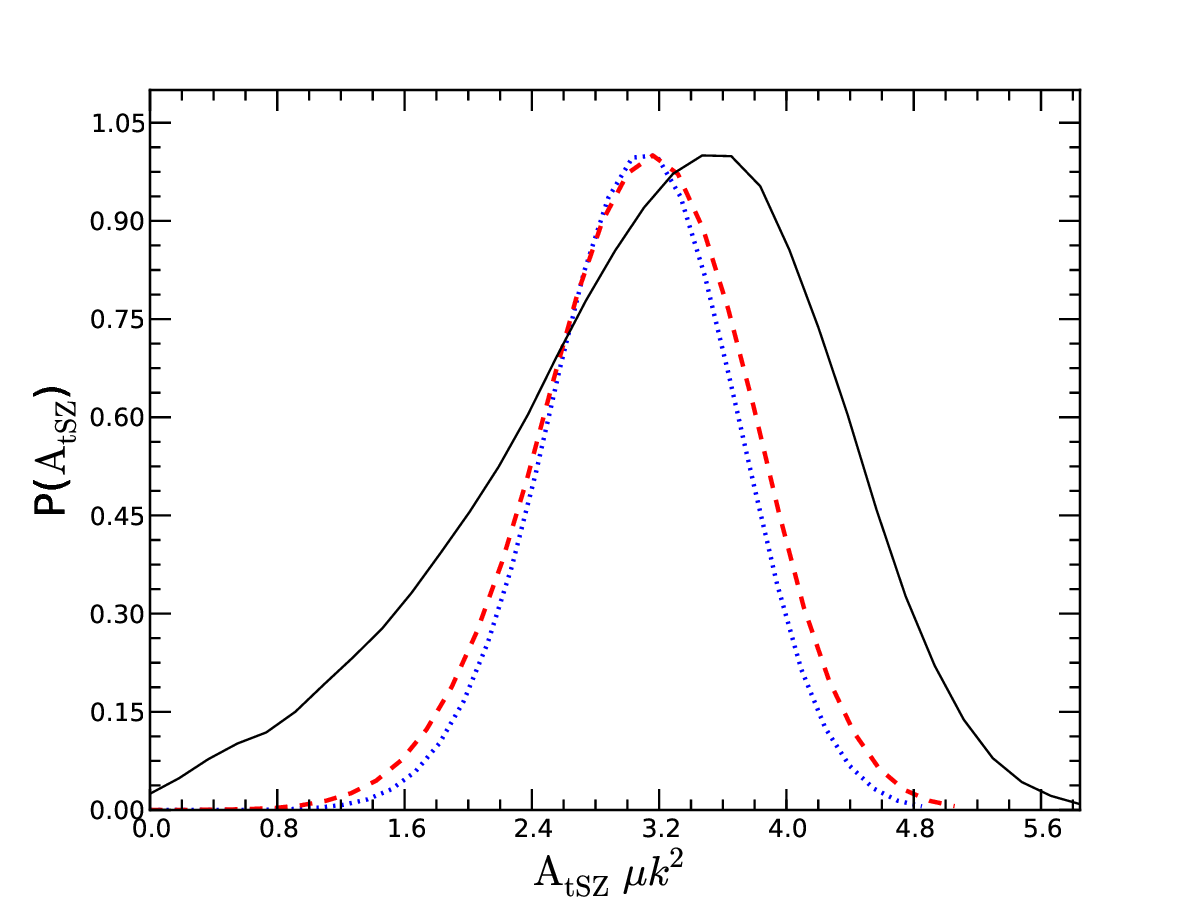}}
{\includegraphics[width=0.48\textwidth]{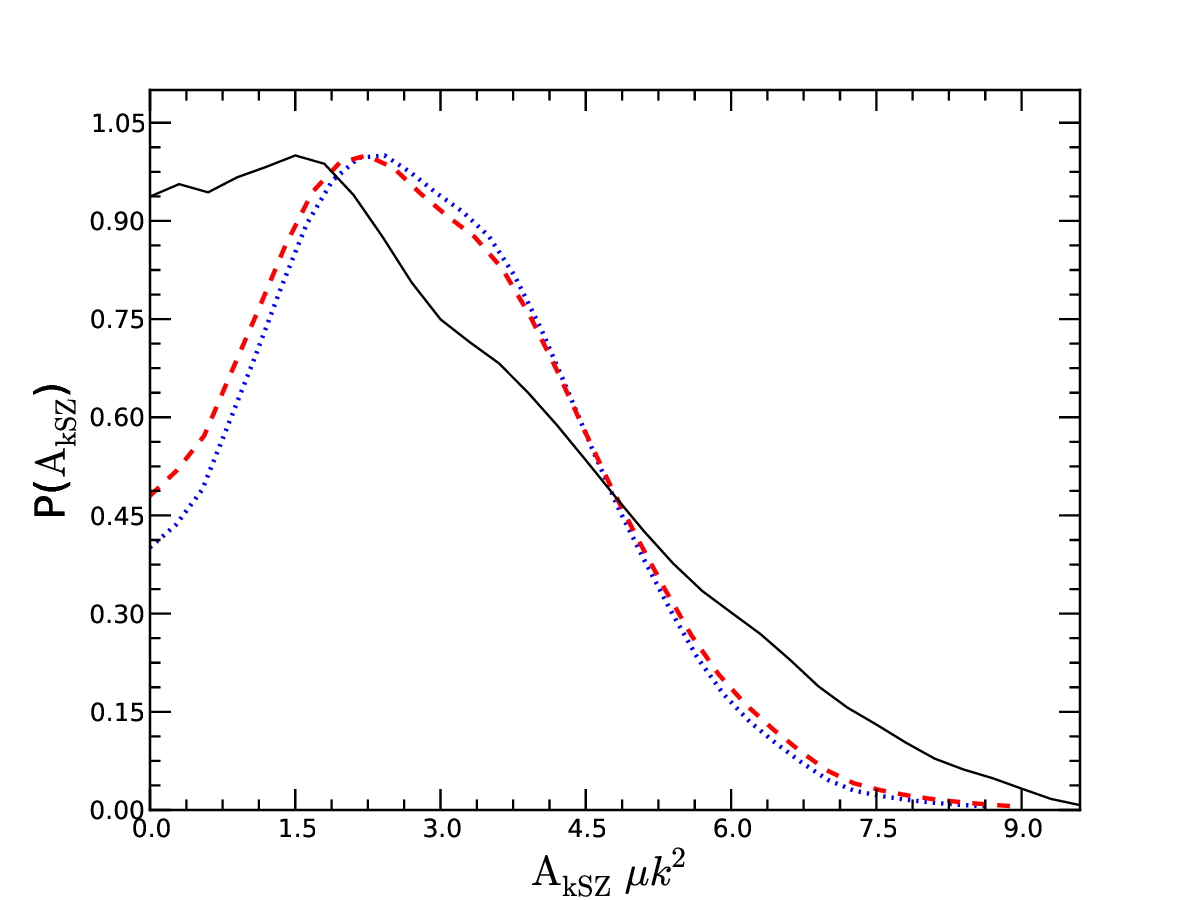}}
\end{center}
\caption{
One-dimensional posterior probability distributions from R12 for \atsz\ (left)
and \aksz\ (right), in the case in which the spatial correlation
between tSZ and CIB was a free parameter in the R12 fits,
before and after applying the bispectrum-based prior 
in Equation \ref{eqn:atszpred}. The black (solid) line shows 
constraints with no bispectrum information added;
the blue (dotted) line shows the constraints assuming the
default 11\% modeling uncertainty in \atsz\ for fixed \btsz; the red (dashed) line shows 
constraints assuming the extreme 18\% modeling uncertainty.
In all cases, adding constraints from the bispectrum 
data improves the power-spectrum-only constraints.
}
\label{fig:ksz}
\end{figure*}

\subsubsection{A \sigeight\ constraint from the thermal SZ bispectrum}
\label{sec:tszsigeight}
In this Section, we translate our measurement of the 
amplitude of the tSZ bispectrum into a constraint on \sigeight.  Section 
\ref{sec:tszmodel}, summarizing B12, describes our model for the 
tSZ bispectrum and its dependence on cosmological parameters.  We 
use this model to determine the cosmological scaling of the tSZ bispectrum 
amplitude as well as the modeling uncertainty associated with our measurement.  
Rather than replicating the full three-dimensional fit as it is performed for the 
data, we work with the one-dimensional quantity $\btsz(\lrad=3000)$ when 
determining the cosmological scaling and modeling uncertainty.  
This allows for a more straightforward generalization to other data 
sets and experiments.  Note that B12 employed a similar approach but used 
the value of the tSZ skewness spectrum at $l=3000$ rather than $\btsz(\lrad=3000)$.  
The choice of $\btsz(\lrad=3000)$ as a proxy for the amplitude estimated from the 
full three-dimensional fit is supported by tests with simulated data showing that 
the two observables track each other with less than 2\% scatter in their ratio.

The modeling uncertainty and cosmological scaling calculated here for \btsz\ are 
slightly different than those quoted in B12.  This is because a different proxy 
observable is used, and the most massive clusters are not included in the analysis 
presented here.
We find a modeling uncertainty of 36\%, compared to the 33\% in B12 for 
the skewness spectrum at $l=3000$ and no cluster cut.
For a six-parameter flat $\Lambda$CDM model, we find the cosmological scaling 
of \btsz(\lrad=3000) with clusters above \bigmass\ cut to be
\begin{eqnarray}
\btsz&&(\lrad=3000; \ \mtwoh < \bigmass) \\
\nonumber \propto && 
\left ( \frac{\sigeight}{0.8} \right )^{9.1} 
\left ( \frac{\Omega_b}{0.045} \right )^{3.82} 
\left ( \frac{h}{0.71} \right )^{2.25} \\
\nonumber \times&& 
\left ( \frac{n_s}{0.97} \right )^{-1.12}
\left ( \frac{\Omega_m}{0.27} \right )^{-0.27},
\end{eqnarray}
(with no measurable dependence on $\tau$). Compared to the 
scaling in B12 (for the skewness spectrum at $l=3000$ and no 
cluster cut), the primary difference is a slightly shallower scaling 
with \sigeight ($\btsz \propto \sigeight^{11.6}$ in B12).

To compare our result to the model predictions, 
we first translate our best-fit amplitude with respect to
the model prediction for the tSZ bispectrum (presented in Section \ref{sec:fitresults} 
and Tables \ref{tab:fitresults}-\ref{tab:fitresults_mask2}) into a 
value of $\btsz(\lrad=3000)$ by multiplying the 
model by our best-fit amplitude and summing the model as in Equation \ref{eqn:blrad}, 
using the weights from the 150~GHz data.
We use the result from our fit with all clusters above
$\mtwoh = \bigmass$ masked, and we use the mask threshold error, 
sample variance, and modeling 
uncertainty estimated for that mass cut.
We marginalize over all cosmological parameters other than \sigeight.
Although the dependence of \btsz\
on \sigeight\ is far stronger than on the other parameters,
the dependence on $\Omega_b$ is strong enough that we place a 
{\it WMAP7} \citep{larson11} prior on 
$\Omega_b h^2$ and a prior on $h$ from \citet{riess11}.

Taking into account the full uncertainty (statistical + systematic + 
mask threshold + sample variance) 
on our measurement, and adding the 36\% modeling uncertainty,
the resulting constraint on \sigeight\ is
\beq
\sigeight = \sigeightmeas \pm \sigeightunc.
\eeq
The uncertainty on our determination of \sigeight\ is dominated by 
the assumed 36\% modeling uncertainty. Given the steep scaling 
of \sigeight\ with \btsz\ and the mild dependence on 
other parameters, in the absence of modeling uncertainty, we 
would expect to achieve a $\sim$2\% constraint on \sigeight\ from 
our 21\% constraint on \btsz, compared 
to the 4\% constraint achieved when modeling uncertainty is included.

This constraint on \sigeight\ from the tSZ bispectrum is comparable in 
significance to, and statistically consistent with, other recent determinations of 
\sigeight\ from tSZ and/or the primary CMB. From the primary CMB, in a flat
$\Lambda$CDM model, \citet{hinshaw12} find $\sigeight = 0.821 \pm 0.023$
from {\it WMAP9} data alone, while \citet{story13} find $\sigeight = 0.795 \pm 0.022$
when adding SPT constraints from the primary CMB damping tail to {\it WMAP7} data. Adding constraints
from the tSZ power spectrum to {\it WMAP7} plus earlier damping-tail constraints, 
\citet{shirokoff11} obtain a constraint on \sigeight\ with
statistical precision at the $\pm 0.01$ level, but which varies in best-fit value from $0.77 < \sigeight < 0.80$
depending on the model template used. Simlarly, \citet{dunkley11}, using only 
tSZ power spectrum data, obtain a $\pm 0.05$ constraint (statistical only) but 
find best-fit values from $0.74 < \sigeight < 0.79$, depending on the choice of model template.
Adding SPT cluster counts to {\it WMAP7} and the \citet{keisler11} SPT measurement of
the damping tail, and marginalizing over uncertainty in the X-ray-calibrated tSZ-mass scaling 
relation from \citet{benson13}, \citet{reichardt13} find 
$\sigeight = 0.798 \pm 0.017.$  Combining {\it WMAP7} with ACT-detected clusters and 
marginalizing over uncertainty in a dynamical-mass-calibrated scaling relation, 
\citet{hasselfield13} find $\sigeight = 0.829 \pm 0.024$. 

Finally, in the analysis most directly comparable to the
one presented here, \citet{wilson12} find $\sigeight = 0.79 \pm 0.03$ from a measurement
of the tSZ real-space three-point function (skewness). \citet{wilson12} do not explicitly 
marginalize over modeling uncertainty, but they obtain \sigeight\ constraints using different
gas model prescriptions and find that, even for the extreme case of turning off 
cooling, feedback, and star formation, the constraint on \sigeight\ changes by less than 
$1 \sigma$. The agreement between the SPT and ACT constraints on \sigeight\ from the
tSZ three-point function is not surprising, given the consistency between the measured
Fourier-domain and real-space three-point amplitudes discussed in Section \ref{sec:tszampl}.
This consistency is worthy of note, however, given the very different analysis techniques leading to the two
constraints and the different regions of the sky measured.

\subsubsection{Predicting \atsz\ and sharpening \aksz}
\label{sec:tszpowspec}
Using the approach of B12, we now convert our constraint on the amplitude 
of the tSZ bispectrum into a prediction for \atsz, the amplitude of the tSZ
power spectrum. We then use that prediction to sharpen the R12 constraint 
on \aksz, the amplitude of the kSZ power spectrum. 

We do not apply any cluster cut to the bispectrum-derived prediction for \atsz\ 
or to the measurement of \atsz\ from SPT power spectrum data (which we take
directly from R12). As detailed in B12 (and references therein), the tSZ power
spectrum is dominated by lower-mass halos, so the mass-function uncertainties
at the high-mass end are not as important for the tSZ power spectrum as they 
are for the tSZ bispectrum. 

As in the previous section, we use the value of $\btsz(\lrad=3000)$
as a proxy for the results of the full, three-dimensional fit of our data to the model
predictions for $\btsz(l_1,l_2,l_3)$.
We find a slightly different scaling
between \atsz\ and $\btsz(\lrad=3000)$ with clusters above \bigmass\ 
cut than between 
$\atsz(\mathrm{no \ cut})$ and the tSZ skewness spectrum at $l=3000$ and 
no clusters cut.
Specifically, B12 found 
\beq
\btsz \propto \atsz^{1.4}
\eeq
using the skewness spectrum at $l=3000$ and no clusters cut, 
while we find 
\beq
\btsz \propto \atsz^{1.14}
\eeq
using $\btsz(\lrad=3000)$ and clusters above \bigmass\ cut.
We also find a slightly different uncertainty in our prediction of 
\atsz\ given \btsz, 
namely 11\% for the default B12 gas physics assumptions and 18\%
for the extreme case (as compared to 7\% and 15\% for 
\atsz\ at fixed \btsz\ using the skewness spectrum at $l=3000$ and no cluster cut).

As detailed in B12, the bispectrum measurement acts to
constrain gas model parameters (including redshift evolution, which is
poorly constrained without the bispectrum measurement).  The effects
of the gas model on predictions of \atsz\ and \btsz\ are highly
correlated, so a bispectrum measurement allows us to reduce the
uncertainty on \atsz\ significantly.

R12 report \atsz\ and \aksz\ in terms of the power---when expressed as $D_l = l (l+1)C_l / 2\pi$---at $l=3000$.
Using the best-fit bispectrum tSZ amplitude measurement with clusters above \bigmass\ masked
(including statistical, systematic, and sample-variance uncertainties), 
the prediction for \atsz\ using the default B12 11\% modeling uncertainty is
\beq
\atsz(\mathrm{predicted} \ \mathrm{from} \ \btsz) = \atszbtsz \pm \datszbtsznom \ \muk^2.
\label{eqn:atszpred}
\eeq
Increasing the modeling uncertainty to 18\% increases this error bar to 
$\pm \datszbtszbig \ \muk^2.$

We create a Gaussian prior from this prediction and use it to importance-sample
the posterior probability distributions from R12. In that work, \atsz\ and \aksz\ 
were estimated using two different assumptions about the spatial correlation 
between tSZ and CIB: 1) assuming zero correlation; 2) assuming a single 
correlation coefficient independent of angular scale and allowing that 
coefficient to be a free parameter.
If we importance-sample the zero-correlation result from R12, the improvement 
from the bispectrum prior is small. 
In the case of tSZ-CIB correlation as a free parameter, however, the bispectrum 
prior reduces the tSZ uncertainty by nearly a factor of two and results in a posterior
\aksz\ distribution with a clear non-zero peak (see Figure \ref{fig:ksz}, right panel).
There is some modeling inconsistency in using the \btsz\ constraint from this work, which 
is derived assuming no tSZ-CIB correlation, to improve the R12 \atsz\ constraint 
derived with tSZ-CIB correlation as a free parameter. However, as detailed in 
Section \ref{sec:cibmodel}, we expect tSZ-CIB correlation to affect the bispectrum
far less than the power spectrum, such that we can ignore the effects of 
tSZ-CIB correlation on our measurement of \btsz.

The bispectrum-informed constraints on \atsz\
and \aksz\ from R12, with tSZ-CIB correlation as a free parameter (and using the 
\citealt{shaw12} cooling + star formation template for kSZ), are
\begin{eqnarray}
\atsz &=& \atszbtszrt \pm \datszbtszrtnom \ \muk^2 \\
\nonumber \aksz &=& \akszbtszrt \pm \dakszbtszrtnom \ \muk^2 \\
\nonumber \aksz &<& \akszninefivebtszrtnom \ \muk^2 \ (95 \%).
\end{eqnarray}
These results are fairly insensitive to modeling uncertainty: 
using the extreme 18\% modeling uncertainty instead of the
default 11\% value increases the 
$1 \sigma$ error
on \atsz\ to $\pm \datszbtszrtbig \ \muk^2$,
while the $1 \sigma$ error on \aksz\ is effectively unchanged.
We note that this result is lower in power than the best-fit \aksz\ 
of $5.3^{+2.2}_{-2.4} \ \muk^2$
found by \citet{addison12b} using combined SPT, ACT, {\it Herschel}-SPIRE, 
and {\it Planck} data, but that the two results are consistent at the $1 \sigma$ level.

\begin{figure}
\begin{center}
{\includegraphics[width=0.48\textwidth]{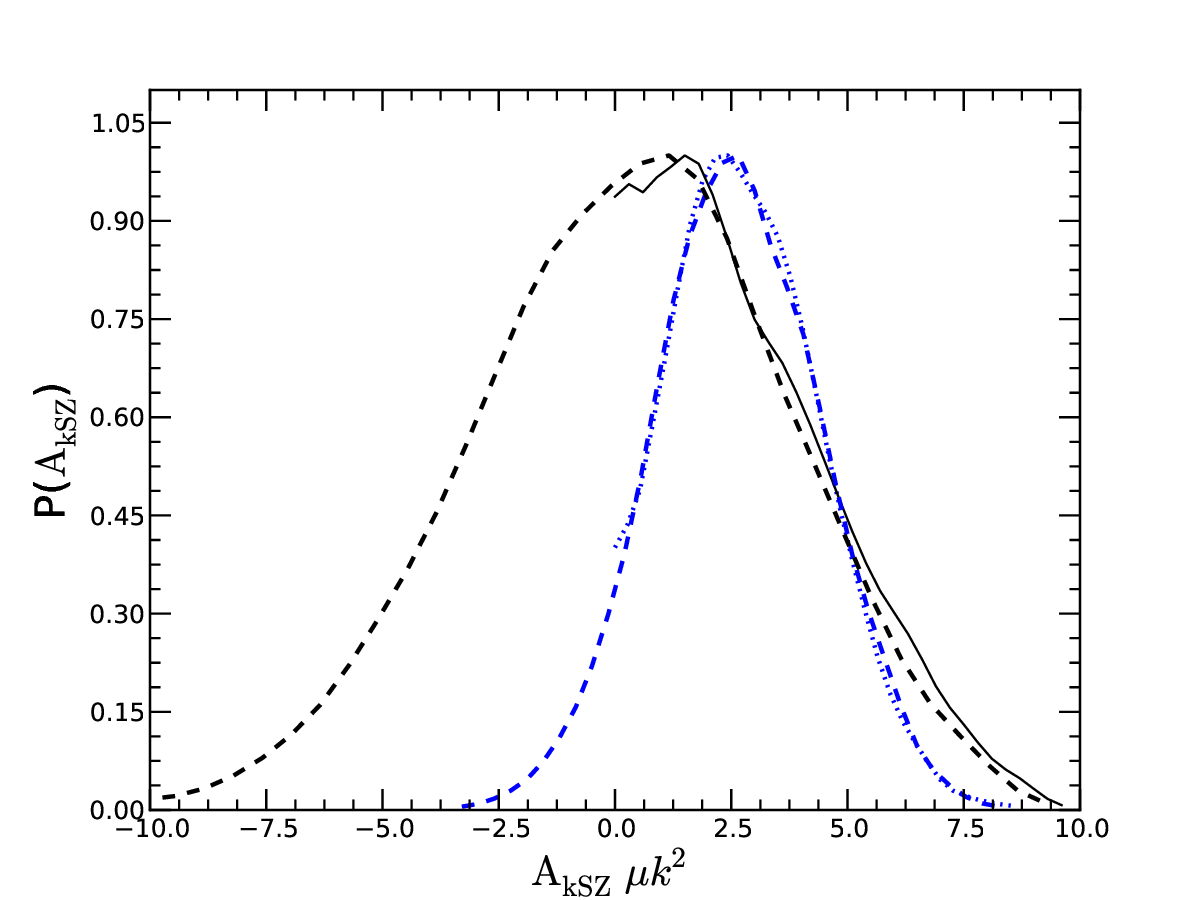}}
\end{center}
\caption{
One-dimensional posterior probability distributions from R12 for 
\aksz, with and without including bispectrum information, and with and 
without applying the $\aksz>0$ prior. 
The solid and dashed black lines show constraints with no bispectrum information added,
with and without the $\aksz>0$ prior. (The black solid line is identical to 
the black solid line in the right panel of Figure \ref{fig:ksz}.) 
The dotted and dashed blue lines show the constraints 
with bispectrum information added, with and without the $\aksz>0$ prior.
(The blue dotted line is identical to blue dotted line in the right panel of Figure \ref{fig:ksz}.)
}
\label{fig:ksz_noprior}
\end{figure}

Note that we have applied a prior of $\aksz > 0$ in obtaining these results.
Under the assumption that we are measuring real sky power in the power
spectrum data in R12, there is no reason to lift
the $\aksz>0$ prior, but one could choose to account for possible
unknown systematics by expanding the prior below zero. The \aksz\ posterior 
with no prior on \aksz\ is plotted (with and without the bispectrum information
included) in Figure \ref{fig:ksz_noprior}.
The no-prior mean and $1 \sigma$ values are 
$\aksz = \akszbtszrtnp \pm \dakszbtszrtnp \ \muk^2$ 
when the bispectrum information is included, as compared to 
$\aksz = \akszrtnp \pm \dakszrtnp \ \muk^2$ 
from the R12 power spectrum only.

Before adding the bispectrum constraint, 
the R12 $1 \sigma$ uncertainty on \atsz\ was $\datszrt \ \muk^2$, 
the 95\% upper limit on \aksz\ was 
$\akszninefive \ \muk^2$, and the peak of the \aksz\ distribution was within
$1 \sigma$ of zero.
The addition of the bispectrum constraint reduces the
error of \atsz\ by a factor of two compared to the power spectrum
constraints alone.  In turn, this improves the constraints on \aksz,
reducing the upper limit by 20\% 
(and the $1 \sigma$ uncertainty by nearly a factor of two in the 
no-kSZ-prior case), and showing a preference
for non-zero kSZ.

\subsubsection{Prospects for the full 2500~\degs\ survey}
\label{sec:prospects}
The full SPT-SZ survey comprises 2500~\degs\ of 95, 150, and 220~GHz data at noise levels
comparable to the 800~\degs\ subset used in this analysis, and work is ongoing 
to produce the data and simulation products necessary to measure the 
small-angular-scale power spectrum and bispectrum in the full survey.
The statistical uncertainty and the sample-variance uncertainty on \btsz\ from the 
full survey should be roughly a factor of $\sqrt{3}$ lower than the corresponding 
values in this work, simply from the larger sky coverage. The systematic
uncertainty is not expected to change, but in the 800~\degs\ result, the 
statistical + systematic + sample variance uncertainty is dominated by the
sample variance contribution in both the no-cluster-masking and $> \bigmass$ 
masking cases; this total uncertainty is also expected to decrease by nearly 
$\sqrt{3}.$ For the $> \bigmass$ masking case, this would result in a $\sim$12\%
constraint on \btsz.

Because the constraint on \sigeight\ from \btsz\ is already limited by the 
assumed 36\% modeling uncertainty, we do not expect a measurable 
improvement in the \sigeight\ constraints from the full 2500~\degs\ survey, 
unless significant progress is made in measuring pressure profiles of the 
clusters responsible for the tSZ bispectrum. To achieve a lower 
statistical + systematic + sample variance
uncertainty using the 2500 \degs\ result, we would need to
reduce the modeling uncertainty by roughly a factor of two.
This is an ambitious goal; however, the amount of X-ray and millimeter-wave data on high-mass clusters
at all redshifts is increasing rapidly, with X-ray programs such as {\it Chandra}
observations of 80 SPT-discovered clusters at $0.4 \le z \le 1.2$ (B. Benson et al., 
in prep.) and millimeter-wave pressure profile measurements from such instruments
as Bolocam \citep{sayers13b}, the Combined Array for Research in Millimeter-wave Astronomy 
\citep[e.g.,][]{plagge13}, and {\it Planck} \citep{planck12-5}.

Even with no improvement in modeling uncertainty, the 2500~\degs\ measurement
of \btsz\ will improve our ability to separate \atsz\ and \aksz\ in the power spectrum.
Because the relationship of \atsz\ to \btsz\ is constrained far better than either one
individually, our current bispectrum-derived constraint on \atsz\ is limited by sample
variance. Reducing the full statistical + systematic + sample variance + cluster mask
threshold uncertainty
on \btsz\ to $\sim$12\% will result in uncertainties of $\sim \datszbtszprojnom \ \muk^2$ on 
\atsz\ and $\sim \dakszbtszprojnom \ \muk^2$ on \aksz, 
assuming the default 11\% modeling uncertainty. 
Achieving this total error budget will also require an improvement
in the systematic uncertainty on our cluster mass determinations, but that is
expected to be achieved with a program of multi-wavelength follow-up of SPT
clusters that is currently underway (see \citealt{benson13} for details).

If the current best-fit value of \aksz\ 
turns out to be correct, these constraints will result in nearly a $3 \sigma$ detection 
of the kSZ effect.
The addition of 100~\degs\ of already collected 
{\it Herschel}-SPIRE submillimeter data (program OT1\_jcarls01\_3, PI: Carlstrom) will provide strong constraints on 
the behavior of the CIB, which in turn will further tighten the tSZ and kSZ
constraints. Full-survey SPT power spectrum + full-survey SPT bispectrum + 
100~\degs\ SPIRE constraints on the
kSZ are expected to be at the $< 0.5 \ \muk^2$ level.
These constraints will
lead to unprecedented limits on the reionization history of the 
universe \citep[e.g.,][]{zahn12}.

\section{Conclusions}
\label{sec:conclusions}
We have used 800~\degs\ of multi-frequency data from the SPT-SZ survey to make a 
high-significance detection of the Fourier-domain angular three-point function, 
or angular bispectrum, of the small-angular-scale 
($1^\prime \lesssim \theta \lesssim 10^\prime$, $1000 \lesssim l \lesssim \num{10000}$)
millimeter-wave sky. A bispectrum signal model that includes contributions from 
the thermal SZ effect, the clustered cosmic infrared background, 
and the spatially uncorrelated (or Poisson) point-source signal in each of the three bands
provides a reasonable fit to the data. The tSZ bispectrum is detected at $>$10$\sigma$, 
the Poisson point-source component is detected in each band individually at 
$\sim$5 to $\sim$11$\sigma$,
and the clustered CIB bispectrum is detected at $>$5$\sigma$.
This is the first detection of the clustered CIB bispectrum.

We have compared the measured Poisson point-source bispectrum in each band
to predictions from source models. We find that no combination of models of radio-loud
and dusty, radio-quiet sources can reproduce the measured Poisson bispectrum
amplitudes, implying that bispectrum measurements can provide interesting new 
constraints on source models.

Applying the methods originally presented in \citet[][B12]{bhattacharya12}, 
we have used the measurement of \btsz, the amplitude of the tSZ bispectrum to 
constrain \sigeight, the normalization of the matter power spectrum, and to 
predict \atsz, the amplitude of the tSZ power spectrum. The constraint on 
\sigeight\ using just SPT bispectrum data and a prior on $\Omega_b$ is 
$\sigeight = 0.786 \pm 0.031$. This constraint is competitive with, and statistically
consistent with, other recent measurements. Our bispectrum-derived
prediction for \atsz, combined with the power spectrum results of \citet[][R12]{reichardt12b},
results in  some evidence for a non-zero kinematic SZ power spectrum, 
with $\aksz = \akszbtszrt \pm \dakszbtszrtnom \ \muk^2$, 
or $\aksz = \akszbtszrtnp \pm \dakszbtszrtnp \ \muk^2$ if the $\aksz > 0$ prior is removed.

In addition to constraining cosmology and models of source emission, these 
measurements of the small-scale, secondary-anisotropy- and foreground-dominated
bispectra provide valuable constraints on potential contamination to measurements
of the primordial CMB bispectrum on larger scales, such as those expected 
soon from the {\it Planck} team.

\clearpage

\begin{acknowledgments}
We thank Blake Sherwin for providing the ACT $l$-space filter function, and 
we thank an anonymous referee for helpful comments.
T.~Crawford and R.~Keisler thank the University of Texas Department of Astronomy and the Texas Cosmology Center,
where much of this work was done,
for their hospitality. 

The SPT is supported by the National Science Foundation through grant ANT-0638937, with partial support provided by NSF grant PHY-1125897, the Kavli Foundation, and the Gordon and Betty Moore Foundation.
The McGill group acknowledges funding from the National Sciences and Engineering Research Council of Canada, Canada Research Chairs program, and the Canadian Institute for Advanced Research. 
Work at Harvard is supported by grant AST-1009012.
S.~Bhattacharya acknowledges support from NSF grant AST- 1009811,
R.~Keisler from NASA Hubble Fellowship grant HF-51275.01,
B.~Benson from a KICP Fellowship,
M.~Dobbs from an Alfred P. Sloan Research Fellowship,
O.~Zahn from a BCCP fellowship, 
and L.~Knox and M.~Millea from NSF grant 0709498.  

Some of the results in this paper have been derived using the HEALPix \citep{gorski05} package. 
This research used resources of the National Energy Research Scientific Computing Center (NERSC), which is supported by the Office of Science of the U.S. Department of Energy under Contract No. DE-AC02-05CH11231, and resources of the University of Chicago Computing Cooperative (UC3), supported in part by the Open Science Grid, NSF grant PHY-1148698.
We acknowledge the use of the Legacy Archive for Microwave Background Data Analysis (LAMBDA). 
Support for LAMBDA is provided by the NASA Office of Space Science.
\end{acknowledgments}

\appendix

 In this work, we predict the tSZ power spectrum amplitude from the measured tSZ bispectrum amplitude, and we combine that prediction with the R12 measurement of the total SZ power spectrum. Our analysis includes the measurement uncertainties of the power spectrum and the bispectrum, but we assume that the covariance between the two observables is negligible. Here we compute the covariance between the power spectrum and the bispectrum and show that our assumption is justified.
 
 The covariance between the bispectrum ($B$) and the power spectrum ($C$) is given by \citep{kayo13}
 \beq
 \mathrm{Cov}[C(l_4)B(l_1,l_2,l_3)] = \delta_{l_4 l_1} \frac{4\pi}{\Omega_sl_1\Delta l_1} C(l_4)B(l_1,l_2,l_3)+
 \text{2 perms.} +\frac{1}{\Omega_s} \int \frac{d\phi}{2\pi} T_5(l_4,-l_4,l_1,l_2,l_3;\phi),
  \eeq
where $\Omega_s$ is the survey area, $T_5$ is the tSZ five-point function, and $\phi$ is the angle between $l_4$ and $l_1$. Since we use the combined measurements of  the power spectrum and the bispectrum at a similar $l$ range, we compute the covariance for the case when $l_4= l_1$. We do not include the 
correlated sample variance term from \citet{kayo13}, because we assume here that the correlation between the sample variance of the power spectrum and the bispectrum is negligible.

\begin{figure}
\begin{center}
\includegraphics[scale=0.5]{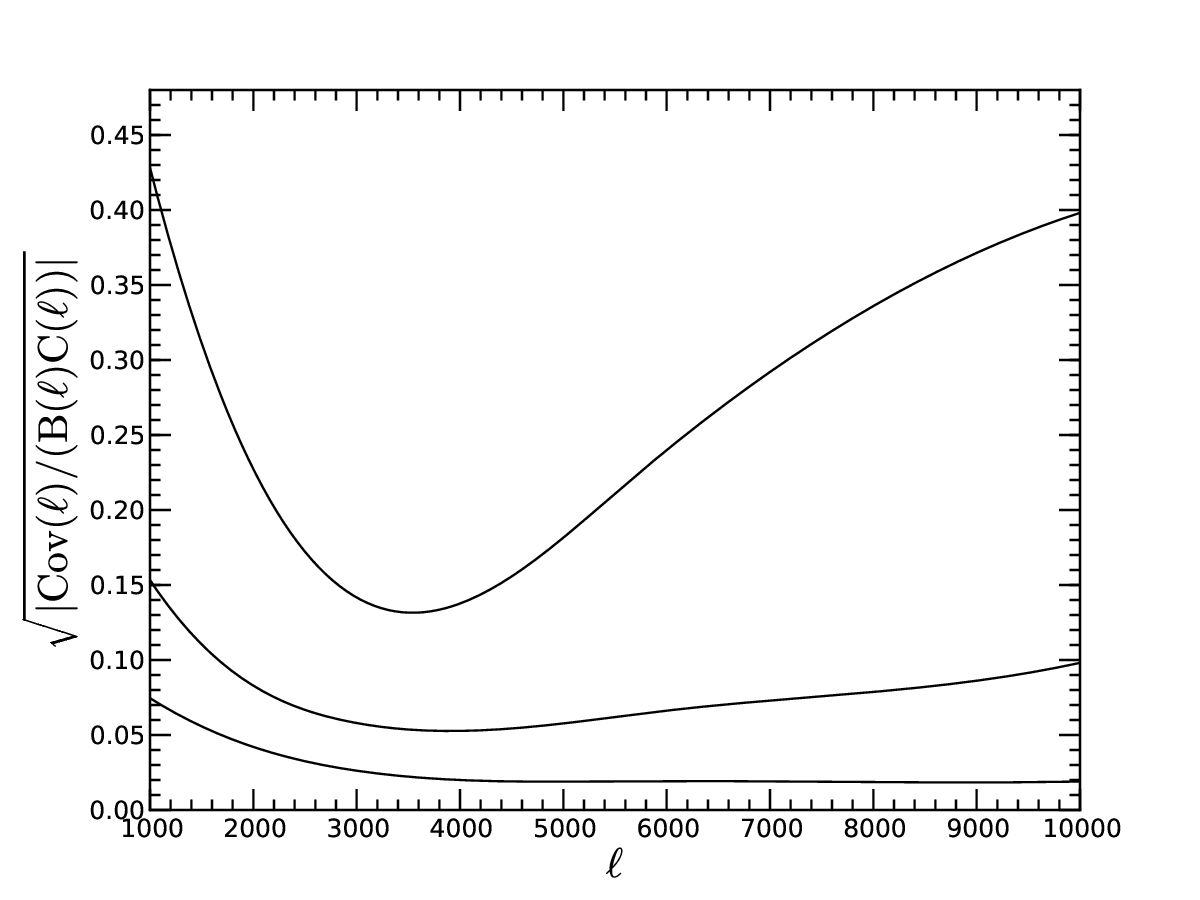}
\end{center}
\caption{The square root of the fractional covariance between the tSZ bispectrum amplitude $B$ and the tSZ power spectrum amplitude $C$ as a function of $l$ for the different mass cuts. The three-dimensional quantities $\mathrm{Cov}[C(l_1) B(l_1,l_2,l_3)]$ and $B(l_1,l_2,l_3)$ are contracted to one dimension as described in the text.
From top to bottom, this quantity is plotted for no cluster cut, $\mtwoh \ge \bigmass$ clusters cut, and $\mtwoh \ge \threshmass$ clusters cut.}
\label{fig:cov}
\end{figure}

We calculate the fractional covariance, $\sqrt{|\mathrm{Cov}(\lrad)| / |{B}(\lrad) C(\lrad)|}$, where $\lrad$ is defined in Equation~\ref{eqn:lrad}, and $\mathrm{Cov}(C B)$ and $B$ are contracted from three dimensions to one dimension as in Equations~\ref{eqn:blrad} and \ref{eqn:wlrad}.  The results are shown in Figure~\ref{fig:cov} for the three different mass cuts used in the bispectrum estimate (no cut, $\mtwoh > \bigmass$ clusters cut, and $\mtwoh > \threshmass$ clusters cut). For a given mass cut, at lower $l$, the covariance increases while the power spectrum and bispectrum decrease slightly. Hence the fractional covariance increases at lower $l$. At higher $l$, the power spectrum and the bispectrum decrease, with the bispectrum decreasing slightly faster than the covariance. This is because the last term in Equation~1 of the Appendix drops quickly at large $l$, while the first 3 terms stay non-zero (as they are the product of the bispectrum and the power spectrum). The net result is that at large $l$, the ratio increases again. The minimum of the ratio appears at $l\sim 3000$ as both the power spectrum and the bispectrum peak in that range. The covariance decreases sharply with mass cut. This occurs because, with more clusters masked out, the last term of Equation~1 drops quickly. As shown in Figure~\ref{fig:cov}, the fractional covariance is $\approx$ 6\% at $l \sim 3000$ for the mass cut $\mtwoh=\bigmass$. This is small compared to the other sources of uncertainty in the $\atsz$-$\btsz$ calcuation.

\bibliography{../../BIBTEX/spt}

\end{document}

%% file: tab_fitresults_all.tex
\begin{table*}[]
\caption{Multi-band fit results, No cluster masking}
\centering
\begin{tabular}{ | l c c c c c c | }
\hline
Template & best-fit & noise              & systematic         & quadrature & detection    & constraint   \\ [-0.03in]
         & value    & error ($1 \sigma$) & error ($1 \sigma$) & sum        & significance & significance \\
\hline
tSZ, relative to analytical prediction &    0.53 &    0.04 &    0.03 &    0.05 &    13 &    10 \\
Clustered CIB [$10^{-9} \muk^3$ at $l=2000$ and $\nu=220$~GHz] &    0.68 &    0.13 &    0.06 &    0.14 &    5.1 &    4.7 \\
Poisson, 95~GHz, [$10^{-10} \muk^3$] &    0.79 &    0.16 &    0.05 &    0.16 &    5.0 &    4.8 \\
Poisson, 150~GHz, [$10^{-11} \muk^3$] &    1.12 &    0.09 &    0.09 &    0.13 &    12 &    8.7 \\
Poisson, 220~GHz, [$10^{-10} \muk^3$] &    1.84 &    0.26 &    0.23 &    0.35 &    7.0 &    5.3 \\
\hline
\end{tabular}
\label{tab:fitresults}
\begin{tablenotes}
\item{Parameter best-fit values and $1 \sigma$ statistical and systematic uncertainties.  
``Detection significance" refers to the best-fit parameter value divided by 
statistical uncertainty only; ``constraint significance" refers to the 
best-fit parameter value divided by the quadrature sum of statistical and systematic uncertainty.
Sample variance is not included in the constraint significance; see Section \ref{sec:tszcosmo}
and Table \ref{tab:tszplussampvar} for tSZ results including sample variance.}
\end{tablenotes}
\end{table*}

%% file: tab_fitresults_8e14.tex
\begin{table*}[]
\caption{Multi-band fit results, $\mtwoh > \bigmass$ clusters masked}
\centering
\begin{tabular}{ | l c c c c c c | }
\hline
Template & best-fit & noise              & systematic         & quadrature & detection    & constraint   \\ [-0.03in]
         & value    & error ($1 \sigma$) & error ($1 \sigma$) & sum        & significance & significance \\
\hline
tSZ, relative to analytical prediction &    0.59 &    0.05 &    0.04 &    0.07 &    11 &    9.0 \\
Clustered CIB [$10^{-9} \muk^3$ at $l=2000$ and $\nu=220$~GHz] &    0.74 &    0.13 &    0.07 &    0.15 &    5.6 &    5.0 \\
Poisson, 95~GHz, [$10^{-10} \muk^3$] &    0.88 &    0.16 &    0.05 &    0.17 &    5.6 &    5.3 \\
Poisson, 150~GHz, [$10^{-11} \muk^3$] &    1.25 &    0.10 &    0.10 &    0.14 &    13 &    9.1 \\
Poisson, 220~GHz, [$10^{-10} \muk^3$] &    1.73 &    0.26 &    0.21 &    0.34 &    6.6 &    5.2 \\
\hline
\end{tabular}
\label{tab:fitresults_mask1}
\begin{tablenotes}
\item{See Table \ref{tab:fitresults} for notes.}
\end{tablenotes}
\end{table*}

%% file: tab_fitresults_3e14.tex
\begin{table*}[]
\caption{Multi-band fit results, $\mtwoh > \threshmass$ clusters masked}
\centering
\begin{tabular}{ | l c c c c c c | }
\hline
Template & best-fit & noise              & systematic         & quadrature & detection    & constraint   \\ [-0.03in]
         & value    & error ($1 \sigma$) & error ($1 \sigma$) & sum        & significance & significance \\
\hline
tSZ, relative to analytical prediction &    0.68 &    0.17 &    0.06 &    0.18 &    4.0 &    3.8 \\
Clustered CIB [$10^{-9} \muk^3$ at $l=2000$ and $\nu=220$~GHz] &    0.75 &    0.13 &    0.07 &    0.15 &    5.6 &    4.9 \\
Poisson, 95~GHz, [$10^{-10} \muk^3$] &    0.93 &    0.16 &    0.06 &    0.17 &    5.7 &    5.4 \\
Poisson, 150~GHz, [$10^{-11} \muk^3$] &    1.27 &    0.10 &    0.10 &    0.14 &    13 &    9.0 \\
Poisson, 220~GHz, [$10^{-10} \muk^3$] &    1.67 &    0.27 &    0.20 &    0.33 &    6.3 &    5.0 \\
\hline
\end{tabular}
\label{tab:fitresults_mask2}
\begin{tablenotes}
\item{See Table \ref{tab:fitresults} for notes.}
\end{tablenotes}
\end{table*}

%% file: tab_fitresults_corrmat.tex
\begin{table}[]
\caption{Parameter correlation matrices for multi-band fits}
\centering
\begin{tabular}{ | l c c c c c | }
\hline
Stat. only & & & & & \\
 & tSZ & CIB & P95 & P150 & P220 \\ 
tSZ &   1.00 &   0.32 &   0.39 &   0.17 &  -0.28 \\
CIB &   0.32 &   1.00 &   0.12 &  -0.61 &  -0.88 \\
P95 &   0.39 &   0.12 &   1.00 &   0.07 &  -0.11 \\
P150 &   0.17 &  -0.61 &   0.07 &   1.00 &   0.54 \\
P220 &  -0.28 &  -0.88 &  -0.11 &   0.54 &   1.00 \\
\hline
Stat.~+ syst. & & & & & \\
 & tSZ & CIB & P95 & P150 & P220 \\ 
tSZ &   1.00 &   0.35 &   0.46 &   0.37 &   0.03 \\
CIB &   0.35 &   1.00 &   0.16 &  -0.35 &  -0.43 \\
P95 &   0.46 &   0.16 &   1.00 &   0.17 &   0.01 \\
P150 &   0.37 &  -0.35 &   0.17 &   1.00 &   0.35 \\
P220 &   0.03 &  -0.43 &   0.01 &   0.35 &   1.00 \\
\hline
\end{tabular}
\label{tab:fitresults_corr}
\begin{tablenotes}
\item{Correlation between parameters for the multi-band fit with 
no clusters masked.  The top table shows the normalized elements 
of the parameter covariance matrix $r_{\psi \omega} = C_{\psi \omega}/\sqrt{C_{\psi \psi} C_{\omega \omega}}$ 
for statistical covariance only; the bottom table shows the same
quantities for the sum of statistical and systematic covariance 
(see Section \ref{sec:systematics} for details on the calculation of 
systematic covariance).  The parameter labels refer to thermal 
SZ amplitude, clustered CIB amplitude, and Poisson point source 
amplitudes in each of the three bands.}
\end{tablenotes}
\end{table}

%% file: tab_fitresults_chisqall.tex
\begin{table}[]
\caption{\chisq\ for multi-band fits}
\centering
\begin{tabular}{ | l c c c c | }
\hline
Cluster masking & \chisq & \chisq, null hypothesis & \dchisq & \chired \\ [0.03in]
\hline
None &  51798.1 &  52792.8 &   -994.8 &    1.10 \\
\bigmass &  51745.8 &  52788.5 &  -1042.7 &    1.10 \\
\threshmass &  50954.8 &  52063.7 &  -1108.9 &    1.08 \\
\hline
\end{tabular}
\label{tab:fitresults_chi}
\begin{tablenotes}
\item{\chisq\ for simultaneous fits to bispectrum data in all three bands
with three levels of cluster masking.  Also shown is the \chisq\ for the 
null hypothesis, the \dchisq\ between the null hypothesis and the full model, and
the reduced \chisq\ for the full model (for $\num{47005}$ degrees of freedom).}
\end{tablenotes}
\end{table}

%% file: tab_poissresults.tex
\begin{table*}[]
\caption{Poisson source component results vs.~model predictions}
\centering
\begin{tabular}{ | l c c c | }
\hline
Measured Poisson bispectrum amplitudes & & & \\ [0.05in]
\hspace{0.35cm} Masking level & 95~GHz value & 150~GHz value & 220~GHz value \\
 & [$10^{-10} \muk^3$] & [$10^{-11} \muk^3$] & [$10^{-10} \muk^3$] \\ [0.02in]
\hspace{0.35cm} no cluster masking & $   0.79 \pm    0.16$ & $   1.12 \pm    0.13$ & $   1.84 \pm    0.35$ \\
\hspace{0.35cm} $\mtwoh > \bigmass$ clusters masked & $   0.88 \pm    0.17$ & $   1.25 \pm    0.14$ & $   1.73 \pm    0.34$ \\
\hspace{0.35cm} $\mtwoh > \threshmass$ clusters masked & $   0.93 \pm    0.17$ & $   1.27 \pm    0.14$ & $   1.67 \pm    0.33$ \\
\hline
Model Predictions & & & \\ [0.05in]
\hspace{0.35cm} Model & 95~GHz value & 150~GHz value & 220~GHz value \\ 
 & [$10^{-10} \muk^3$] & [$10^{-11} \muk^3$] & [$10^{-10} \muk^3$] \\ [0.02in]
\hspace{0.35cm} de Zotti radio &    0.685 &    0.563 &    0.018 \\
\hspace{0.35cm} Tucci radio &    0.484 &    0.385 &    0.011 \\
\hspace{0.35cm} B\'{e}thermin (2011) dusty &    0.023 &    3.285 &    2.264 \\
\hspace{0.35cm} B\'{e}thermin (2012) dusty &    0.016 &    1.878 &    2.900 \\
\hspace{0.35cm} de Zotti + B\'{e}thermin (2011) &    0.708 &    3.849 &    2.282 \\
\hspace{0.35cm} de Zotti + B\'{e}thermin (2012) &    0.701 &    2.441 &    2.918 \\
\hspace{0.35cm} Tucci + B\'{e}thermin (2011) &    0.507 &    3.670 &    2.275 \\
\hspace{0.35cm} Tucci + B\'{e}thermin (2012) &    0.500 &    2.263 &    2.911 \\
\hline
\end{tabular}
\label{tab:poiss}
\begin{tablenotes}
\item{Comparison of measured 
single-band Poisson point-source bispectrum amplitudes with predictions from source count models.  
In the upper section, measured values of the Poisson 
source-component bispectrum amplitudes---and $1 \sigma$ statistical-plus-systematic errors 
on those values---are shown for the three levels of cluster masking.
In the lower section, predicted Poisson bispectrum amplitudes are shown 
for two models of radio-loud source counts (\citealt{dezotti10}
and \citealt{tucci11}), two models of dusty, radio-quiet source counts
(\citealt{bethermin11} and \citealt{bethermin12}), and combinations thereof.}
\end{tablenotes}
\end{table*}

%% file: tab_tszplussampvar.tex
\begin{table*}[]
\caption{Thermal SZ bispectrum error budget}
\centering
\begin{tabular}{ | l c c c c c | }
\hline
Masking  & stat. & syst. & mask thresh. & sample-variance & stat. + syst. + mask+ \\ [-0.03in]
         & error & error & error        & error           & sample-variance \\
\hline
No cluster masking &     0.08 &     0.06 &     0.00 &     0.20 &     0.22 \\
$\mtwoh > \bigmass$ clusters masked &     0.09 &     0.06 &     0.11 &     0.15 &     0.21 \\
$\mtwoh > \threshmass$ clusters masked &     0.25 &     0.09 &     0.32 &     0.06 &     0.42 \\
\hline
\end{tabular}
\label{tab:tszplussampvar}
\begin{tablenotes}
\item{Fractional $1 \sigma$ uncertainty on the amplitude of the tSZ bispectrum, including statistical, 
systematic, cluster-mask-threshold, and sample-variance contributions, and the combination of these in quadrature.
Sample-variance errors are estimated as described in Section \ref{sec:tszsampvar}.
Errors due to the uncertainty in cluster mask threshold are calculated assuming a 15\% 
systematic uncertainty in cluster mass estimation (see Section \ref{sec:tszmodel} for details).
Values are given for each of the three cluster mask thresholds 
(no masking, clusters with $\mtwoh > \bigmass$ masked, clusters with $\mtwoh > \threshmass$ masked).}
\end{tablenotes}
\end{table*}

%% file: Crawford14.bbl
\begin{thebibliography}{88}
\expandafter\ifx\csname natexlab\endcsname\relax\def\natexlab#1{#1}\fi

\bibitem[{{Acquaviva} {et~al.}(2003){Acquaviva}, {Bartolo}, {Matarrese}, \&
  {Riotto}}]{acquaviva03}
{Acquaviva}, V., {Bartolo}, N., {Matarrese}, S., \& {Riotto}, A. 2003, Nuclear
  Physics B, 667, 119

\bibitem[{{Addison} {et~al.}(2012{\natexlab{a}}){Addison}, {Dunkley}, \&
  {Bond}}]{addison12b}
{Addison}, G.~E., {Dunkley}, J., \& {Bond}, J.~R. 2012{\natexlab{a}}, ArXiv
  e-prints, 1210.6697

\bibitem[{{Addison} {et~al.}(2012{\natexlab{b}}){Addison}, {Dunkley}, {Hajian},
  {Viero}, {Bond}, {Das}, {Devlin}, {Halpern}, {Hincks}, {Hlozek}, {Marriage},
  {Moodley}, {Page}, {Reese}, {Scott}, {Spergel}, {Staggs}, \&
  {Wollack}}]{addison12a}
{Addison}, G.~E., {et~al.} 2012{\natexlab{b}}, \apj, 752, 120

\bibitem[{{Allen} {et~al.}(2008){Allen}, {Rapetti}, {Schmidt}, {Ebeling},
  {Morris}, \& {Fabian}}]{allen08}
{Allen}, S.~W., {Rapetti}, D.~A., {Schmidt}, R.~W., {Ebeling}, H., {Morris},
  R.~G., \& {Fabian}, A.~C. 2008, \mnras, 383, 879

\bibitem[{{Arnaud} {et~al.}(2010){Arnaud}, {Pratt}, {Piffaretti},
  {B{\"o}hringer}, {Croston}, \& {Pointecouteau}}]{arnaud10}
{Arnaud}, M., {Pratt}, G.~W., {Piffaretti}, R., {B{\"o}hringer}, H., {Croston},
  J.~H., \& {Pointecouteau}, E. 2010, \aap, 517, A92+

\bibitem[{{Benson} {et~al.}(2013){Benson}, {de Haan}, {Dudley}, {Reichardt},
  {Aird}, {Andersson}, {Armstrong}, {Ashby}, {Bautz}, {Bayliss}, {Bazin},
  {Bleem}, {Brodwin}, {Carlstrom}, {Chang}, {Cho}, {Clocchiatti}, {Crawford},
  {Crites}, {Desai}, {Dobbs}, {Foley}, {Forman}, {George}, {Gladders},
  {Gonzalez}, {Halverson}, {Harrington}, {High}, {Holder}, {Holzapfel},
  {Hoover}, {Hrubes}, {Jones}, {Joy}, {Keisler}, {Knox}, {Lee}, {Leitch},
  {Liu}, {Lueker}, {Luong-Van}, {Mantz}, {Marrone}, {McDonald}, {McMahon},
  {Mehl}, {Meyer}, {Mocanu}, {Mohr}, {Montroy}, {Murray}, {Natoli}, {Padin},
  {Plagge}, {Pryke}, {Rest}, {Ruel}, {Ruhl}, {Saliwanchik}, {Saro}, {Sayre},
  {Schaffer}, {Shaw}, {Shirokoff}, {Song}, {Spieler}, {Stalder},
  {Staniszewski}, {Stark}, {Story}, {Stubbs}, {Suhada}, {van Engelen},
  {Vanderlinde}, {Vieira}, {Vikhlinin}, {Williamson}, {Zahn}, \&
  {Zenteno}}]{benson13}
{Benson}, B.~A., {et~al.} 2013, \apj, 763, 147

\bibitem[{{B{\'e}thermin} {et~al.}(2012){B{\'e}thermin}, {Daddi}, {Magdis},
  {Sargent}, {Hezaveh}, {Elbaz}, {Le Borgne}, {Mullaney}, {Pannella}, {Buat},
  {Charmandaris}, {Lagache}, \& {Scott}}]{bethermin12}
{B{\'e}thermin}, M., {et~al.} 2012, \apjl, 757, L23

\bibitem[{{B{\'e}thermin} {et~al.}(2011){B{\'e}thermin}, {Dole}, {Lagache}, {Le
  Borgne}, \& {Penin}}]{bethermin11}
{B{\'e}thermin}, M., {Dole}, H., {Lagache}, G., {Le Borgne}, D., \& {Penin}, A.
  2011, \aap, 529, A4+

\bibitem[{{Bhattacharya} {et~al.}(2011){Bhattacharya}, {Heitmann}, {White},
  {Luki{\'c}}, {Wagner}, \& {Habib}}]{bhattacharya10}
{Bhattacharya}, S., {Heitmann}, K., {White}, M., {Luki{\'c}}, Z., {Wagner}, C.,
  \& {Habib}, S. 2011, \apj, 732, 122

\bibitem[{{Bhattacharya} {et~al.}(2012){Bhattacharya}, {Nagai}, {Shaw},
  {Crawford}, \& {Holder}}]{bhattacharya12}
{Bhattacharya}, S., {Nagai}, D., {Shaw}, L., {Crawford}, T., \& {Holder}, G.~P.
  2012, \apj, 760, 5

\bibitem[{{Butcher} \& {Oemler}(1984)}]{butcher84}
{Butcher}, H., \& {Oemler}, Jr., A. 1984, \apj, 285, 426

\bibitem[{{Carlstrom} {et~al.}(2011){Carlstrom}, {Ade}, {Aird}, {Benson},
  {Bleem}, {Busetti}, {Chang}, {Chauvin}, {Cho}, {Crawford}, {Crites}, {Dobbs},
  {Halverson}, {Heimsath}, {Holzapfel}, {Hrubes}, {Joy}, {Keisler}, {Lanting},
  {Lee}, {Leitch}, {Leong}, {Lu}, {Lueker}, {Luong-van}, {McMahon}, {Mehl},
  {Meyer}, {Mohr}, {Montroy}, {Padin}, {Plagge}, {Pryke}, {Ruhl}, {Schaffer},
  {Schwan}, {Shirokoff}, {Spieler}, {Staniszewski}, {Stark}, {Tucker},
  {Vanderlinde}, {Vieira}, \& {Williamson}}]{carlstrom11}
{Carlstrom}, J.~E., {et~al.} 2011, \pasp, 123, 568

\bibitem[{{Coles} \& {Barrow}(1987)}]{coles87}
{Coles}, P., \& {Barrow}, J.~D. 1987, \mnras, 228, 407

\bibitem[{{Das} {et~al.}(2013){Das}, {Louis}, {Nolta}, {Addison},
  {Battistelli}, {Bond}, {Calabrese}, {Devlin}, {Dicker}, {Dunkley},
  {D{\"u}nner}, {Fowler}, {Gralla}, {Hajian}, {Halpern}, {Hasselfield},
  {Hilton}, {Hincks}, {Hlozek}, {Huffenberger}, {Hughes}, {Irwin}, {Kosowsky},
  {Lupton}, {Marriage}, {Marsden}, {Menanteau}, {Moodley}, {Niemack}, {Page},
  {Partridge}, {Reese}, {Schmitt}, {Sehgal}, {Sherwin}, {Sievers}, {Spergel},
  {Staggs}, {Swetz}, {Switzer}, {Thornton}, {Trac}, \& {Wollack}}]{das13}
{Das}, S., {et~al.} 2013, ArXiv e-prints, 1301.1037

\bibitem[{{Das} {et~al.}(2011{\natexlab{a}}){Das}, {Marriage}, {Ade},
  {Aguirre}, {Amiri}, {Appel}, {Barrientos}, {Battistelli}, {Bond}, {Brown},
  {Burger}, {Chervenak}, {Devlin}, {Dicker}, {Bertrand Doriese}, {Dunkley},
  {D{\"u}nner}, {Essinger-Hileman}, {Fisher}, {Fowler}, {Hajian}, {Halpern},
  {Hasselfield}, {Hern{\'a}ndez-Monteagudo}, {Hilton}, {Hilton}, {Hincks},
  {Hlozek}, {Huffenberger}, {Hughes}, {Hughes}, {Infante}, {Irwin}, {Baptiste
  Juin}, {Kaul}, {Klein}, {Kosowsky}, {Lau}, {Limon}, {Lin}, {Lupton},
  {Marsden}, {Martocci}, {Mauskopf}, {Menanteau}, {Moodley}, {Moseley},
  {Netterfield}, {Niemack}, {Nolta}, {Page}, {Parker}, {Partridge}, {Reid},
  {Sehgal}, {Sherwin}, {Sievers}, {Spergel}, {Staggs}, {Swetz}, {Switzer},
  {Thornton}, {Trac}, {Tucker}, {Warne}, {Wollack}, \& {Zhao}}]{das11b}
------. 2011{\natexlab{a}}, \apj, 729, 62

\bibitem[{{Das} {et~al.}(2011{\natexlab{b}}){Das}, {Sherwin}, {Aguirre},
  {Appel}, {Bond}, {Carvalho}, {Devlin}, {Dunkley}, {D{\"u}nner},
  {Essinger-Hileman}, {Fowler}, {Hajian}, {Halpern}, {Hasselfield}, {Hincks},
  {Hlozek}, {Huffenberger}, {Hughes}, {Irwin}, {Klein}, {Kosowsky}, {Lupton},
  {Marriage}, {Marsden}, {Menanteau}, {Moodley}, {Niemack}, {Nolta}, {Page},
  {Parker}, {Reese}, {Schmitt}, {Sehgal}, {Sievers}, {Spergel}, {Staggs},
  {Swetz}, {Switzer}, {Thornton}, {Visnjic}, \& {Wollack}}]{das11}
------. 2011{\natexlab{b}}, Physical Review Letters, 107, 021301

\bibitem[{{De Zotti} {et~al.}(2010){De Zotti}, {Massardi}, {Negrello}, \&
  {Wall}}]{dezotti10}
{De Zotti}, G., {Massardi}, M., {Negrello}, M., \& {Wall}, J. 2010, \aapr, 18,
  1

\bibitem[{{Dudley}(2012)}]{dudley12}
{Dudley}, J. 2012, PhD thesis, McGill University

\bibitem[{{Duffy} {et~al.}(2008){Duffy}, {Schaye}, {Kay}, \& {Dalla
  Vecchia}}]{duffy08}
{Duffy}, A.~R., {Schaye}, J., {Kay}, S.~T., \& {Dalla Vecchia}, C. 2008,
  \mnras, 390, L64

\bibitem[{{Dunkley} {et~al.}(2011){Dunkley}, {Hlozek}, {Sievers}, {Acquaviva},
  {Ade}, {Aguirre}, {Amiri}, {Appel}, {Barrientos}, {Battistelli}, {Bond},
  {Brown}, {Burger}, {Chervenak}, {Das}, {Devlin}, {Dicker}, {Bertrand
  Doriese}, {D{\"u}nner}, {Essinger-Hileman}, {Fisher}, {Fowler}, {Hajian},
  {Halpern}, {Hasselfield}, {Hern{\'a}ndez-Monteagudo}, {Hilton}, {Hilton},
  {Hincks}, {Huffenberger}, {Hughes}, {Hughes}, {Infante}, {Irwin}, {Juin},
  {Kaul}, {Klein}, {Kosowsky}, {Lau}, {Limon}, {Lin}, {Lupton}, {Marriage},
  {Marsden}, {Mauskopf}, {Menanteau}, {Moodley}, {Moseley}, {Netterfield},
  {Niemack}, {Nolta}, {Page}, {Parker}, {Partridge}, {Reid}, {Sehgal},
  {Sherwin}, {Spergel}, {Staggs}, {Swetz}, {Switzer}, {Thornton}, {Trac},
  {Tucker}, {Warne}, {Wollack}, \& {Zhao}}]{dunkley11}
{Dunkley}, J., {et~al.} 2011, \apj, 739, 52

\bibitem[{{Fergusson} \& {Shellard}(2009)}]{fergusson09}
{Fergusson}, J.~R., \& {Shellard}, E.~P.~S. 2009, \prd, 80, 043510

\bibitem[{{Finkbeiner} {et~al.}(1999){Finkbeiner}, {Davis}, \&
  {Schlegel}}]{finkbeiner99}
{Finkbeiner}, D.~P., {Davis}, M., \& {Schlegel}, D.~J. 1999, \apj, 524, 867

\bibitem[{{Fowler} {et~al.}(2010){Fowler}, {Acquaviva}, {Ade}, {Aguirre},
  {Amiri}, {Appel}, {Barrientos}, {Battistelli}, {Bond}, {Brown}, {Burger},
  {Chervenak}, {Das}, {Devlin}, {Dicker}, {Doriese}, {Dunkley}, {D{\"u}nner},
  {Essinger-Hileman}, {Fisher}, {Hajian}, {Halpern}, {Hasselfield},
  {Hern{\'a}ndez-Monteagudo}, {Hilton}, {Hilton}, {Hincks}, {Hlozek},
  {Huffenberger}, {Hughes}, {Hughes}, {Infante}, {Irwin}, {Jimenez}, {Juin},
  {Kaul}, {Klein}, {Kosowsky}, {Lau}, {Limon}, {Lin}, {Lupton}, {Marriage},
  {Marsden}, {Martocci}, {Mauskopf}, {Menanteau}, {Moodley}, {Moseley},
  {Netterfield}, {Niemack}, {Nolta}, {Page}, {Parker}, {Partridge}, {Quintana},
  {Reid}, {Sehgal}, {Sievers}, {Spergel}, {Staggs}, {Swetz}, {Switzer},
  {Thornton}, {Trac}, {Tucker}, {Verde}, {Warne}, {Wilson}, {Wollack}, \&
  {Zhao}}]{fowler10}
{Fowler}, J.~W., {et~al.} 2010, \apj, 722, 1148

\bibitem[{{G{\'o}rski} {et~al.}(2005){G{\'o}rski}, {Hivon}, {Banday},
  {Wandelt}, {Hansen}, {Reinecke}, \& {Bartelmann}}]{gorski05}
{G{\'o}rski}, K.~M., {Hivon}, E., {Banday}, A.~J., {Wandelt}, B.~D., {Hansen},
  F.~K., {Reinecke}, M., \& {Bartelmann}, M. 2005, \apj, 622, 759

\bibitem[{{Gruzinov} \& {Hu}(1998)}]{gruzinov98}
{Gruzinov}, A., \& {Hu}, W. 1998, \apj, 508, 435

\bibitem[{{Haiman} {et~al.}(2001){Haiman}, {Mohr}, \& {Holder}}]{haiman01}
{Haiman}, Z., {Mohr}, J.~J., \& {Holder}, G.~P. 2001, \apj, 553, 545

\bibitem[{{Hall} {et~al.}(2010){Hall}, {Knox}, {Reichardt}, {Ade}, {Aird},
  {Benson}, {Bleem}, {Carlstrom}, {Chang}, {Cho}, {Crawford}, {Crites}, {de
  Haan}, {Dobbs}, {George}, {Halverson}, {Holder}, {Holzapfel}, {Hrubes},
  {Joy}, {Keisler}, {Lee}, {Leitch}, {Lueker}, {McMahon}, {Mehl}, {Meyer},
  {Mohr}, {Montroy}, {Padin}, {Plagge}, {Pryke}, {Ruhl}, {Schaffer}, {Shaw},
  {Shirokoff}, {Spieler}, {Staniszewski}, {Stark}, {Switzer}, {Vanderlinde},
  {Vieira}, {Williamson}, \& {Zahn}}]{hall10}
{Hall}, N.~R., {et~al.} 2010, \apj, 718, 632

\bibitem[{{Hashimoto} {et~al.}(1998){Hashimoto}, {Oemler}, {Lin}, \&
  {Tucker}}]{hashimoto98}
{Hashimoto}, Y., {Oemler}, Jr., A., {Lin}, H., \& {Tucker}, D.~L. 1998, \apj,
  499, 589

\bibitem[{{Hasselfield} {et~al.}(2013){Hasselfield}, {Hilton}, {Marriage},
  {Addison}, {Barrientos}, {Battaglia}, {Battistelli}, {Bond}, {Crichton},
  {Das}, {Devlin}, {Dicker}, {Dunkley}, {Dunner}, {Fowler}, {Gralla}, {Hajian},
  {Halpern}, {Hincks}, {Hlozek}, {Hughes}, {Infante}, {Irwin}, {Kosowsky},
  {Marsden}, {Menanteau}, {Moodley}, {Niemack}, {Nolta}, {Page}, {Partridge},
  {Reese}, {Schmitt}, {Sehgal}, {Sherwin}, {Sievers}, {Sif{\'o}n}, {Spergel},
  {Staggs}, {Swetz}, {Switzer}, {Thornton}, {Trac}, \&
  {Wollack}}]{hasselfield13}
{Hasselfield}, M., {et~al.} 2013, ArXiv e-prints, 1301.0816

\bibitem[{{Hauser} \& {Dwek}(2001)}]{hauser01}
{Hauser}, M.~G., \& {Dwek}, E. 2001, \araa, 39, 249

\bibitem[{{Hill} \& {Sherwin}(2013)}]{hill13}
{Hill}, J.~C., \& {Sherwin}, B.~D. 2013, \prd, 87, 023527

\bibitem[{{Hinshaw} {et~al.}(2012){Hinshaw}, {Larson}, {Komatsu}, {Spergel},
  {Bennett}, {Dunkley}, {Nolta}, {Halpern}, {Hill}, {Odegard}, {Page}, {Smith},
  {Weiland}, {Gold}, {Jarosik}, {Kogut}, {Limon}, {Meyer}, {Tucker}, {Wollack},
  \& {Wright}}]{hinshaw12}
{Hinshaw}, G., {et~al.} 2012, ArXiv e-prints, 1212.5226

\bibitem[{{Hinshaw} {et~al.}(2003){Hinshaw}, {Spergel}, {Verde}, {Hill},
  {Meyer}, {Barnes}, {Bennett}, {Halpern}, {Jarosik}, {Kogut}, {Komatsu},
  {Limon}, {Page}, {Tucker}, {Weiland}, {Wollack}, \& {Wright}}]{hinshaw03}
------. 2003, \apjs, 148, 135

\bibitem[{{Hivon} {et~al.}(2002){Hivon}, {G{\' o}rski}, {Netterfield}, {Crill},
  {Prunet}, \& {Hansen}}]{hivon02}
{Hivon}, E., {G{\' o}rski}, K.~M., {Netterfield}, C.~B., {Crill}, B.~P.,
  {Prunet}, S., \& {Hansen}, F. 2002, \apj, 567, 2

\bibitem[{{Holder}(2012)}]{holder12}
{Holder}, G. 2012, ArXiv e-prints, 1207.0856

\bibitem[{{Holder}(2002)}]{holder02b}
{Holder}, G.~P. 2002, \apj, 580, 36

\bibitem[{{Hou} {et~al.}(2014){Hou}, {Reichardt}, {Story}, {Follin}, {Keisler},
  {Aird}, {Benson}, {Bleem}, {Carlstrom}, {Chang}, {Cho}, {Crawford}, {Crites},
  {de Haan}, {de Putter}, {Dobbs}, {Dodelson}, {Dudley}, {George}, {Halverson},
  {Holder}, {Holzapfel}, {Hoover}, {Hrubes}, {Joy}, {Knox}, {Lee}, {Leitch},
  {Lueker}, {Luong-Van}, {McMahon}, {Mehl}, {Meyer}, {Millea}, {Mohr},
  {Montroy}, {Padin}, {Plagge}, {Pryke}, {Ruhl}, {Sayre}, {Schaffer}, {Shaw},
  {Shirokoff}, {Spieler}, {Staniszewski}, {Stark}, {van Engelen},
  {Vanderlinde}, {Vieira}, {Williamson}, \& {Zahn}}]{hou14}
{Hou}, Z., {et~al.} 2014, \apj, 782, 74

\bibitem[{{Hu}(2000)}]{hu00}
{Hu}, W. 2000, \apj, 529, 12

\bibitem[{{Kayo} {et~al.}(2013){Kayo}, {Takada}, \& {Jain}}]{kayo13}
{Kayo}, I., {Takada}, M., \& {Jain}, B. 2013, \mnras, 429, 344

\bibitem[{{Keisler} {et~al.}(2011){Keisler}, {Reichardt}, {Aird}, {Benson},
  {Bleem}, {Carlstrom}, {Chang}, {Cho}, {Crawford}, {Crites}, {de Haan},
  {Dobbs}, {Dudley}, {George}, {Halverson}, {Holder}, {Holzapfel}, {Hoover},
  {Hou}, {Hrubes}, {Joy}, {Knox}, {Lee}, {Leitch}, {Lueker}, {Luong-Van},
  {McMahon}, {Mehl}, {Meyer}, {Millea}, {Mohr}, {Montroy}, {Natoli}, {Padin},
  {Plagge}, {Pryke}, {Ruhl}, {Schaffer}, {Shaw}, {Shirokoff}, {Spieler},
  {Staniszewski}, {Stark}, {Story}, {van Engelen}, {Vanderlinde}, {Vieira},
  {Williamson}, \& {Zahn}}]{keisler11}
{Keisler}, R., {et~al.} 2011, \apj, 743, 28

\bibitem[{{Knox} {et~al.}(1998){Knox}, {Scoccimarro}, \& {Dodelson}}]{knox98}
{Knox}, L., {Scoccimarro}, R., \& {Dodelson}, S. 1998, Physical Review Letters,
  81, 2004

\bibitem[{{Komatsu} {et~al.}(2009){Komatsu}, {Dunkley}, {Nolta}, {Bennett},
  {Gold}, {Hinshaw}, {Jarosik}, {Larson}, {Limon}, {Page}, {Spergel},
  {Halpern}, {Hill}, {Kogut}, {Meyer}, {Tucker}, {Weiland}, {Wollack}, \&
  {Wright}}]{komatsu09}
{Komatsu}, E., {et~al.} 2009, \apjs, 180, 330

\bibitem[{{Komatsu} \& {Seljak}(2002)}]{komatsu02}
{Komatsu}, E., \& {Seljak}, U. 2002, \mnras, 336, 1256

\bibitem[{{Komatsu} {et~al.}(2011){Komatsu}, {Smith}, {Dunkley}, {Bennett},
  {Gold}, {Hinshaw}, {Jarosik}, {Larson}, {Nolta}, {Page}, {Spergel},
  {Halpern}, {Hill}, {Kogut}, {Limon}, {Meyer}, {Odegard}, {Tucker}, {Weiland},
  {Wollack}, \& {Wright}}]{komatsu11}
{Komatsu}, E., {et~al.} 2011, \apjs, 192, 18

\bibitem[{{Komatsu} \& {Spergel}(2001)}]{komatsu01}
{Komatsu}, E., \& {Spergel}, D.~N. 2001, \prd, 63, 063002

\bibitem[{{Lacasa}(2012)}]{lacasa12}
{Lacasa}, F. 2012, ArXiv e-prints, 1204.1480

\bibitem[{{Larson} {et~al.}(2011){Larson}, {Dunkley}, {Hinshaw}, {Komatsu},
  {Nolta}, {Bennett}, {Gold}, {Halpern}, {Hill}, {Jarosik}, {Kogut}, {Limon},
  {Meyer}, {Odegard}, {Page}, {Smith}, {Spergel}, {Tucker}, {Weiland},
  {Wollack}, \& {Wright}}]{larson11}
{Larson}, D., {et~al.} 2011, \apjs, 192, 16

\bibitem[{{Lueker} {et~al.}(2010){Lueker}, {Reichardt}, {Schaffer}, {Zahn},
  {Ade}, {Aird}, {Benson}, {Bleem}, {Carlstrom}, {Chang}, {Cho}, {Crawford},
  {Crites}, {de Haan}, {Dobbs}, {George}, {Hall}, {Halverson}, {Holder},
  {Holzapfel}, {Hrubes}, {Joy}, {Keisler}, {Knox}, {Lee}, {Leitch}, {McMahon},
  {Mehl}, {Meyer}, {Mohr}, {Montroy}, {Padin}, {Plagge}, {Pryke}, {Ruhl},
  {Shaw}, {Shirokoff}, {Spieler}, {Stalder}, {Staniszewski}, {Stark},
  {Vanderlinde}, {Vieira}, \& {Williamson}}]{lueker10}
{Lueker}, M., {et~al.} 2010, \apj, 719, 1045

\bibitem[{{Mocanu} {et~al.}(2013){Mocanu}, {Crawford}, {Vieira}, {Aird},
  {Aravena}, {Austermann}, {Benson}, {B{\'e}thermin}, {Bleem}, {Bothwell},
  {Carlstrom}, {Chang}, {Chapman}, {Cho}, {Crites}, {de Haan}, {Dobbs},
  {Everett}, {George}, {Halverson}, {Harrington}, {Hezaveh}, {Holder},
  {Holzapfel}, {Hoover}, {Hrubes}, {Keisler}, {Knox}, {Lee}, {Leitch},
  {Lueker}, {Luong-Van}, {Marrone}, {McMahon}, {Mehl}, {Meyer}, {Mohr},
  {Montroy}, {Natoli}, {Padin}, {Plagge}, {Pryke}, {Rest}, {Reichardt}, {Ruhl},
  {Sayre}, {Schaffer}, {Shirokoff}, {Spieler}, {Spilker}, {Stalder},
  {Staniszewski}, {Stark}, {Story}, {Switzer}, {Vanderlinde}, \&
  {Williamson}}]{mocanu13}
{Mocanu}, L.~M., {et~al.} 2013, \apj, 779, 61

\bibitem[{{Navarro} {et~al.}(1996){Navarro}, {Frenk}, \& {White}}]{navarro96}
{Navarro}, J.~F., {Frenk}, C.~S., \& {White}, S.~D.~M. 1996, \apj, 462, 563

\bibitem[{{Nozawa} {et~al.}(2000){Nozawa}, {Itoh}, {Kawana}, \&
  {Kohyama}}]{nozawa00}
{Nozawa}, S., {Itoh}, N., {Kawana}, Y., \& {Kohyama}, Y. 2000, \apj, 536, 31

\bibitem[{{Ostriker} \& {Vishniac}(1986)}]{ostriker86}
{Ostriker}, J.~P., \& {Vishniac}, E.~T. 1986, \apjl, 306, L51

\bibitem[{{Plagge} {et~al.}(2013){Plagge}, {Marrone}, {Abdulla}, {Bonamente},
  {Carlstrom}, {Gralla}, {Greer}, {Joy}, {Lamb}, {Leitch}, {Mantz}, {Muchovej},
  \& {Woody}}]{plagge13}
{Plagge}, T.~J., {et~al.} 2013, \apj, 770, 112

\bibitem[{{Planck Collaboration} {et~al.}(2012){Planck Collaboration}, {Ade},
  {Aghanim}, {Arnaud}, {Ashdown}, {Atrio-Barandela}, {Aumont}, {Baccigalupi},
  {Balbi}, {Banday}, \& et~al.}]{planck12-5}
{Planck Collaboration}, {et~al.} 2012, ArXiv e-prints, 1207.4061

\bibitem[{{Planck Collaboration} {et~al.}(2011){Planck Collaboration}, {Ade},
  {Aghanim}, {Arnaud}, {Ashdown}, {Aumont}, {Baccigalupi}, {Balbi}, {Banday},
  {Barreiro}, \& et~al.}]{planck11-5.1a}
------. 2011, \aap, 536, A8

\bibitem[{{Press} {et~al.}(1986){Press}, {Flannery}, {Teukolsky}, \&
  {Vetterling}}]{press86}
{Press}, W., {Flannery}, B., {Teukolsky}, S., \& {Vetterling}, W. 1986,
  Numerical Recipes: The Art of Scientific Computing (Cambridge University
  Press)

\bibitem[{{Reichardt} {et~al.}(2009){Reichardt}, {Ade}, {Bock}, {Bond},
  {Brevik}, {Contaldi}, {Daub}, {Dempsey}, {Goldstein}, {Holzapfel}, {Kuo},
  {Lange}, {Lueker}, {Newcomb}, {Peterson}, {Ruhl}, {Runyan}, \&
  {Staniszewski}}]{reichardt09a}
{Reichardt}, C.~L., {et~al.} 2009, \apj, 694, 1200

\bibitem[{{Reichardt} {et~al.}(2012){Reichardt}, {Shaw}, {Zahn}, {Aird},
  {Benson}, {Bleem}, {Carlstrom}, {Chang}, {Cho}, {Crawford}, {Crites}, {de
  Haan}, {Dobbs}, {Dudley}, {George}, {Halverson}, {Holder}, {Holzapfel},
  {Hoover}, {Hou}, {Hrubes}, {Joy}, {Keisler}, {Knox}, {Lee}, {Leitch},
  {Lueker}, {Luong-Van}, {McMahon}, {Mehl}, {Meyer}, {Millea}, {Mohr},
  {Montroy}, {Natoli}, {Padin}, {Plagge}, {Pryke}, {Ruhl}, {Schaffer},
  {Shirokoff}, {Spieler}, {Staniszewski}, {Stark}, {Story}, {van Engelen},
  {Vanderlinde}, {Vieira}, \& {Williamson}}]{reichardt12b}
------. 2012, \apj, 755, 70

\bibitem[{{Reichardt} {et~al.}(2013){Reichardt}, {Stalder}, {Bleem}, {Montroy},
  {Aird}, {Andersson}, {Armstrong}, {Ashby}, {Bautz}, {Bayliss}, {Bazin},
  {Benson}, {Brodwin}, {Carlstrom}, {Chang}, {Cho}, {Clocchiatti}, {Crawford},
  {Crites}, {de Haan}, {Desai}, {Dobbs}, {Dudley}, {Foley}, {Forman}, {George},
  {Gladders}, {Gonzalez}, {Halverson}, {Harrington}, {High}, {Holder},
  {Holzapfel}, {Hoover}, {Hrubes}, {Jones}, {Joy}, {Keisler}, {Knox}, {Lee},
  {Leitch}, {Liu}, {Lueker}, {Luong-Van}, {Mantz}, {Marrone}, {McDonald},
  {McMahon}, {Mehl}, {Meyer}, {Mocanu}, {Mohr}, {Murray}, {Natoli}, {Padin},
  {Plagge}, {Pryke}, {Rest}, {Ruel}, {Ruhl}, {Saliwanchik}, {Saro}, {Sayre},
  {Schaffer}, {Shaw}, {Shirokoff}, {Song}, {Spieler}, {Staniszewski}, {Stark},
  {Story}, {Stubbs}, {{\v S}uhada}, {van Engelen}, {Vanderlinde}, {Vieira},
  {Vikhlinin}, {Williamson}, {Zahn}, \& {Zenteno}}]{reichardt13}
------. 2013, \apj, 763, 127

\bibitem[{{Riess} {et~al.}(2011){Riess}, {Macri}, {Casertano}, {Lampeitl},
  {Ferguson}, {Filippenko}, {Jha}, {Li}, \& {Chornock}}]{riess11}
{Riess}, A.~G., {et~al.} 2011, \apj, 730, 119

\bibitem[{{Santos} {et~al.}(2002){Santos}, {Balbi}, {Borrill}, {Ferreira},
  {Hanany}, {Jaffe}, {Lee}, {Magueijo}, {Rabii}, {Richards}, {Smoot},
  {Stompor}, {Winant}, \& {Wu}}]{santos02}
{Santos}, M.~G., {et~al.} 2002, Physical Review Letters, 88, 241302

\bibitem[{{Sayers} {et~al.}(2013){Sayers}, {Czakon}, {Mantz}, {Golwala},
  {Ameglio}, {Downes}, {Koch}, {Lin}, {Maughan}, {Molnar}, {Moustakas},
  {Mroczkowski}, {Pierpaoli}, {Shitanishi}, {Siegel}, {Umetsu}, \& {Van der
  Pyl}}]{sayers13b}
{Sayers}, J., {et~al.} 2013, \apj, 768, 177

\bibitem[{{Schaffer} {et~al.}(2011){Schaffer}, {Crawford}, {Aird}, {Benson},
  {Bleem}, {Carlstrom}, {Chang}, {Cho}, {Crites}, {de Haan}, {Dobbs}, {George},
  {Halverson}, {Holder}, {Holzapfel}, {Hoover}, {Hrubes}, {Joy}, {Keisler},
  {Knox}, {Lee}, {Leitch}, {Lueker}, {Luong-Van}, {McMahon}, {Mehl}, {Meyer},
  {Mohr}, {Montroy}, {Padin}, {Plagge}, {Pryke}, {Reichardt}, {Ruhl},
  {Shirokoff}, {Spieler}, {Stalder}, {Staniszewski}, {Stark}, {Story},
  {Vanderlinde}, {Vieira}, \& {Williamson}}]{schaffer11}
{Schaffer}, K.~K., {et~al.} 2011, \apj, 743, 90

\bibitem[{{Sehgal} {et~al.}(2010){Sehgal}, {Bode}, {Das},
  {Hernandez-Monteagudo}, {Huffenberger}, {Lin}, {Ostriker}, \&
  {Trac}}]{sehgal10}
{Sehgal}, N., {Bode}, P., {Das}, S., {Hernandez-Monteagudo}, C.,
  {Huffenberger}, K., {Lin}, Y., {Ostriker}, J.~P., \& {Trac}, H. 2010, \apj,
  709, 920

\bibitem[{{Shaw} {et~al.}(2010){Shaw}, {Nagai}, {Bhattacharya}, \&
  {Lau}}]{shaw10}
{Shaw}, L.~D., {Nagai}, D., {Bhattacharya}, S., \& {Lau}, E.~T. 2010, \apj,
  725, 1452

\bibitem[{{Shaw} {et~al.}(2012){Shaw}, {Rudd}, \& {Nagai}}]{shaw12}
{Shaw}, L.~D., {Rudd}, D.~H., \& {Nagai}, D. 2012, \apj, 756, 15

\bibitem[{{Shaw} {et~al.}(2009){Shaw}, {Zahn}, {Holder}, \&
  {Dor{\'e}}}]{shaw09}
{Shaw}, L.~D., {Zahn}, O., {Holder}, G.~P., \& {Dor{\'e}}, O. 2009, \apj, 702,
  368

\bibitem[{{Shirokoff} {et~al.}(2011){Shirokoff}, {Reichardt}, {Shaw}, {Millea},
  {Ade}, {Aird}, {Benson}, {Bleem}, {Carlstrom}, {Chang}, {Cho}, {Crawford},
  {Crites}, {de Haan}, {Dobbs}, {Dudley}, {George}, {Halverson}, {Holder},
  {Holzapfel}, {Hrubes}, {Joy}, {Keisler}, {Knox}, {Lee}, {Leitch}, {Lueker},
  {Luong-Van}, {McMahon}, {Mehl}, {Meyer}, {Mohr}, {Montroy}, {Padin},
  {Plagge}, {Pryke}, {Ruhl}, {Schaffer}, {Spieler}, {Staniszewski}, {Stark},
  {Story}, {Vanderlinde}, {Vieira}, {Williamson}, \& {Zahn}}]{shirokoff11}
{Shirokoff}, E., {et~al.} 2011, \apj, 736, 61

\bibitem[{{Sievers} {et~al.}(2013){Sievers}, {Hlozek}, {Nolta}, {Acquaviva},
  {Addison}, {Ade}, {Aguirre}, {Amiri}, {Appel}, {Barrientos}, {Battistelli},
  {Battaglia}, {Bond}, {Brown}, {Burger}, {Calabrese}, {Chervenak}, {Crichton},
  {Das}, {Devlin}, {Dicker}, {Bertrand Doriese}, {Dunkley}, {D{\"u}nner},
  {Essinger-Hileman}, {Faber}, {Fisher}, {Fowler}, {Gallardo}, {Gordon},
  {Gralla}, {Hajian}, {Halpern}, {Hasselfield}, {Hern{\'a}ndez-Monteagudo},
  {Hill}, {Hilton}, {Hilton}, {Hincks}, {Holtz}, {Huffenberger}, {Hughes},
  {Hughes}, {Infante}, {Irwin}, {Jacobson}, {Johnstone}, {Baptiste Juin},
  {Kaul}, {Klein}, {Kosowsky}, {Lau}, {Limon}, {Lin}, {Louis}, {Lupton},
  {Marriage}, {Marsden}, {Martocci}, {Mauskopf}, {McLaren}, {Menanteau},
  {Moodley}, {Moseley}, {Netterfield}, {Niemack}, {Page}, {Page}, {Parker},
  {Partridge}, {Plimpton}, {Quintana}, {Reese}, {Reid}, {Rojas}, {Sehgal},
  {Sherwin}, {Schmitt}, {Spergel}, {Staggs}, {Stryzak}, {Swetz}, {Switzer},
  {Thornton}, {Trac}, {Tucker}, {Uehara}, {Visnjic}, {Warne}, {Wilson},
  {Wollack}, {Zhao}, \& {Zuncke}}]{sievers13}
{Sievers}, J.~L., {et~al.} 2013, ArXiv e-prints, 1301.0824

\bibitem[{{Smith} {et~al.}(2004){Smith}, {Rocha}, {Challinor}, {Battye},
  {Carreira}, {Cleary}, {Davies}, {Davis}, {Dickinson}, {Genova-Santos},
  {Grainge}, {Guti{\'e}rrez}, {Hafez}, {Hobson}, {Jones}, {Kneissl},
  {Lancaster}, {Lasenby}, {Leahy}, {Maisinger}, {Pooley}, {Rajguru}, {Rebolo},
  {Rubi{\~n}o-Martin}, {Sosa Molina}, {Saunders}, {Savage}, {Scott}, {Slosar},
  {Taylor}, {Titterington}, {Waldram}, \& {Watson}}]{smith04}
{Smith}, S., {et~al.} 2004, \mnras, 352, 887

\bibitem[{{Stanek} {et~al.}(2010){Stanek}, {Rasia}, {Evrard}, {Pearce}, \&
  {Gazzola}}]{stanek10}
{Stanek}, R., {Rasia}, E., {Evrard}, A.~E., {Pearce}, F., \& {Gazzola}, L.
  2010, \apj, 715, 1508

\bibitem[{{Story} {et~al.}(2013){Story}, {Reichardt}, {Hou}, {Keisler}, {Aird},
  {Benson}, {Bleem}, {Carlstrom}, {Chang}, {Cho}, {Crawford}, {Crites}, {de
  Haan}, {Dobbs}, {Dudley}, {Follin}, {George}, {Halverson}, {Holder},
  {Holzapfel}, {Hoover}, {Hrubes}, {Joy}, {Knox}, {Lee}, {Leitch}, {Lueker},
  {Luong-Van}, {McMahon}, {Mehl}, {Meyer}, {Millea}, {Mohr}, {Montroy},
  {Padin}, {Plagge}, {Pryke}, {Ruhl}, {Sayre}, {Schaffer}, {Shaw}, {Shirokoff},
  {Spieler}, {Staniszewski}, {Stark}, {van Engelen}, {Vanderlinde}, {Vieira},
  {Williamson}, \& {Zahn}}]{story13}
{Story}, K.~T., {et~al.} 2013, \apj, 779, 86

\bibitem[{Sunyaev \& Zel'dovich(1980)}]{sunyaev80}
Sunyaev, R., \& Zel'dovich, Y. 1980, ARAA, 18, 537

\bibitem[{{Sunyaev} \& {Zel'dovich}(1970)}]{sunyaev70}
{Sunyaev}, R.~A., \& {Zel'dovich}, Y.~B. 1970, Comments on Astrophysics and
  Space Physics, 2, 66

\bibitem[{{Tinker} {et~al.}(2008){Tinker}, {Kravtsov}, {Klypin}, {Abazajian},
  {Warren}, {Yepes}, {Gottl{\"o}ber}, \& {Holz}}]{tinker08}
{Tinker}, J., {Kravtsov}, A.~V., {Klypin}, A., {Abazajian}, K., {Warren}, M.,
  {Yepes}, G., {Gottl{\"o}ber}, S., \& {Holz}, D.~E. 2008, \apj, 688, 709

\bibitem[{{Tucci} {et~al.}(2011){Tucci}, {Toffolatti}, {de Zotti}, \&
  {Mart{\'{\i}}nez-Gonz{\'a}lez}}]{tucci11}
{Tucci}, M., {Toffolatti}, L., {de Zotti}, G., \&
  {Mart{\'{\i}}nez-Gonz{\'a}lez}, E. 2011, \aap, 533, A57

\bibitem[{{van Engelen} {et~al.}(2012){van Engelen}, {Keisler}, {Zahn}, {Aird},
  {Benson}, {Bleem}, {Carlstrom}, {Chang}, {Cho}, {Crawford}, {Crites}, {de
  Haan}, {Dobbs}, {Dudley}, {George}, {Halverson}, {Holder}, {Holzapfel},
  {Hoover}, {Hou}, {Hrubes}, {Joy}, {Knox}, {Lee}, {Leitch}, {Lueker},
  {Luong-Van}, {McMahon}, {Mehl}, {Meyer}, {Millea}, {Mohr}, {Montroy},
  {Natoli}, {Padin}, {Plagge}, {Pryke}, {Reichardt}, {Ruhl}, {Sayre},
  {Schaffer}, {Shaw}, {Shirokoff}, {Spieler}, {Staniszewski}, {Stark}, {Story},
  {Vanderlinde}, {Vieira}, \& {Williamson}}]{vanengelen12}
{van Engelen}, A., {et~al.} 2012, \apj, 756, 142

\bibitem[{{Vanderlinde} {et~al.}(2010){Vanderlinde}, {Crawford}, {de Haan},
  {Dudley}, {Shaw}, {Ade}, {Aird}, {Benson}, {Bleem}, {Brodwin}, {Carlstrom},
  {Chang}, {Crites}, {Desai}, {Dobbs}, {Foley}, {George}, {Gladders}, {Hall},
  {Halverson}, {High}, {Holder}, {Holzapfel}, {Hrubes}, {Joy}, {Keisler},
  {Knox}, {Lee}, {Leitch}, {Loehr}, {Lueker}, {Marrone}, {McMahon}, {Mehl},
  {Meyer}, {Mohr}, {Montroy}, {Ngeow}, {Padin}, {Plagge}, {Pryke}, {Reichardt},
  {Rest}, {Ruel}, {Ruhl}, {Schaffer}, {Shirokoff}, {Song}, {Spieler},
  {Stalder}, {Staniszewski}, {Stark}, {Stubbs}, {van Engelen}, {Vieira},
  {Williamson}, {Yang}, {Zahn}, \& {Zenteno}}]{vanderlinde10}
{Vanderlinde}, K., {et~al.} 2010, \apj, 722, 1180

\bibitem[{{Vieira} {et~al.}(2010){Vieira}, {Crawford}, {Switzer}, {Ade},
  {Aird}, {Ashby}, {Benson}, {Bleem}, {Brodwin}, {Carlstrom}, {Chang}, {Cho},
  {Crites}, {de Haan}, {Dobbs}, {Everett}, {George}, {Gladders}, {Hall},
  {Halverson}, {High}, {Holder}, {Holzapfel}, {Hrubes}, {Joy}, {Keisler},
  {Knox}, {Lee}, {Leitch}, {Lueker}, {Marrone}, {McIntyre}, {McMahon}, {Mehl},
  {Meyer}, {Mohr}, {Montroy}, {Padin}, {Plagge}, {Pryke}, {Reichardt}, {Ruhl},
  {Schaffer}, {Shaw}, {Shirokoff}, {Spieler}, {Stalder}, {Staniszewski},
  {Stark}, {Vanderlinde}, {Walsh}, {Williamson}, {Yang}, {Zahn}, \&
  {Zenteno}}]{vieira10}
{Vieira}, J.~D., {et~al.} 2010, \apj, 719, 763

\bibitem[{{Vieira} {et~al.}(2013){Vieira}, {Marrone}, {Chapman}, {De Breuck},
  {Hezaveh}, {Wei{$\beta$}}, {Aguirre}, {Aird}, {Aravena}, {Ashby}, {Bayliss},
  {Benson}, {Biggs}, {Bleem}, {Bock}, {Bothwell}, {Bradford}, {Brodwin},
  {Carlstrom}, {Chang}, {Crawford}, {Crites}, {de Haan}, {Dobbs}, {Fomalont},
  {Fassnacht}, {George}, {Gladders}, {Gonzalez}, {Greve}, {Gullberg},
  {Halverson}, {High}, {Holder}, {Holzapfel}, {Hoover}, {Hrubes}, {Hunter},
  {Keisler}, {Lee}, {Leitch}, {Lueker}, {Luong-van}, {Malkan}, {McIntyre},
  {McMahon}, {Mehl}, {Menten}, {Meyer}, {Mocanu}, {Murphy}, {Natoli}, {Padin},
  {Plagge}, {Reichardt}, {Rest}, {Ruel}, {Ruhl}, {Sharon}, {Schaffer}, {Shaw},
  {Shirokoff}, {Spilker}, {Stalder}, {Staniszewski}, {Stark}, {Story},
  {Vanderlinde}, {Welikala}, \& {Williamson}}]{vieira13}
------. 2013, \nat, 495, 344

\bibitem[{{Viero} {et~al.}(2012){Viero}, {Wang}, {Zemcov}, {Addison},
  {Amblard}, {Arumugam}, {Aussel}, {Bethermin}, {Bock}, {Boselli}, {Buat},
  {Burgarella}, {Casey}, {Clements}, {Conley}, {Conversi}, {Cooray}, {De
  Zotti}, {Dowell}, {Farrah}, {Franceschini}, {Glenn}, {Griffin},
  {Hatziminaoglou}, {Heinis}, {Ibar}, {Ivision}, {Lagache}, {Levenson},
  {Marchetti}, {Marsden}, {Nguyen}, {O'Halloran}, {Oliver}, {Omont}, {Page},
  {Papageorgiou}, {Peason}, {Perez-Fournon}, {Pohlen}, {Rigopoulou},
  {Roseboom}, {Rowan-Robinson}, {Scott}, {Seymour}, {Schulz}, {Shupe}, {Smith},
  {Symeonidis}, {Vaccari}, {Valtchanov}, {Vieira}, {Wardlow}, \&
  {Xu}}]{viero12b}
{Viero}, M.~P., {et~al.} 2012, ArXiv e-prints, 1208.5049

\bibitem[{{Wang} \& {Steinhardt}(1998)}]{wang98}
{Wang}, L., \& {Steinhardt}, P.~J. 1998, \apj, 508, 483

\bibitem[{{Williamson} {et~al.}(2011){Williamson}, {Benson}, {High},
  {Vanderlinde}, {Ade}, {Aird}, {Andersson}, {Armstrong}, {Ashby}, {Bautz},
  {Bazin}, {Bertin}, {Bleem}, {Bonamente}, {Brodwin}, {Carlstrom}, {Chang},
  {Chapman}, {Clocchiatti}, {Crawford}, {Crites}, {de Haan}, {Desai}, {Dobbs},
  {Dudley}, {Fazio}, {Foley}, {Forman}, {Garmire}, {George}, {Gladders},
  {Gonzalez}, {Halverson}, {Holder}, {Holzapfel}, {Hoover}, {Hrubes}, {Jones},
  {Joy}, {Keisler}, {Knox}, {Lee}, {Leitch}, {Lueker}, {Luong-Van}, {Marrone},
  {McMahon}, {Mehl}, {Meyer}, {Mohr}, {Montroy}, {Murray}, {Padin}, {Plagge},
  {Pryke}, {Reichardt}, {Rest}, {Ruel}, {Ruhl}, {Saliwanchik}, {Saro},
  {Schaffer}, {Shaw}, {Shirokoff}, {Song}, {Spieler}, {Stalder}, {Stanford},
  {Staniszewski}, {Stark}, {Story}, {Stubbs}, {Vieira}, {Vikhlinin}, \&
  {Zenteno}}]{williamson11}
{Williamson}, R., {et~al.} 2011, \apj, 738, 139

\bibitem[{{Wilson} {et~al.}(2012){Wilson}, {Sherwin}, {Hill}, {Addison},
  {Battaglia}, {Bond}, {Das}, {Devlin}, {Dunkley}, {D{\"u}nner}, {Fowler},
  {Gralla}, {Hajian}, {Halpern}, {Hilton}, {Hincks}, {Hlozek}, {Huffenberger},
  {Hughes}, {Kosowsky}, {Louis}, {Marriage}, {Marsden}, {Menanteau}, {Moodley},
  {Niemack}, {Nolta}, {Page}, {Partridge}, {Reese}, {Sehgal}, {Sievers},
  {Spergel}, {Staggs}, {Swetz}, {Switzer}, {Trac}, \& {Wollack}}]{wilson12}
{Wilson}, M.~J., {et~al.} 2012, \prd, 86, 122005

\bibitem[{{Yadav} {et~al.}(2007){Yadav}, {Komatsu}, \& {Wandelt}}]{yadav07}
{Yadav}, A.~P.~S., {Komatsu}, E., \& {Wandelt}, B.~D. 2007, \apj, 664, 680

\bibitem[{{Yadav} \& {Wandelt}(2008)}]{yadav08}
{Yadav}, A.~P.~S., \& {Wandelt}, B.~D. 2008, Physical Review Letters, 100,
  181301

\bibitem[{{Yadav} \& {Wandelt}(2010)}]{yadav10}
------. 2010, Advances in Astronomy, 2010, 181301

\bibitem[{{Zahn} {et~al.}(2012){Zahn}, {Reichardt}, {Shaw}, {Lidz}, {Aird},
  {Benson}, {Bleem}, {Carlstrom}, {Chang}, {Cho}, {Crawford}, {Crites}, {de
  Haan}, {Dobbs}, {Dor{\'e}}, {Dudley}, {George}, {Halverson}, {Holder},
  {Holzapfel}, {Hoover}, {Hou}, {Hrubes}, {Joy}, {Keisler}, {Knox}, {Lee},
  {Leitch}, {Lueker}, {Luong-Van}, {McMahon}, {Mehl}, {Meyer}, {Millea},
  {Mohr}, {Montroy}, {Natoli}, {Padin}, {Plagge}, {Pryke}, {Ruhl}, {Schaffer},
  {Shirokoff}, {Spieler}, {Staniszewski}, {Stark}, {Story}, {van Engelen},
  {Vanderlinde}, {Vieira}, \& {Williamson}}]{zahn12}
{Zahn}, O., {et~al.} 2012, \apj, 756, 65

\end{thebibliography}
